\documentclass[11pt,a4paper]{article}
\usepackage[margin=25mm]{geometry}
\usepackage{amsmath,amssymb,amsfonts,amsthm}
\usepackage{bm}
\usepackage{color}
\usepackage{hyperref}
\usepackage{float}
\usepackage{booktabs,tabularx}
\usepackage{caption}
\usepackage{graphicx}
\usepackage{tikz}
\usepackage{circuitikz}
\usepackage{array}
\usepackage{mathrsfs}

\hypersetup{colorlinks=true,linkcolor=red,citecolor=blue,urlcolor=cyan}

\newcommand{\aver}[1]{\langle{#1}\rangle}
\newcommand{\be}{\begin{equation}}
\newcommand{\ee}{\end{equation}}
\newcommand{\FP}{\mbox{FP}}
\newcommand{\DeltaV}{\Delta}
\newcommand{\Tr}{\mbox{Tr}}
\newcommand{\PV}{\mbox{PV}}
\newcommand{\Pf}{\mbox{Pf}}
\newcommand{\Ai}{\mbox{Ai}}
\newcommand{\Bi}{\mbox{Bi}}
\newcommand{\Imn}{\mbox{Im}}
\newcommand{\Rea}{\mbox{Re}}
\newcommand{\sech}{\mbox{sech}}
\newcommand{\cF}{\mathcal F}
\newcommand{\cJ}{\mathcal J}
\newcommand{\cA}{\mathcal A}
\newcommand{\cH}{\mathcal H}

\newcommand{\cS}{\mathcal S}
\newcommand{\cZ}{\mathcal Z}
\newcommand{\Zh}{\mathsf Z}
\newcommand{\cE}{\mathcal E}

\newcommand{\cT}{\mathcal T}
\newcommand{\cC}{\mathfrak C}

\newcommand{\fs}{\mathfrak s}
\newcommand{\dd}{\mathrm d}

\newcommand{\ii}{\mathrm i}
\newcommand{\R}{\mathbb R}
\newcommand{\C}{\mathbb C}
\newcommand{\Hq}{\mathbb H}
\newcommand{\F}{\mathbb F}
\newcommand{\bml}{\bm\lambda}

\newcommand{\Herm}{\mathrm{Herm}}

\newtheorem{proposition}{Proposition}
\newtheorem{lemma}{Lemma}

\renewcommand{\thesubsection}{\arabic{section}\Alph{subsection}}

\makeatletter
\renewcommand{\p@subsection}{}
\makeatother

\begin{document}

\title{The $N$-channel white-noise Schr\"odinger operator
in all three Dyson classes.\\
I. Explicit stationary admittance distribution,
density of states and outgoing Kontsevich matrix Airy integrals}

\author{Yan V. Fyodorov\\[0.5ex]
\small Department of Mathematics, King's College London\\
\small Strand, London WC2R 2LS, United Kingdom}

\date{ \today}

\maketitle

\begin{abstract}
We obtain an explicit stationary distribution of the
matrix Riccati variable---the boundary admittance---for
an $N$-channel Schr\"odinger operator with Gaussian
white-noise disorder, in all three Dyson symmetry
classes $\beta=1,2,4$.
Uniform absorption regularizes the Riccati dynamics,
and the distribution's normalization factorizes into
outgoing Kontsevich matrix Airy integrals.
This gives a common route to the density of states
and the sum of Lyapunov exponents, extending
Halperin's scalar solution to coupled channels.

Exact finite-$N$ formulas connect two spectral regimes:
the Lifshitz tail, including its symmetry-dependent
prefactor, and the collective large-$N$ edge.
We reduce the $\beta=4$ and even-$N$ $\beta=1$ problems
to unitary orthogonal polynomials and an auxiliary
recursion. Their critical double-scaling limit yields edge profiles
on the $N^{-4/5}$ energy scale, governed by a complex
Painlev\'e-I tritronqu\'ee solution and a linear
auxiliary equation. We conjecture that these profiles
describe a universal edge regime of random band matrices.
\end{abstract}
\tableofcontents

\section{Introduction and summary}

The one-channel white-noise Schr\"odinger operator is the textbook example of a disordered
spectral problem allowing a very detailed statistical analysis for arbitrary disorder beyond perturbation theory. In particular, Frisch and Lloyd and then Halperin obtained its eigenvalue density in the infinite-length limit through Airy functions~\cite{FrischLLoyd1960,Halperin1965}, and Thouless~\cite{Thouless1972}  related that density to the Lyapunov exponent characterizing the localization length in the system. These two quantities are the real and
imaginary parts of a single ``complex characteristic exponent'', see e.g. \cite{ComtetTourigny2017}. Many early results have been  reviewed in the book \cite{LifshitsGredeskulPastur},  with subsequent work providing a detailed characterisation of eigenfunction statistics 
 \cite{Kolokolov1993,IvanovOstrovskySkvortsov2009,IvanovSkvortsovOstrovskyFominov2012}. Complementary mathematical work has revealed a rich rigorous probabilistic picture of localized eigenfunctions and spectral fluctuations ~\cite{GoldsheidMolchanovPastur1977,DumazLabbe2020Localization,DumazLabbe2023Delocalized,DumazLabbe2024WhiteNoise,DumazLabbe2024Crossover}.

Physically, strictly one-dimensional systems are highly idealised models, raising the natural question: how much of the solvability survives for an $N$-channel, matrix-valued microscopic Schr\"odinger operator which describes with growing $N$ the gradual transition to quasi-one-dimensional rather than truly one-dimensional localization? Investigating this problem as fully as possible remains not only a challenging theoretical undertaking, but is also motivated by continuing experimental relevance of systems displaying multichannel localization, see e.g.  \cite{ShiGenack2012,Pena_etal_2014,DavyShiParkTianGenack2015, DavyShiWangChengGenack2015, HuangMaGenackGopar2021}

In the recent paper Ref.~\cite{Letter} it was shown that in the unitary class $\beta=2$  the $N$-channel problem in many aspects remains exactly integrable for any length $L$ of the sample. The method of solution exploited the matrix Riccati variable and discovered that the associated Dyson gas of Riccati eigenvalues can be exactly fermionized. In the limit $L\to \infty$ the fermionization produced the analogue of complex characteristic exponent exactly for every $N$, through an Airy Hankel $\tau$-function, which has been demonstrated to be equivalent to an outgoing-contour version of the Kontsevich
matrix Airy integral~\cite{Kontsevich1992} (the reason for calling the contour ``outgoing'' will be justified below). Such a representation turned out to be very powerful for analysing the  associated Lifshitz tails 
and further helped to identify the large-N eigenvalue density edge profile conjectured to be universal across a class of $\beta=2$ quasi-one-dimensional systems including random band matrices.  

The mapping to Kontsevich integral through fermionization relied on cancellations specific to $\beta=2$ and not available for the other two standard symmetry classes $\beta=1,4$.  The aim of the present paper is to suggest an alternative method which does not depend on fermionization and hence is equally applicable to all three classes, with the price of being implementable only in the limit of infinite sample length.
 The organizing object is the {\it full stationary distribution} of
the Riccati matrix at {\it complex energy}. The Kontsevich integral then appears as the holomorphic factor of its
normalization, most useful in the limit of real energy.

Complex energy is not just a mathematical device:
$E+\ii\eta$ describes a quasi-one-dimensional wire
with uniform absorption. Physically, the Riccati matrix is the admittance matrix of a semi-infinite absorbing wire, whose probability distribution is given by our stationary law. The underlying geometry is a matrix analogue of
the hyperbolic half-plane, expressed in terms of
a Siegel domain.
The connections with waveguide scattering, hyperbolic diffusion
and transport theory are discussed in
Appendix~\ref{app:admittance-scattering}.

In the present paper, the first in a planned series,
we concentrate mainly on the spectral characteristics of $N$-channel system of infinite length, namely its normalization factor
and holomorphic sector, which yields the density
of states and the sum of Lyapunov exponents for all classical values $\beta=1,2,4$.
Scattering applications of the full distribution are deferred to
Paper~II~\cite{PaperII}. Appendix~\ref{app:admittance-scattering}
collects the relevant physical interpretation and boundary identities,
and outlines that programme.

\subsection{Outline of the main results}

Let $\cE=E+\ii\eta$ with $\eta>0$ and $\F=\R,\C,\Hq$ for $\beta=1,2,4$. We consider $\psi:[0,L]\to\F^N$ and the (quasi-)one-dimensional
multichannel Schr\"odinger operator with Dirichlet boundary conditions:
\be
\cH_L=-\partial_x^2\,I_N+W(x),\qquad
\cH_L\psi=E\psi,\quad \psi(0)=\psi(L)=0,
\label{eq:H}
\ee
where $I_N$ is the $N\times N$ identity matrix and
$W(x)=W(x)^\dagger$ is real symmetric, complex Hermitian or
quaternion self-dual Gaussian white noise with the weight
defined in \eqref{eq:noise}. Its variance is controlled by
the parameter $T=J^2/(2N)>0$.

Let $\Psi(x)$ be the $N\times N$ ($\F$-valued) fundamental solution of the $N$-channel Schr\"odinger equation:  $-\Psi''+W\Psi=E\Psi$ with
$\Psi(0)=0$, $\Psi'(0)=I_N$. The Riccati matrix $Z=\Psi'\Psi^{-1}$ obeys
\be
Z'=-E-Z^2+W(x).
\label{eq:Riccati}
\ee

Write the Riccati matrix as $Z=A-\ii P$ with $A,P$ Hermitian of class $\beta$, and set $Y=\eta P^{-1}>0$.

\paragraph*{1. The stationary joint probability density (Sec.~\ref{sec:mechanism}).}
In the limit of a semi-infinite sample, $L\to\infty$, the pair of matrices $(A,Y)$ acquires a
stationary joint probability density, independent of the initial conditions. It is given
explicitly by 
\be
\begin{gathered}
\mathscr P_{E,\eta}(A,Y)=\frac{1}{\cZ_N(E,\eta)}\,
\exp\Big\{\frac1T\Big[\Tr U_E(A-Y)-\Tr U_E(A)-\eta^2\Tr Y^{-1}\Big]\Big\},\\
U_E(\lambda)=E\lambda+\frac{\lambda^3}3,
\end{gathered}
\label{eq:law}
\ee
defined for $A\in\Herm_N(\F)$ and $Y>0$. The normalization is fixed by
$\int\mathscr P_{E,\eta}(A,Y)\,\dd A\,\dd Y=1$, where $\dd A$ and $\dd Y$ are the flat
measures, i.e.\ products of the differentials of all independent real components of $A$ and $Y$. For $N=1$ it reduces to the explicit
scalar density of Sec.~\ref{sec:N1}.

\paragraph*{2. Factorization (Sec.~\ref{sec:kontsevich}).} Let $s=-E/T^{2/3}$ and $\fs=-\cE/T^{2/3}$, and let $\mathcal C_-$ be the
{\it outgoing} (see the explanation after \eqref{eq:ZN}) Airy contour, running from $e^{\ii\pi/3}\infty$ to
$e^{\ii\pi}\infty$, which produces $\cA_-=\Ai-\ii\Bi$. Define
\be
\tau_N^{(\beta)}(s)=\int_{w_1\prec w_2\prec\dots\prec w_N\in\mathcal C_-}
\ \prod_{i<j}(w_j-w_i)^\beta\ \prod_{j=1}^N e^{\frac{w_j^3}{3}-sw_j}\,\dd w_j ,
\label{eq:tau}
\ee
where $\prec$ denotes the order along the contour. 

A contour deformation, justified by the physical argument in
Sec.~\ref{sec:kontsevich}, gives 
\be
\cZ_N(E,\eta)=C_{N,\beta}\,T^{\kappa_{N,\beta}}\,\big|\tau_N^{(\beta)}(\fs)\big|^2 .
\label{eq:factor}
\ee
Here $C_{N,\beta}$ and $\kappa_{N,\beta}$ are constants. In words, the normalization of the stationary distribution~\eqref{eq:law}
is proportional to the squared modulus of the outgoing Kontsevich-type
integral with Dyson index $\beta$, defined in~\eqref{eq:tau}. Note that although we use it only for $\beta=1,2,4$, 
the definition makes sense for every $\beta>0$: for ordered points $w_j-w_i\neq 0$,
and we take the continuous branch of $(w_j-w_i)^\beta$ on the ordered configuration
space (on the deformed contour $e^{\ii\pi/3}R\to0\to-R$ one has
$\arg(w_j-w_i)\in[\pi,\tfrac{4\pi}3]$). The remaining global phase drops out of
$|\tau_N^{(\beta)}|^2$ and of $\partial_s\log\tau_N^{(\beta)}$.

\paragraph*{3. Holomorphic statistics.}
At fixed $Y$, the cubic terms in $A$ cancel in the exponent
of~\eqref{eq:law}, leaving a Gaussian density in $A$.
Averaging $e^{-\ii k\Tr Z}$, with $Z=A-\ii\eta Y^{-1}$ and
$k=q/T^{1/3}\ge0$, over this Gaussian distribution at fixed $Y$ leaves an
integral over $Y$ of exactly the same form as the normalization
integral $\cZ_N(E,\eta)$ as in \eqref{eq:factor}, but with shifted parameters
\[
E'=E+\frac{\ii kT}{2},
\qquad
\eta'=\eta+\frac{kT}{2},
\]
so that
$\big\langle e^{-\ii k\Tr Z}\big\rangle_{E,\eta}
=\cZ_N(E',\eta')/\cZ_N(E,\eta)$,
where $\langle\cdot\rangle_{E,\eta}$ denotes averaging
with respect to the stationary density~\eqref{eq:law}.
Here the complex value of $E'$ denotes analytic continuation of
the normalization integral, not a change of the physical real
energy. To apply the factorization~\eqref{eq:factor}, write it as
\[
\cZ_N(E,\eta)\propto
\tau_N^{(\beta)}\!\left(-\frac{E+\ii\eta}{T^{2/3}}\right)
\tau_N^{(\beta)\sharp}\!\left(-\frac{E-\ii\eta}{T^{2/3}}\right),
\qquad
\tau_N^{(\beta)\sharp}(z):=
\overline{\tau_N^{(\beta)}(\bar z)}.
\]
For real $E$ and $\eta$ this product is
$|\tau_N^{(\beta)}(\fs)|^2$; for complex $E'$ it is understood
as the product of the two analytically continued factors.
Since
\[
E'+\ii\eta'=E+\ii\eta+\ii kT,
\qquad
E'-\ii\eta'=E-\ii\eta,
\]
the argument of the first factor shifts from $\fs$ to $\fs-\ii q$,
while the second factor remains unchanged and cancels in the ratio.
Hence (Sec.~\ref{sec:kontsevich})
\be
\big\langle e^{-\ii q\Tr Z/T^{1/3}}\big\rangle_{E,\eta}
=\frac{\tau_N^{(\beta)}(\fs-\ii q)}
       {\tau_N^{(\beta)}(\fs)},
\qquad q\ge0 .
\label{eq:charfun}
\ee

\paragraph*{4. Real energy (Sec.~\ref{sec:realE}).} As $\eta\to0^+$, for every $N\ge1$ and
$\beta\in\{1,2,4\}$,
\be
\Xi_N^{(\beta)}(E):=\Gamma_N(E)-\ii\pi\nu_N(E)=T^{1/3}\,\partial_s\log\tau_N^{(\beta)}(s)\Big|_{s=-E/T^{2/3}},\qquad
\rho_N(E)=\frac{1}{\pi T^{1/3}}\,\Imn\partial_s^2\log\tau_N^{(\beta)}(s).
\label{eq:main}
\ee
Here $\nu_N$ is the integrated density of states (IDOS) per unit length
of the multichannel Schr\"odinger operator~\eqref{eq:H},
$\rho_N=\nu_N'$ is the density of states, and
$\Gamma_N=\sum_{j=1}^N\gamma_j$ is the sum of the positive
Lyapunov exponents.

In the zero-absorption limit, the stationary joint probability
density takes the form of a constrained two-matrix model with weight
$e^{-\Tr U_E(\Lambda)/T}e^{+\Tr U_E(M)/T}$ and
$\Lambda-M>0$. Further integrating out $M$ yields a matrix generalization of
Halperin's stationary Riccati density; the IDOS $\nu_N$ and the sum of
Lyapunov exponents $\Gamma_N$ admit then explicit representations
as ratios of positive integrals, Eqs.~\eqref{eq:nupos} and~\eqref{eq:Gammapos}.

Rescaling $w_j=-z_j/T^{1/3}$ turns Eq.~\eqref{eq:tau} into
the partition function of a $\beta$-ensemble in the cubic
potential $U_E$, with all integration variables on the
{\it outgoing contour}
$\Gamma_-:e^{-2\pi\ii/3}\infty\to+\infty$:
\be
\begin{gathered}
\Zh_N^{(\beta)}(E)
=
\int_{\Gamma_-^N,\ \rm ordered}
\prod_{i<j}(z_j-z_i)^\beta
\prod_{j=1}^N e^{-U_E(z_j)/T}\,\dd z_j
\ \propto\
\tau_N^{(\beta)}\big(-E/T^{2/3}\big),
\\
\Xi_N^{(\beta)}(E)
=
-T\,\partial_E\log\Zh_N^{(\beta)}(E).
\end{gathered}
\label{eq:ZN}
\ee
The Riccati observables are thus obtained from the energy
derivative of the free energy of a cubic $\beta$-ensemble:
its formal Gibbs weight is not normalizable on the real axis,
but its partition function becomes convergent on the outgoing
contour. The positive physical normalization is proportional
to $|\tau_N^{(\beta)}|^2$, with $\tau_N^{(\beta)}$ as its
holomorphic factor.

\begin{figure}[t]
\centering
\begin{tikzpicture}[scale=1.1, >=stealth]
  \foreach \ang in {0,120,240} {
    \fill[gray!15]
      (0,0) -- (\ang-30:2.6)
      arc[start angle=\ang-30,
          end angle=\ang+30,
          radius=2.6] -- cycle;
  }

  \foreach \ang in {-150,-90,-30,30,90,150} {
    \draw[gray!60,dashed] (0,0) -- (\ang:2.6);
  }

  \draw[->] (-2.9,0) -- (3.0,0)
    node[right] {$\Re\lambda$};
  \draw[->] (0,-2.8) -- (0,2.9)
    node[above] {$\Im\lambda$};

  \draw[thick]
    (-1.3,-2.25) -- (-0.65,-1.125);
  \draw[thick,->]
    (-0.65,-1.125)
    .. controls (-0.15,-0.28) and (0.35,-0.12)
    .. (1.0,-0.14);
  \draw[thick,->]
    (1.0,-0.14) -- (2.65,-0.37);

  \node[below left] at (-1.3,-2.25)
    {$\arg\lambda=-2\pi/3$};
  \node[below] at (2.05,-0.45)
    {$\arg\lambda=-\varepsilon$};
  \node at (0.65,-0.75) {$\Gamma_-$};
  \node[above right] at (0,0) {$0$};
\end{tikzpicture}
\caption{Decay sectors of $e^{-\lambda^3/(3T)}$
(shaded) and a schematic outgoing contour $\Gamma_-$.
The contour runs from infinity along
$\arg\lambda=-2\pi/3$ to infinity just below
the positive real axis.}
\label{fig:outgoing-contour}
\end{figure}
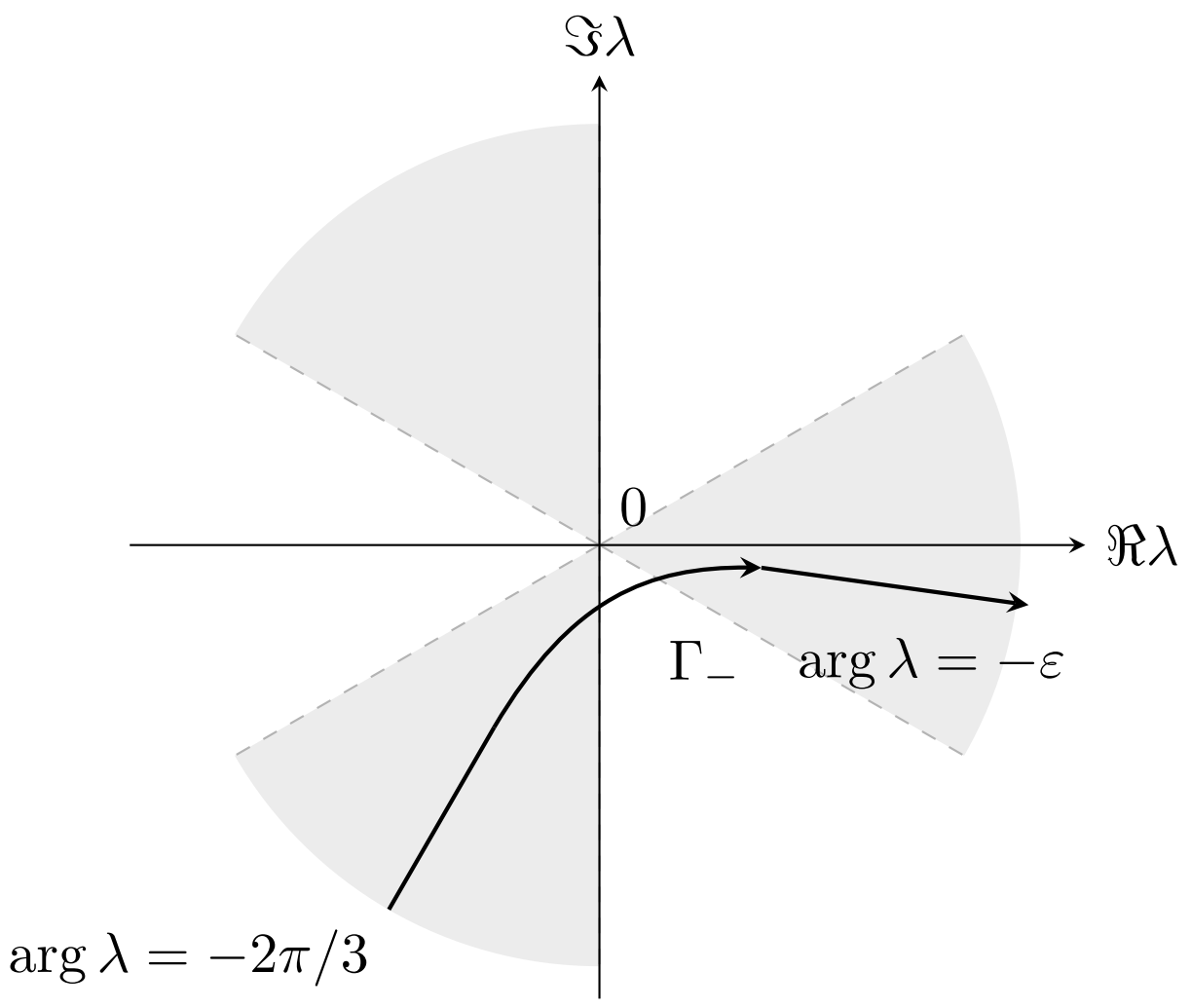

\paragraph*{Why ``outgoing''.}
The name can be understood in several equivalent ways.
In terms of waves, the cubic-exponential integral along $\Gamma_-$,
after the rescaling $w=-z/T^{1/3}$, is proportional to the Airy
combination $\cA_-=\Ai-\ii\Bi$. Whereas $\Ai$ and $\Bi$ are standing waves on the
oscillatory side, $\Ai\mp\ii\Bi$ are the two travelling waves, the Airy
analogues of the Hankel functions $H^{(1,2)}=J\pm\ii Y$; for
$x\to+\infty$,
$\cA_-(-x)\simeq-\ii\pi^{-1/2}x^{-1/4}e^{\ii(\frac23x^{3/2}+\frac\pi4)}$.
Physically, the choice is made by absorption: at $E+\ii\eta$ with
$\eta>0$, the wave entering the semi-infinite wire from its end is
absorbed inside and never comes back. This is the radiation condition of
scattering theory, i.e.\ the retarded prescription $E+\ii0$. The
conjugate contour $\overline{\Gamma_-}$ gives the incoming combination
$\cA_+=\Ai+\ii\Bi$ and the advanced prescription $E-\ii0$. In terms of
observables, the two choices differ in the sign of the imaginary part of
$\Xi_N=\Gamma_N-\ii\pi\nu_N$. Only the outgoing one gives a positive
integrated density of states, i.e.\ a winding current of Riccati
eigenvalues running in the physical direction, from $+\infty$ down to
$-\infty$. Finally, from the point of view of the matrix integral, the
cubic Gibbs weight $e^{-\Tr U_E/T}$ cannot be normalized on the real
axis, and among the contours on which it can be, $\Gamma_-$ is the one
that runs out to infinity along the positive real axis from below, in
the lower half-plane where $Z$ lives; see Fig.~\ref{fig:outgoing-contour}. Its complex conjugate supplies the
second factor of the positive normalization $|\tau_N|^2$.

\vspace{0.3cm}

The tables below provides a compact summary of content of the most essential formulas of the paper.
\begin{table}[H]
\centering
\caption{Summary. Here $S=\frac{4(-E)^{3/2}}{3T}$ is the Halperin action,
$K_\beta(N)=\frac{3N}{4}[4\beta^2N^2-9\beta^2N+5\beta^2+18\beta N-18\beta+20]$ and
$c_N^{(\beta)}=-\frac{21\beta^2N^2-60\beta^2N+54\beta N+39\beta^2-54\beta+20}{96}$.}
\label{tab:summary}
\begingroup
\footnotesize
\setlength{\tabcolsep}{3pt}
\renewcommand{\arraystretch}{1.7}
\begin{tabularx}{\linewidth}{@{}>{\raggedright\arraybackslash}p{0.20\linewidth}*{3}{>{\raggedright\arraybackslash}X}@{}}
\toprule
 & $\beta=1$ & $\beta=2$ & $\beta=4$\\ \midrule
$\tau_N$ explicitly & Pfaffian of double Airy integrals;\newline even $N$: unitary data at $T/2$, App.~\ref{app:beta1reduction} & Hankel $\det[\cA_-^{(j+k)}]$ & $\Pf[(k-j)\cA_-^{(j+k-1)}]$\newline $(2N\times2N)$; unitary data, Sec.~\ref{sec:b4exact}\\
$\Gamma_N$, $E\gg T^{2/3}$ & $\frac{N(N+1)T}{8E}$ & $\frac{N^2T}{4E}$ & $\frac{N(2N-1)T}{4E}$\\
$\frac{2\pi\sqrt E\rho_N}{N}-1$, $E\gg T^{2/3}$ & $-\frac{5(4N^2+9N+7)}{128}\frac{T^2}{E^3}$ & $-\frac{5(4N^2+1)}{32}\frac{T^2}{E^3}$ & $-\frac{5(16N^2-18N+7)}{32}\frac{T^2}{E^3}$\\
\bottomrule
\end{tabularx}
\endgroup
\end{table}

\begin{table}[H]
\centering
\caption{Lifshitz tails, large-$N$ spectral edges, and
the conjectured correspondence with random band matrices
(Appendix~\ref{app:RBM1} and Ref.~\cite{Letter}). Here
$\xi=(2-\lambda)W^{4/5}$, $\sigma_\beta(\xi)$ is the cumulative edge
intensity of band matrices of bandwidth $W$, and $\delta\sigma_\beta$ is
$\sigma_\beta$ with its allowed-side terms of the first band row removed.
The $\beta=1$ entries agree with the band-matrix diagram expansion, the
$\beta=4$ entries are predictions.}
\begingroup
\footnotesize
\setlength{\tabcolsep}{3pt}
\renewcommand{\arraystretch}{1.7}
\begin{tabularx}{\linewidth}{@{}>{\raggedright\arraybackslash}p{0.20\linewidth}*{3}{>{\raggedright\arraybackslash}X}@{}}
\toprule
 & $\beta=1$ & $\beta=2$ & $\beta=4$\\ \midrule
Lifshitz prefactor $\nu_N/\nu_1$ & $\frac{\sqrt\pi}{\Gamma(N/2)}(6S)^{\frac{N-1}2}$ & $\frac{1}{(N-1)!}(6S)^{N-1}$ & $\frac{1}{(2N-1)!}(6S)^{2(N-1)}$\\
Lifshitz correction $c_N^{(\beta)}$ & $-\frac{21N^2-6N+5}{96}$ & $-\frac{21N^2-33N+17}{24}$ & $-\frac{84N^2-186N+107}{24}$\\
Large-$N$ edge & \multicolumn{3}{l}{$E^*_\beta=-\frac34(\beta J^2)^{2/3}$, window $N^{-4/5}$}\\
Edge profile & P$_{\rm I}$ and $q''=\frac34yq$;\newline $q$ decaying at $t\to+\infty$ (even $N$) & P$_{\rm I}$ tritronqu\'ee~\eqref{eq:yphys} & P$_{\rm I}$ and $q''=\frac34yq$;\newline $q$ decaying at $t\to-\infty$\\
Band edge, $\xi\to+\infty$:\newline $\sigma_\beta-\frac{2}{3\pi}\xi^{3/2}$ & $+\frac{\sqrt6}{2\pi}\,\xi^{1/4}+O(\xi^{-9/4})$ & $O(\xi^{-7/2})$ & $-\frac{\sqrt6}{4\pi}\,\xi^{1/4}+O(\xi^{-9/4})$\\
Band edge moment\newline $\int\delta\sigma_\beta\,\dd\xi$ & $\frac{13}{48}$ & $\frac{1}{48}$ & $\frac{13}{192}$\\
Band edge tail, $\xi\to-\infty$:\newline $-\log\sigma_\beta/|\xi|^{5/4}$ & $\frac{4\sqrt3}{5}$ & $\frac{8\sqrt3}{5}$ & $\frac{16\sqrt3}{5}$\\
\bottomrule
\end{tabularx}
\endgroup
\end{table}

In the deep tail,
\be
\nu^{(\beta)}_N(E)\simeq\frac{N\,\Gamma(1+\beta/2)}{\pi\,\Gamma(1+N\beta/2)}\,(-E)^{1/2}
\Big(\frac{8(-E)^{3/2}}T\Big)^{\frac{\beta(N-1)}2}e^{-\frac{4(-E)^{3/2}}{3T}}
\Big[1+c_N^{(\beta)}\frac{T}{(-E)^{3/2}}+\dots\Big].
\label{eq:Lif}
\ee
The leading exponential factor is independent of $\beta$ and
of $N$ at fixed $T$. The reason is that the optimal rare fluctuation (instanton)
producing a deeply bound state has the rank-one form
$W_*(x)=w_*(x)uu^\dagger$, where $u$ is a unit vector
in channel space. The bound-state wavefunction is proportional
to $u$, so the binding problem reduces to the scalar one
and has the same optimal-fluctuation action.

The dependence on $N$ and $\beta$ enters the leading prefactor
through the freedom to orient $u$ in channel space and through
fluctuations about the optimal potential. Since $uu^\dagger$
is unchanged by multiplying $u$ by a sign, a complex phase,
or a unit quaternion, the distinct orientations form
$\mathbb{RP}^{N-1}$, $\mathbb{CP}^{N-1}$, or
$\mathbb{HP}^{N-1}$ for $\beta=1,2,4$, respectively.
Integration over this orientation space, together with the
collective-coordinate Jacobians and transverse fluctuation
determinants, reproduces the $N$- and $\beta$-dependent
leading prefactor in Eq.~\eqref{eq:Lif}
(Sec.~\ref{sec:Lifshitz}.

\subsection{The mechanisms behind the approach }
We now summarize the physical mechanisms behind these results.

 At real energy, the eigenvalues $\lambda_1,\ldots,\lambda_N$
of the Riccati matrix $Z$ perform Dyson Brownian motion in the
non-confining cubic potential $U_E(\lambda)=E\lambda+\lambda^3/3$.
They run to $-\infty$ and are reinjected at $+\infty$, with each
such winding counting one level. The formal Gibbs weight $|\Delta(\bm\lambda)|^\beta
e^{-\sum_{i=1}^N U_E(\lambda_i)/T}$, where
$\Delta(\bm\lambda)=\prod_{i<j}(\lambda_j-\lambda_i)$,
is not normalizable.

In contrast, at complex energy, i.e. for $\eta>0$, the Riccati matrix $Z=A-\ii P$ remains in the
Siegel half-space $P>0$, so the real-energy passages through
infinity are regularized. This confinement does not imply
detailed balance: the stationary state supports circulating
probability currents. Absorption also makes solutions driven
by the same realization of the random potential contract
towards one another, providing the mechanism for loss of
memory of initial conditions and uniqueness of the stationary law.

An important role in finding the exact stationary solution is played by a hidden gradient structure. 
Namely, the variables $(A,Y=\eta P^{-1})$, the noise acts only on $A$,
while the equation for $Y$ contains no noise term.
The drift has the form $(D+\mathcal A)\nabla\Phi$,
where $D$ is the diffusion matrix, $\mathcal A$ is a
{\it constant antisymmetric} matrix, and $\Phi$ is the exponent
in~\eqref{eq:law}. This structure gives a stationary density
proportional to $e^\Phi$, with divergence-free probability
current proportional to $\mathcal A\nabla e^\Phi$ and no
probability flux through the boundary.
The two identities underlying this representation of the drift
are trace identities, valid for non-commuting matrices
and for all three values of $\beta$.

The central role in the large-$N$ analysis is palyed by revealing the role of Kontsevich integrals as normalizing factors of the stationary density. The product $\tau(\fs)\overline{\tau(\fs)}$ turn out to be a double
matrix-contour integral of a Gibbs weight for one matrix and an anti-Gibbs weight for the
other. The anti-Gibbs weight appears because $U_E$ is odd. Changing to centre-of-mass and relative coordinates and
completing the square in the centre-of-mass coordinate reproduces  the stationary joint probability density~\eqref{eq:law} after a contour deformation.
 On the deformed contour, the Gibbs weight is
$e^{-\Tr U_{\cE}(Z)/T}$, so holomorphic observables of the
Riccati matrix depend only on the factor $\tau$. The outgoing contour is selected by the requirement that
the contour integral and its complex conjugate reproduce the positive stationary probability density after deformation
of the double integration contour.

After the revealing detour via complex energies one may come back to real energy, setting $\eta\to0^+$. Then the stationary probability density approaches the real-energy stationary density with nonzero probability current.
Taking this limit also yields the characteristic function of $\Tr Z$, whose one-sided expansion as $q\to0^+$ gives
$\Gamma_N-\ii\pi\nu_N$.

Finally, it is worth commenting on the mathematical status of the results. Most of what follows consists of exact calculations, done by hand
or in exact symbolic algebra. Three steps go beyond that.\\[1ex]
({\bf i}) Uniqueness of the stationary probability density. At finite absorption
it follows from the exact contraction identity~\eqref{eq:contract},
valid for each realization of the random potential,
combined with an ergodic argument based on the loss of memory
of initial conditions. At real energy the same loss of memory
follows, at the physics level, from the positivity of all $N$ Lyapunov
exponents (Sec.~\ref{sec:justify}). We do not give a complete
mathematical argument for the present matrix dynamics in either case.\\[1ex]
({\bf ii}) The factorization~\eqref{eq:factor} rests on a statement about integration surfaces.
We give a physics-level argument for it (Sec.~\ref{sec:kontsevich}): the deformed integration surface is an
integer combination of products of single-matrix contours, and matching at high energy selects
the outgoing one.\\[1ex]
({\bf iii}) The identification of the zero-absorption distribution~\eqref{eq:Pst} with the current-carrying
stationary state at real energy is exact for $\beta=2$, see
Appendix~\ref{app:matrix-fermion} where it is shown algebraically that \eqref{eq:Pst} coincides with the
fermionic stationary state of Ref.~\cite{Letter}, so only the uniqueness question of ({\bf i}) remains.
For $\beta=1,4$ it is argued at the physics level (Sec.~\ref{sec:justify}): the limit
$\eta\to0$ is followed through the Riccati poles, and the flux leaving at $-\infty$ is checked
to equal the flux reinjected at $+\infty$. The real-energy formulas
\eqref{eq:main}, \eqref{eq:Gammapos} and~\eqref{eq:nupos} do not depend on this
identification.

\vspace{0.2cm}

\subsection {Main reasoning line and layout of the paper} 

The derivation follows through the following steps:\\[-2ex]
\[
\text{explicit JPD}\,\, P_{E,\eta}(A,Y)\ \longrightarrow\ \cZ_N\propto\big|\tau_N^{(\beta)}\big|^2\ \longrightarrow\
\Xi_N=T^{1/3}\partial_s\log\tau_N^{(\beta)}, 
\]
and finally the evaluation of $\tau_N^{(\beta)}$,
which completes this chain of reasoning.
The outgoing Kontsevich integral is the central object of this
construction: it follows from the stationary probability density
of the complex Riccati matrix and provides all the explicit results
for the holomorphic observables considered here.
The Dyson parameter $\beta$ enters only through the flat matrix
integration measures.

The paper is organized as follows.  Section~\ref{sec:N1} demonstrates the derivation for $N=1$, where every formula is easy to check.
Sections~\ref{sec:mechanism}--\ref{sec:realE} extend the derivation to general $(N,\beta)$: the stationary probability density
(Sec.~\ref{sec:mechanism}), its normalization as an outgoing Kontsevich integral
(Sec.~\ref{sec:kontsevich}), and the passage to real energy (Sec.~\ref{sec:realE}).
 Sections~\ref{sec:explicit}--\ref{sec:beta4edge} evaluate $\tau_N^{(\beta)}$: determinant and
Pfaffian formulas at finite $N$, the high-energy genus expansion, the Lifshitz tail of the Schr\"odinger spectral density, the leading large-$N$ (planar) limit, and the $\beta=4$ and $\beta=1$ spectral edges.
 Section~\ref{sec:discussion} summarizes the present paper and outlines Paper~II.
 The appendices contain material that illustrates and completes the main line without being part of it:
an algebraic comparison with the $\beta=2$ fermionic stationary state
(Appendix~\ref{app:matrix-fermion}), the complete even-$N$ reduction for $\beta=1$
(Appendix~\ref{app:beta1reduction}), and the connection to intersection numbers,
large-order asymptotics, and random surface gluings (Appendix~\ref{app:intersection}).
Appendix~\ref{app:admittance-scattering} collects the scattering
interpretation and time-delay identities as preparation for Paper~II.

\section{Model, Riccati gas and observables}
\label{sec:model}

\subsection{Three symmetry classes with a common normalization}

We recall that $W(x)=W(x)^\dagger$ is real symmetric, complex Hermitian or quaternion self-dual,
matrix Gaussian white noise with weight explicitly defined as
\be
\mathcal P[W]\propto\exp\Big[-\frac1{4T}\int_0^L\Tr W(x)^2\,\dd x\Big],\qquad T=\frac{J^2}{2N}.
\label{eq:noise}
\ee
For $\beta=4$ we write quaternions as $2\times2$ complex blocks, and $\Tr$ denotes
{\it half} the trace of the resulting $2N\times2N$ matrix (the ``quaternion trace''), so
that $\Tr I_N=N$ in all three classes. Eq.~\eqref{eq:noise} means, in all three classes each diagonal element $W_{ii}$ is real, with
$\aver{W_{ii}(x)W_{ii}(x')}=2T\,\delta(x-x')$. Further, each off-diagonal element $W_{ij}$ ($i<j$) has
$\beta$ real components, each with covariance
$T\,\delta(x-x')$. Hence
$\aver{W_{ij}(x)\overline{W_{ij}(x')}}=\beta T\,\delta(x-x')$. For $\beta=2$ this is
$\aver{W_{ij}(x)W_{kl}(x')}=
2T\,\delta_{il}\delta_{jk}\delta(x-x')$,
the convention of \cite{Letter}. For $\beta=1$ it is
$\aver{W_{ij}(x)W_{kl}(x')}=
T(\delta_{ik}\delta_{jl}+\delta_{il}\delta_{jk})\delta(x-x')$.
For $N=1$ all three classes reduce to the scalar white noise
$\aver{W(x)W(x')}=2T\,\delta(x-x')$
of Halperin's problem.

{\it Matrix coordinates and the trace metric.}
To write the Fokker--Planck equation in components we use real
coordinates on $\Herm_N(\F)$. Let $e_{ij}$ denote the matrix unit with
entries $(e_{ij})_{kl}=\delta_{ik}\delta_{jl}$. The standard real basis
$\{e_a\}$ consists of the $N$ diagonal matrices $e_{ii}$ together with,
for each pair $i<j$, the $\beta$ matrices $e_{ij}u+e_{ji}\bar u$, where
$u$ runs over the standard real basis of $\F$. Altogether there are
$d=N+\beta N(N-1)/2$ basis elements. They are orthogonal with respect to
the trace metric $\langle X,X'\rangle=\Tr XX'$, but not all of the same
length: $g_a:=\Tr e_a^2$ equals $1$ for diagonal and $2$ for off-diagonal
basis elements.

These unequal lengths appear in two places, and they cancel. The first is
the noise. In the real coordinates the white-noise matrix has covariance
per unit length $C_{ab}=2T\,g_a^{-1}\delta_{ab}$. Different coordinates
are therefore uncorrelated, and each diagonal entry diffuses with variance
$2T$ per unit length, while each real component of an off-diagonal entry
diffuses with variance $T$. The second is the gradient. We define the
matrix gradient $G=\nabla_X\Phi$ by $\delta\Phi=\Tr(G\,\delta X)$ for
every Hermitian variation $\delta X$. Expanding $X=\sum_a X_a e_a$ and
$G=\sum_a G_a e_a$ and using $\Tr(e_a e_b)=g_a\delta_{ab}$, we find
$\partial_{X_a}\Phi=g_aG_a$, so a coordinate derivative carries the same
factor $g_a$ that the diffusion coefficient divides out. Combining the
two,
\be
\frac12\,C_{aa}\,\partial_{X_a}\Phi=T\,G_a,
\qquad\text{with no summation over }a.
\label{eq:FDT}
\ee
Thus, when the probability density is proportional to $e^\Phi$,
the diffusive contribution to its probability current is
$-T e^\Phi\nabla_A\Phi$ in matrix notation, independently of
the basis normalization. This compatibility between the noise
covariance and the trace metric is the identity used repeatedly
below.
In eigenvalue--eigenvector coordinates,
$X=U\,\mathrm{diag}(\lambda_1,\ldots,\lambda_N)U^\dagger$,
the flat measure $\dd X=\prod_a\dd X_a$ takes the form
\[
\dd X\propto
|\Delta(\bml)|^\beta
\prod_{j=1}^N\dd\lambda_j\,\dd U_{\rm Haar}.
\]

{\it Kramers degeneracy ($\beta=4$).} Every eigenvalue  of the Schr\"odinger operator ~\eqref{eq:H} (which we will call energy levels to distinguish from eigenvalues of the Riccati matrix) is then doubly
degenerate. Throughout we count {\it Kramers pairs}: $\nu_N$ is the number of distinct
levels per unit length, and $\Gamma_N=\sum_{j=1}^N\gamma_j$ runs over the $N$ distinct
positive exponents. With this convention the $N=1$ problem is literally Halperin's in all
three classes.

For $\beta=4$, logarithmic determinants are normalized
consistently with this counting convention:
$\log\det\Psi$ denotes
$\frac12\log\det_{\mathbb C}\Psi$ in the
$2N\times2N$ complex representation, with the logarithm
continued along the Riccati evolution. Thus
$\partial_x\log\det\Psi=\Tr Z$ in all three classes.

\subsection{Riccati matrix and its eigenvalue gas}

At real energies $E$, conservation of the Wronskian
$\Psi^\dagger\Psi'-\Psi'^\dagger\Psi$, see~\eqref{eq:wronskian}, which vanishes at
$x=0$, implies that $Z=\Psi'\Psi^{-1}$ is Hermitian wherever
$\Psi$ is invertible.
Its $N$ real eigenvalues (counting each Kramers pair once
for $\beta=4$) are ordered as $\lambda_1<\dots<\lambda_N$. Their dynamics follows from second-order
perturbation theory, as for Dyson's Brownian motion~\cite{Dyson1962}. Viewing the spatial coordinate $x$ as ``time'', the Riccati matrix
receives the increment $-(E+Z^2)\dd x+\dd W$ over $\dd x$.
In the eigenbasis of $Z$, the eigenvalue increments satisfy
\[
\dd\lambda_i=-(E+\lambda_i^2)\dd x+\dd W_{ii}
+\sum_{j\neq i}\frac{|\dd W_{ij}|^2}{\lambda_i-\lambda_j}.
\]
and $|\dd W_{ij}|^2=\beta T\,\dd x$. Hence 
\be
\dd\lambda_i=\Big[-(E+\lambda_i^2)+\beta T\sum_{j\neq i}\frac1{\lambda_i-\lambda_j}\Big]\dd x+\sqrt{2T}\,\dd B_i ,
\label{eq:DBM}
\ee
This is Dyson's gas at inverse temperature $\beta$ in the {\it cubic} potential
$U_E(\lambda)=E\lambda+\lambda^3/3$, with drift $-U_E'(\lambda_i)$. It is exactly the same $\beta$-Dyson gas
which appeared in Allez and Dumaz~\cite{AllezDumaz2015} without any reference to underlying matrix Schr\"odinger equation; the same model also underlies the large-deviation analysis of Ref.~\cite{FLDO2025}. Its finite-$N$ stationary density with reinjection is constructed for every $\beta>0$ in Ref.~\cite{FyodorovAD}. Treatment of this gas displays the following subtlety: the cubic potential does not confine.
The drift $-\lambda^2$ sends the smallest eigenvalue to $-\infty$ in finite ``time'' $x$.
This ``explosion'' is however benign: in the projective (Pr\"ufer) variable $\lambda=\cot\theta$ the point
$\lambda=-\infty$ is identified with $\lambda=+\infty$, and the eigenvalue simply reappears at
the top of the spectrum. In the language of the $\F^N$-valued solutions, $\det\Psi$ passes
through zero. By matrix oscillation theory~\cite{BaurKratz1989} each such event counts one
Dirichlet level below $E$. The Riccati eigenvalue gas is thus a gas on a circle, with a nonzero
stationary current  around it.   Such a current has profound consequences, see more on this in \cite{Letter}.


For a confining potential the stationary probability density of Dyson's diffusion~\eqref{eq:DBM} would be the Gibbs distribution
$|\Delta(\bml)|^\beta\prod_je^{-U_E(\lambda_j)/T}$, with
$\Delta(\bml)=\prod_{i<j}(\lambda_j-\lambda_i)$. Here it is not normalizable: $e^{-\lambda^3/3T}$
blows up as $\lambda\to-\infty$. The true stationary state $P_{\rm st}$ instead carries a
probability current $J^{\rm wind}_{\rm st}$ of Riccati eigenvalues flowing around the projective circle.  The mean integrated density of states $\nu_N(E)$ for the random Schr\"odinger operator equals the stationary
winding current and determines the asymptotic tails of the mean
Riccati eigenvalue density, see Ref.~\cite{Letter}:
\be
\nu_N(E)=J^{\rm wind}_{\rm st},\qquad
n(\lambda):=
\aver{\textstyle\sum_j\delta(\lambda-\lambda_j)}_{\rm st}
\simeq\frac{\nu_N(E)}{\lambda^2}
\quad(\lambda\to\pm\infty).
\label{eq:current}
\ee
Indeed, for large $|\lambda|$ the Riccati eigenvalue motion is
dominated by the drift, $\dd\lambda/\dd x\simeq-\lambda^2$; the
interaction and diffusion terms give corrections of relative order
$|\lambda|^{-3}$. For large $|\lambda|$, the mean number of Riccati eigenvalues
crossing the value $\lambda$ per unit length is therefore
$|\dd\lambda/\dd x|\,n(\lambda)=\lambda^2n(\lambda)$. In the
stationary state this flux does not depend on $\lambda$, and every
such crossing is a passage through infinity, so it equals the winding
rate $\nu_N(E)$. Reinjection at $+\infty$
after escape to $-\infty$ gives the same flux, and hence
the same tail coefficient, at both ends of the real axis.

For $\beta=2$ the paper\cite{Letter} solved the current-carrying problem exactly. The gauge transformation by
the Vandermonde determinant removes the Calogero interaction and, thanks to an identity special
to the quadratic drift, the system can be mapped to free fermions. For $\beta\ne2$, the same transformation leaves an
inverse-square interaction with coupling proportional to
$\beta(\beta-2)$. The transformed problem therefore remains
an interacting many-particle problem in a quartic
single-particle potential, rather than reducing to free fermions.

What replaces the Gibbs state is found by going {\it off} the real axis. At complex energy the
Riccati matrix becomes a complex matrix with a negative imaginary part, the state space doubles
in dimension, and the stationary distribution of the enlarged dynamical system is explicit for every $\beta$
(Secs.~\ref{sec:N1} and~\ref{sec:mechanism}). The real-energy state is then recovered as a limit
(Sec.~\ref{sec:realE}), Eq.~\eqref{eq:Pst}.

\subsection{Observables and the complex characteristic exponent}
Let $\Psi(x)$ be an $N\times N$ matrix of independent
Schr\"odinger solutions, with generic initial data, and
let $Z=\Psi'\Psi^{-1}$ wherever $\Psi$ is invertible.
Jacobi's formula gives, wherever $\Psi(x)$ is invertible,
\[
\partial_x\log|\det\Psi(x)|
=\Rea\Tr\bigl(\Psi'(x)\Psi(x)^{-1}\bigr)
=\Tr Z(x)
=\sum_{i=1}^N\lambda_i(x),
\]
where we used the Hermiticity of $Z$ at real energy.
Thus the trace of the Riccati matrix is the local
logarithmic growth rate of the volume spanned by the
columns of $\Psi$. For generic initial conditions, the long-distance growth rate
of this volume is the sum $\Gamma_N(E)$ of the $N$ positive
Lyapunov exponents.

This identity requires a prescription at the Riccati
windings. At a simple zero $x_*$ of $\det\Psi$, one has
$\det\Psi(x)\sim c(x-x_*)$, and hence
$\Tr Z(x)\sim (x-x_*)^{-1}$. The divergent contributions
on the two sides of the zero cancel in the
principal-value sense. Accordingly, integration of the
logarithmic derivative through the windings is
understood with this regularization.

In the stationary ergodic regime, the regularized
spatial average can be expressed through the one-point
density
\[
n(\lambda)
=\left\langle\sum_{i=1}^N
\delta\bigl(\lambda-\lambda_i\bigr)\right\rangle,
\qquad
\int_{\mathbb R}n(\lambda)\,\dd\lambda=N.
\]
With the principal-value prescription inherited from
the Riccati windings, this yields
\be
\Gamma_N(E)
=\lim_{L\to\infty}\frac{1}{L}\log|\det\Psi(L)|
=\PV\!\int_{\mathbb R}\lambda\,n(\lambda)\,\dd\lambda,
\label{eq:Gamma}
\ee
where
\[
\PV\!\int_{\mathbb R}\lambda\,n(\lambda)\,\dd\lambda
:=\lim_{R\to\infty}\int_{-R}^{R}
\lambda\,n(\lambda)\,\dd\lambda.
\]
The principal value is essential: the stationary
Riccati density has tails associated with the flux
through infinity, so its first moment need not exist
as an absolutely convergent integral.

Both IDOS $\nu_N$ and $\Gamma_N$ are contained in the one-sided characteristic function of the
Riccati trace,
\be
\cF_N(q;E)=\big\langle e^{-\ii q\Tr Z/T^{1/3}}\big\rangle_{\rm st},\qquad q\ge0 .
\label{eq:F}
\ee
At this point it is worth discussing the subtlety of this characteristic
function. A large value of $\Tr Z$ comes from a single large eigenvalue,
so by~\eqref{eq:current} the distribution of $\Tr Z$ inherits the tails
$\nu_N/\lambda^2$ of $n(\lambda)$. Hence $\Tr Z$ has no absolutely
convergent first moment, and $\cF_N$ is not differentiable at $q=0$.  Its
one-sided expansion is nevertheless meaningful and well-defined. At first order in $q$ only configurations with
one large eigenvalue matter, so
$1-\cF_N\simeq\int n(\lambda)\,(1-e^{-\ii q\lambda/T^{1/3}})\,\dd\lambda$. The even part of the
tail gives $\nu_N\int(1-\cos u)u^{-2}\dd u=\pi\nu_N$ per unit $q/T^{1/3}$. The odd part, thanks
to the {\it equal} tails at $\pm\infty$ in~\eqref{eq:current}, reduces to the principal-value
mean. Thus, at the physics level,
\be
\cF_N(q;E)=1-\frac{q}{T^{1/3}}\big[\pi\nu_N+\ii\,\Gamma_N\big]+o(q),\qquad q\to0^+ .
\label{eq:smallq}
\ee
Any closed formula for $\cF_N$ therefore gives a possibility to evaluate the complex characteristic exponent
$\Xi_N:=\Gamma_N-\ii\pi\nu_N$. This step is identical for all $\beta$. 

The complex characteristic exponent $\Xi_N$ has a direct
spectral interpretation. For $\Imn\cE>0$, $
\Xi_N(\cE)=
\lim_{L\to\infty}\frac1L\log\det\Psi(L;\cE),
$
with the branch of the logarithm fixed by continuation along $x$.
Up to the normalization associated with the chosen determinant
regularization, $\det\Psi(L;\cE)$ is the Dirichlet spectral
determinant of $\cH_L-\cE$, as follows from the matrix
Gelfand--Yaglom construction
\cite{Forman1987,Ossipov2018}.
Thus $\Xi_N$ is the self-averaging logarithm of the spectral
determinant per unit length, or equivalently a spectral free
energy. At the real-energy boundary,
$-\pi^{-1}\Imn\Xi_N(E+\ii0)=\nu_N(E)$ counts the levels below
$E$ per unit length, while $\Rea\Xi_N(E+\ii0)=\Gamma_N(E)$. The two
are not independent. Since $\det\Psi(L;\cE)$ is the spectral
determinant, $\partial_\cE\Xi_N$ is the mean trace of the resolvent
per unit length,
\be
\partial_\cE\Xi_N(\cE)=\int\frac{\rho_N(E')\,\dd E'}{\cE-E'},
\qquad \Imn\cE>0 .
\label{eq:Thouless}
\ee
Integrating once gives the multichannel Thouless
formula~\cite{Thouless1972,KotaniSimon1988},
\[
\Gamma_N(E)-\Gamma_N(E_0)=\int\log\Big|\frac{E-E'}{E_0-E'}\Big|\,
\rho_N(E')\,\dd E' , 
\]
i.e. the sum of the positive Lyapunov exponents is the logarithmic
potential of the density of states. The subtraction at a reference
energy $E_0$ is needed because $\nu_N(E')\simeq N\sqrt{E'}/\pi$ grows
at high energy. Equivalently,
\[
\Gamma_N'(E)=\PV\!\int_{\mathbb R}
\frac{\rho_N(E')}{E-E'}\,\dd E',
\]
so $\Gamma_N$ and $-\pi\nu_N$ form a
Kramers--Kronig pair: they are the real and imaginary
parts of the boundary value of a single function
holomorphic in $\Imn\cE>0$, with the dispersion relation
understood with the appropriate high-energy subtraction.

\section{Warm-up: the scalar case $N=1$ in detail}
\label{sec:N1}

For one channel all three $\beta=1,2,4$ classes coincide, and the final real-energy answer is one of
Halperin~\cite{Halperin1965}. It is however instructive to go through the complete construction for
$N=1$ for two reasons. First, every formula is elementary and can be checked by hand, including the full
two-dimensional stationary probability density at complex energy, which seems to be new as a closed formula. 
Second, the general case (Secs.~\ref{sec:mechanism}--\ref{sec:realE}) repeats the same six steps
in the same order. Only two things change: the variables become matrices, and $\beta$ enters
through flat matrix measures. Table~\ref{tab:dictionary} in Section~\ref{sec:dictionary} gives the dictionary.

For $N=1$ we write the Riccati variable at complex energy
$E+\ii\eta$, $\eta>0$, in terms of the real variables
$u$ and $v$ as
\be
Z=T^{1/3}(u-\ii v),\qquad
\epsilon=\frac E{T^{2/3}},\qquad
a=\frac\eta{T^{2/3}},\qquad t=T^{1/3}x .
\ee
Then $\dd Z=-(E+\ii\eta+Z^2)\,\dd x+\sqrt{2T}\,\dd B_x$ becomes
\be
\dd u=(-\epsilon-u^2+v^2)\,\dd t+\sqrt2\,\dd B_t,\qquad \dd v=(a-2uv)\,\dd t .
\label{eq:uv}
\ee 
All densities in this section are dimensionless. We also use $\fs=-\epsilon-\ii a=-(E+\ii\eta)/T^{2/3}$,
$U_\epsilon(u)=\epsilon u+u^3/3$ and $\cA_\pm=\Ai\pm\ii\Bi$.

\subsection{Step 1: complex energy, absorption and contraction}
\label{sec:N1step2}

\begin{itemize}
 \item {\it Confinement and absence of poles.}
 At $v=0$, Eq.~\eqref{eq:uv} gives $\dd v=a\,\dd t>0$,
so the dynamics preserves $v>0$ and $Z$ remains in the
\emph{lower half-plane}, provided $Z$ stays finite. At real energy
$Z$ can blow up at finite $x$: these Riccati poles, where $Z$ escapes
to $-\infty$ and re-enters from $+\infty$, are the windings of
Sec.~\ref{sec:mechanism}. To show that at $\eta>0$ no such pole occurs
at any finite position $x>0$ along the wire, consider the Dirichlet
solution $\psi(x)$ at energy $E+\ii\eta$, with $\psi(0)=0$ and
$\psi'(0)=1$. Since the random potential is real, its Wronskian satisfies
\[
\frac{\dd}{\dd x}
\big(\bar\psi\,\psi'-\bar\psi'\,\psi\big)
=-2\ii\eta\,|\psi|^2.
\]
Integrating from $0$ to $x$ gives
\[
\bar\psi(x)\psi'(x)-\bar\psi'(x)\psi(x)
=-2\ii\eta\int_0^x|\psi(s)|^2\,\dd s.
\]
If $\psi(x_0)=0$ at some $x_0>0$, the left-hand side would
vanish, whereas the right-hand side is nonzero:
$\eta>0$ and the solution is nontrivial. Thus, for every finite $x>0$, $\psi(x)\ne0$ and
$Z(x)=\psi'(x)/\psi(x)$ is finite. Absorption therefore
removes the Riccati poles at finite positive positions:
there is no escape to infinity and reinjection of the kind
occurring at real energy. This does not preclude circulating
probability currents in the stationary distribution of $Z$.

\item {\it Physical meaning.}
The parameter $\eta>0$ describes uniform absorption.
The logarithmic derivative $Z=\psi'/\psi$ relates the
wave amplitude at the boundary to its derivative and
therefore plays the role of a boundary admittance.

More explicitly, at a fixed complex energy $E+\ii\eta$,
consider the solution satisfying $\psi(0)=0$ and
$\psi(L)=b\ne0$. This solution is a rescaling of the
fundamental solution used above, leaving $Z$ unchanged.
These conditions define an inhomogeneous boundary-value
problem, not an eigenvalue problem. The prescribed boundary
amplitude fixes the normalization, so the integrated wavefunction
intensity is not constrained to unity.
The imaginary part $\Im Z(x)$ determines the probability current.
Indeed, using $\psi'=Z\psi$,
\[
j(x)=\frac{\hbar}{m}\Im\!\left[\psi^*(x)\psi'(x)\right]
=\frac{\hbar}{m}|\psi(x)|^2\Im Z(x)
=-\frac{\hbar}{m}T^{1/3}|\psi(x)|^2v(x).
\]
This local expression is consistent with the integrated
flux balance: the Wronskian identity above gives
\[
|\psi(x)|^2\Im Z(x)
=-\eta\int_0^x|\psi(s)|^2\,\dd s.
\]
Thus $v$ measures the flux per unit boundary intensity,
with the sign fixed by the orientation of the boundary.
The conversion from boundary admittance to scattering observables
is discussed in Appendix~\ref{app:admittance-scattering}.

\item \emph{Loss of memory.}
Consider two solutions $z_1(t)$ and $z_2(t)$ with different
initial conditions but driven by the same realization of
the noise:
\[
\dd z_j=-(\epsilon+\ii a+z_j^2)\,\dd t+\sqrt2\,\dd B_t,
\qquad j=1,2.
\]
The common noise cancels in their difference $\delta=z_1-z_2$,
giving $\delta'=-(z_1+z_2)\delta$. For an infinitesimal separation
about a trajectory $z=u-\ii v$, this becomes $\delta'=-2z\delta$, and
the real part of $\delta'/\delta$ gives
$\frac{\dd}{\dd t}\log|\delta|=-2u$. The $v$-equation
in~\eqref{eq:uv} contains no noise and gives directly
$\frac{\dd}{\dd t}\log v=\frac{a}{v}-2u$. Subtracting the two gives
\be
\frac{\dd}{\dd t}\log\frac{|\delta|}{v}=-\frac{a}{v}<0\quad(a>0),
\qquad\text{hence}\qquad
\frac{|\delta(t)|}{v(t)}
=\frac{|\delta(0)|}{v(0)}
\exp\Big[-a\int_0^t\frac{\dd s}{v(s)}\Big],
\label{eq:contractN1}
\ee
with $v(s)$ taken along the reference trajectory.
Here $|\delta|/v$ is the infinitesimal length in the
Poincar\'e metric of the lower half-plane,
\[
\dd\ell^2=\frac{\dd u^2+\dd v^2}{v^2},
\]
which measures separations relative to the distance from the real
axis. In this metric the real-energy part of the dynamics drops out
(the terms $-2u$ cancel), and only absorption shrinks
separations, at the instantaneous rate $a/v$. Thus,
infinitesimal separations contract exponentially whenever
$1/v$ has a strictly positive long-time average.
This loss of sensitivity to initial conditions provides
the physical mechanism underlying uniqueness of the
stationary probability density. A rigorous proof also
requires control of the long-time dynamics, including
existence of a stationary state; we do not pursue these
steps here.
\end{itemize}

\subsection{Step 2: hidden gradient structure}
\label{sec:lemma}

\begin{lemma}\label{lem:gradient}
Given the It\^o diffusion
\be
\dd X_t=b(X_t)\,\dd t+\sigma\,\dd B_t
\ee
on an open domain $\Omega\subset\R^n$, where $B_t$ is a
standard Brownian motion and $\sigma$ is a constant matrix,
consider the corresponding Fokker--Planck equation
\be 
\partial_t\rho=-\nabla\cdot J,
\qquad
J=b\rho-D\nabla\rho,
\qquad
D=\frac12\sigma\sigma^{\mathsf T}.
\ee Suppose that the drift has the form
\be b=(D+\mathcal A)\nabla\Phi,
\label{eq:lemma}
\ee where $\mathcal A$ is a constant antisymmetric matrix.
Then $\rho_{\rm st}\propto e^\Phi$ solves the stationary
Fokker--Planck equation in the interior of $\Omega$,
with probability current
\be
J_{\rm st}=\mathcal A\nabla\rho_{\rm st}.
\ee
It defines a stationary probability density provided that
it is normalizable and satisfies the boundary conditions
of the dynamics.
\end{lemma}

\emph{Derivation.}
Since $\nabla\rho_{\rm st}=\rho_{\rm st}\nabla\Phi$,
the diffusive contribution cancels the symmetric part
of the drift:
\[
J_{\rm st}
=(D+\mathcal A)\rho_{\rm st}\nabla\Phi
-D\nabla\rho_{\rm st}
=\mathcal A\nabla\rho_{\rm st}.
\]
The remaining current is divergence-free, $\nabla\cdot J=\sum_{ij}\mathcal A_{ij}\partial_i\partial_j\rho=0$, because $\mathcal A$ is
antisymmetric. $\square$

A few remarks are due.\\[0.5ex]
({\bf i}.) For $\mathcal A=0$ this is the familiar gradient (detailed-balance) case. For $\mathcal A\ne0$
the stationary state carries divergence-free {\it circulating} currents. So the absence of
detailed balance does not prevent an explicit solution.\\[0.5ex]
({\bf ii}.) The lemma establishes stationarity in the interior of the domain.
To define a stationary probability density, $\rho_{\rm st}\propto e^\Phi$
must also be normalizable and satisfy the boundary conditions
of the dynamics. For a boundary through which no probability
can pass, the required condition is
\[
\boldsymbol n\cdot J_{\rm st}
=
\boldsymbol n\cdot\mathcal A\nabla\rho_{\rm st}
=0
\qquad\text{on }\partial\Omega,
\]
where $\boldsymbol n$ is the outward unit normal.
Other boundary prescriptions, such as periodic identification
or reinjection, require the corresponding matching conditions
rather than zero normal current.
For an unbounded domain, probability leakage at infinity
must also be excluded. These conditions must be checked
in each application.\\[0.5ex]
({\bf iii}.) Condition~\eqref{eq:lemma} is a property of the drift alone: $(D+\mathcal A)^{-1}b$ must be a
gradient. It is a strong condition, and it is exactly what the Riccati flow satisfies in the right
variables.

\subsection{Step 3: the explicit stationary density}
\label{sec:N1law}

Set $y=a/v>0$. Since $\dd v=(a-2uv)\,\dd t$ contains
no Brownian increment, $v$ is differentiable along each
realization of the noise. The ordinary chain rule therefore
applies to $y=a/v$, with no It\^o correction.  Then 
\be
\dd u=\Big(-\epsilon-u^2+\frac{a^2}{y^2}\Big)\dd t+\sqrt2\,\dd B_t,\qquad \dd y=(2uy-y^2)\,\dd t .
\label{eq:uy}
\ee
Recalling $U_\epsilon(u)=\epsilon u+u^3/3$ define
\be
\Phi_a(u,y)=U_\epsilon(u-y)-U_\epsilon(u)-\frac{a^2}y=-y\Big(u-\frac y2\Big)^2-\epsilon y-\frac{y^3}{12}-\frac{a^2}y .
\label{eq:PhiN1}
\ee
Its derivatives are $\partial_u\Phi_a=-2uy+y^2$ and $\partial_y\Phi_a=-\epsilon-(u-y)^2+a^2/y^2$. Hence, exactly,
\be
b_u=-\epsilon-u^2+\frac{a^2}{y^2}=\partial_u\Phi_a+\partial_y\Phi_a, \quad b_y=2uy-y^2=-\partial_u\Phi_a.
\ee
This is~\eqref{eq:lemma} with $D=\mathrm{diag}(1,0)$ and $\mathcal A=\begin{pmatrix}0&1\\-1&0\end{pmatrix}$.
By Lemma~\ref{lem:gradient}, the stationary joint probability
density is given by
\be
\rho_a(u,y)=\frac{1}{\cZ(\epsilon,a)}\,e^{\Phi_a(u,y)},
\qquad u\in\R,\quad y>0,
\label{eq:R}
\ee
with probability current
\[
J_u=\partial_y\rho_a,
\qquad
J_y=-\partial_u\rho_a.
\]
It remains to check that there is no probability flux through the boundary. Indeed, as $a>0$ the factor $e^{-a^2/y}$ suppresses the density
as $y\to0$, the factor $e^{-y^3/12}$ suppresses it as
$y\to\infty$, and the Gaussian factor $e^{-y(u-y/2)^2}$ suppresses it as
$|u|\to\infty$ at fixed $y>0$.
The decay after Gaussian integration over $u$ also excludes
probability flux in joint limits where $|u|\to\infty$ while $y\to0$ or $y\to\infty$.

 In the original variables $(u,v)$ the same density takes the form
\be
P_a(u,v)=\frac{a}{\cZ(\epsilon,a)\,v^2}
\exp\Big[-\frac{a(u^2+\epsilon)}{v}+\frac{a^2u}{v^2}-\frac{a^3}{3v^3}-av\Big],\qquad v>0 .
\label{eq:Puv}
\ee
The factor $v^{-2}$ is the hyperbolic area density.
Thus, relative to the invariant volume element
$\dd u\,\dd v/v^2$ on the lower half-plane, the stationary
probability density is proportional to $e^{\Phi_a(u,a/v)}$. The loss-of-memory argument of Step~1 identifies this stationary
distribution with that of the absorbing-wire admittance, subject to
the long-distance relaxation assumptions stated there.

The relation of this geometry to waveguide scattering and
hyperbolic diffusion is discussed in
Appendix~\ref{app:admittance-scattering}.

\subsection{Step 4: Gaussian structure and the Airy normalization}
\label{sec:N1norm}

The exponent~\eqref{eq:PhiN1} is quadratic in $u$.
Consequently, at fixed $y$, the Gaussian density of $u$ is
\[
 p(u\mid y)=\sqrt{\frac y\pi}\,\exp\!\left[-y\left(u-\frac y2\right)^2\right],
\]
with mean $y/2$ and variance $1/(2y)$.
Integrating out the $u$-variable gives the marginal
density of $y$:
\be
f_a(y)=\frac{\sqrt\pi}{\cZ(\epsilon,a)}\,y^{-1/2}e^{-\epsilon y-y^3/12-a^2/y},\qquad
\cZ(\epsilon,a)=\sqrt\pi\int_0^\infty y^{-1/2}e^{-\epsilon y-\frac{y^3}{12}-\frac{a^2}y}\,\dd y ,
\label{eq:ZN1}
\ee
and the integral for $\cZ$ turns out to be {\it an Airy modulus squared}:
\be
\cZ(\epsilon,a)=\pi^2\,\big|\cA_-(\fs)\big|^2,\qquad \fs=-\epsilon-\ii a .
\label{eq:AiryZ}
\ee
We derive this by writing $\cZ$ as a double contour integral. The same
steps give the general-$N$ factorization of Sec.~\ref{sec:kontsevich}.

{\it Derivation.}
Let $\Gamma_-$ run from $e^{-2\pi\ii/3}\infty$ to
$+\infty$, below the real axis, and put
\[
U_{\cE}(\lambda)=\cE\lambda+\lambda^3/3,
\qquad \cE=\epsilon+\ii a,\qquad \fs=-\cE.
\]
Under $\lambda=-w$, the induced contour $\mathcal C_-$
runs from $e^{\pi\ii/3}\infty$ to $-\infty$, giving
\be
\cJ(\cE)
:=\int_{\Gamma_-}e^{-U_{\cE}(\lambda)}\,\dd\lambda
=-\int_{\mathcal C_-}e^{w^3/3-\fs w}\,\dd w
=\ii\pi\,\cA_-(\fs),
\ee
where $\cA_-(\fs)=\Ai(\fs)-\ii\Bi(\fs)$.

In the conjugate integral, substitute $\lambda'=-\mu$
and use the oddness of $U$. Define $\Gamma_+$ to run
from $-\infty$ to $e^{-\pi\ii/3}\infty$ in the lower
half-plane: its orientation is opposite to that induced
on $-\bar\Gamma_-$, absorbing the minus sign from
$\dd\lambda'=-\dd\mu$. Thus
\be
|\cJ(\cE)|^2
=\int_{\Gamma_-}\dd\lambda
 \int_{\Gamma_+}\dd\mu\;
e^{-U_{\cE}(\lambda)+U_{\bar\cE}(\mu)}.
\ee

We deform the product contour onto
\[
\Sigma_a
=\Big\{(\lambda,\mu)=(z,z-y):
\ z=u-\frac{\ii a}{y},\quad
u\in\mathbb R,\quad y>0\Big\}.
\]
The justification for this deformation is given for
general $N$ in Sec.~\ref{sec:kontsevich}
(step~4 of Result~\ref{prop:factor}).
The Jacobian satisfies
$\dd\lambda\wedge\dd\mu=\dd y\wedge\dd u$.
Writing $c=u-y/2$, the exponent on $\Sigma_a$ becomes
\[
-U_{\cE}(\lambda)+U_{\bar\cE}(\mu)
=-\epsilon y-2\ii a\left(c-\frac{\ii a}{y}\right)
-y\left(c-\frac{\ii a}{y}\right)^2-\frac{y^3}{12}
=-y\left(u-\frac y2\right)^2-\epsilon y
-\frac{y^3}{12}-\frac{a^2}{y}
=\Phi_a(u,y).
\]
With the inherited orientation, we therefore obtain
\[
|\cJ(\cE)|^2
=\int_0^\infty\dd y\int_{\mathbb R}\dd u\;
e^{\Phi_a(u,y)}.
\]
Performing the Gaussian integral
\[
\int_{\mathbb R}
e^{-y(u-y/2)^2}\,\dd u=\sqrt{\frac{\pi}{y}},
\]
we find
\[
|\cJ(\cE)|^2
=\sqrt{\pi}\int_0^\infty\frac{\dd y}{\sqrt y}\,
\exp\left(-\epsilon y-\frac{y^3}{12}-\frac{a^2}{y}\right)
=\cZ(\epsilon,a),
\]
which proves Eq.~\eqref{eq:AiryZ}.

 At $a=0$, Gaussian integration over $u$ recovers
the classical representation
\[
\Ai^2(x)+\Bi^2(x)
=\pi^{-3/2}\int_0^\infty
t^{-1/2}e^{xt-t^3/12}\,\dd t
\]
\cite[Eq.~9.11.4]{DLMFAiry}, with $x=-\epsilon$,
providing an independent check of the factor $\pi^2$.

The derivation also identifies the variables. On $\Sigma_a$ the first
coordinate $z=u-\ii a/y$ is the Riccati variable itself, and the
density~\eqref{eq:R} is precisely the normalized integrand of $|\cJ|^2$:
a Gibbs weight in $z$ times an anti-Gibbs weight in $z-y$. At $a=0$,
$\Sigma_a$ becomes the real region
$\{(\lambda,\mu)\in\R^2:\lambda>\mu\}$ appearing in Step~6.

\subsection{Step 5: holomorphic statistics from the stationary distribution}
\label{sec:N1holo}

Take $z=u-\ii v=u-\ii a/y$ and $q\ge0$.
Throughout this subsection, $\aver{\cdot}$ denotes
averaging with respect to the stationary
density~\eqref{eq:R}:
\[
 \aver{F(z)}
 =
 \int_0^\infty\!\dd y\int_{\R}\!\dd u\,
 \rho_a(u,y)\,F(u-\ii a/y).
\]

Then the conditional Gaussian average over $u$ at fixed
$y$ gives
\[
 \aver{e^{-\ii qu}\mid y}
 =
 e^{-\ii qy/2-q^2/(4y)}.
\]
 Multiplying by $e^{-qa/y}$ and by the weight $e^{-\epsilon y-y^3/12-a^2/y}$ yields
\[
e^{-(\epsilon+\ii q/2)y-\frac{y^3}{12}-\frac{(a+q/2)^2}{y}}.
\]
This is the same weight at the shifted parameters $\epsilon'=\epsilon+\ii q/2$ and $a'=a+q/2$.
Now,  \eqref{eq:AiryZ} gives $\cZ(\epsilon,a)=\pi^2\cA_-(-\epsilon-\ii a)\,\cA_+(-\epsilon+\ii a)$ which continues analytically in
$(\epsilon,a)$. Under the shift, $-\epsilon'-\ii a'=\fs-\ii q$, while $-\epsilon'+\ii a'=-\epsilon+\ii a$ is
unchanged.
Therefore 
\be
\aver{e^{-\ii qz}}=\frac{\cA_-(\fs-\ii q)}{\cA_-(\fs)},\qquad
\aver{z}=\frac{\cA_-'(\fs)}{\cA_-(\fs)} .
\label{eq:FN1}
\ee
In physical units $\Xi_1(E+\ii\eta)=\aver{Z}=T^{1/3}\cA_-'(\fs)/\cA_-(\fs)$. The outgoing Airy function is
the holomorphic factor of the positive normalization~\eqref{eq:AiryZ}.

\subsection{Step 6: zero absorption, the energy derivative and Halperin's formula}
\label{sec:N1step6}

Put $w=v/a=1/y$. From $\dd v=(a-2uv)\dd t$, the variable $w$ obeys
$\dd w=(1-2uw)\,\dd t$, which is also the equation of $-\partial_\epsilon u$ at real energy. So as
$a\to0$, $w$ becomes the energy derivative of the Riccati variable, $w=-\partial_\epsilon z$, a positive
energy-derivative variable. To see this directly, let $\psi$ solve the real-energy
Schr\"odinger equation
\[
-\psi''+V\psi=\epsilon\psi,
\qquad \psi(0)=0,\qquad \psi'(0)=1,
\]
where primes denote derivatives with respect to $t$.
Differentiating with respect to $\epsilon$ gives
\[
\frac{\dd}{\dd t}
\left[\psi\,\partial_\epsilon\psi'
-\psi'\,\partial_\epsilon\psi\right]
=-\psi^2.
\]
The expression in brackets vanishes at $t=0$, since
the initial data are independent of energy. Hence,
wherever $\psi(t)\ne0$,
\[
w(t)=-\partial_\epsilon
\left(\frac{\psi'(t)}{\psi(t)}\right)
=\frac{\displaystyle\int_0^t\psi^2(s)\,\dd s}
       {\psi^2(t)}>0.
\]
The corresponding matrix identity is derived below
in Eqs.~\eqref{eq:wronskian}--\eqref{eq:Sdef}.

Further, Eq.~\eqref{eq:R} for the joint probability density at $a=0$ in the variables
$(\lambda,\mu)=(u,u-1/w)$ takes the form of Gibbs weight for $\lambda$ times an anti-Gibbs weight for $\mu$, with an order constraint:
\be
\;\rho_0=\frac1{\cZ(\epsilon,0)}\,e^{-U_\epsilon(\lambda)}\,e^{+U_\epsilon(\mu)},\,\quad \mu<\lambda .\;
\label{eq:twopointN1}
\ee
Integrating over $\mu$ yields \emph{Halperin's
stationary density}\cite{Halperin1965}:
\be
p_{\rm H}(u)=j\,e^{-U_\epsilon(u)}
\int_{-\infty}^u e^{U_\epsilon(v')}\,\dd v',
\qquad
j=\frac1{\cZ(\epsilon,0)}
=\frac1{\pi^2[\Ai(-\epsilon)^2+\Bi(-\epsilon)^2]} .
\label{eq:Halperin}
\ee
Here $j>0$ is the magnitude of the stationary
probability current flowing towards $u=-\infty$.
Thus integrating out $\mu$ in the two-point
Gibbs/anti-Gibbs distribution gives Halperin's formula,
with the auxiliary variable $\mu=u-1/w$ expressed
in terms of the energy derivative.

We now extract the complex characteristic exponent,
whose real and imaginary parts determine the
Lyapunov exponent and the integrated density of states.
In the dimensionless variables used here, it is
\[
 \Omega_1(\epsilon)
 =
 \lim_{a\rightarrow 0_+}\aver{z}
 =
 \Gamma_1(\epsilon)-\ii\pi\nu_1(\epsilon).
\]
Recalling that $v=a/y$ and that the conditional Gaussian
mean is $\aver{u\mid y}=y/2$, we obtain from~\eqref{eq:ZN1}
\[
 \aver{v}=a\aver{y^{-1}}
 =-\frac12\partial_a\log\cZ,
 \qquad
 \aver{u}=\frac12\aver{y}
 =-\frac12\partial_\epsilon\log\cZ.
\]
The term linear in $a$ arises from the region
$y\to0$:
\[
 \int_0^\infty y^{-1/2}
 \bigl(e^{-a^2/y}-1\bigr)\,\dd y
 =-2\sqrt{\pi}\,a,
 \qquad a>0.
\]
Consequently,
\[
 \cZ(\epsilon,a)
 =
 \cZ(\epsilon,0)-2\pi a+O(a^2),
\]
and hence
\[
 \nu_1(\epsilon)=j=\frac1{\cZ(\epsilon,0)},
 \qquad
 \Gamma_1(\epsilon)
 =
 -\frac12\partial_\epsilon\log\cZ(\epsilon,0),
\]
where $\Gamma_1$ is the principal-value mean of
the real-energy Riccati variable.
Together these expressions reproduce the classical
Airy formula for the complex characteristic exponent
\cite{Halperin1965,ComtetTourigny2017}:
\[
 \Omega_1(\epsilon)
 =
 \frac{\cA_-'(s)}{\cA_-(s)},
 \qquad s=-\epsilon.
\]

\subsection{Dictionary to the general case}
\label{sec:dictionary}

Below we summarize in the table the modifications required to extend the $N=1$ case to general number of channels $N\ge 1$.

\begin{table}[htbp]
\centering
\caption{The six steps for $N=1$ and their counterparts for general $(N,\beta)$.}
\label{tab:dictionary}
\begingroup
\footnotesize
\setlength{\tabcolsep}{3pt}
\renewcommand{\arraystretch}{1.7}
\begin{tabularx}{\linewidth}{@{}>{\raggedright\arraybackslash}p{0.05\linewidth}*{2}{>{\raggedright\arraybackslash}X}@{}}
\toprule
Step & $N=1$ (this section) & general $(N,\beta)$ (Secs.~\ref{sec:mechanism}--\ref{sec:realE})\\\midrule
1 & lower half-plane; Eq.~\eqref{eq:contractN1} & Siegel half-space; Eq.~\eqref{eq:contract}\\
2 & Lemma~\ref{lem:gradient} & the same lemma, in matrix coordinates\\
3 & $y=a/v$; $\Phi_a=U(u-y)-U(u)-a^2/y$ & $Y=\eta P^{-1}$; Eq.~\eqref{eq:law}; trace identities~\eqref{eq:gradients}\\
4 & $\cZ=\pi^2|\cA_-|^2$ by a double contour & $\cZ_N\propto|\tau_N^{(\beta)}|^2$, Eq.~\eqref{eq:factor}; flat measures give $\beta$\\
5 & $\aver{e^{-\ii qz}}=\cA_-(\fs-\ii q)/\cA_-(\fs)$ & $\tau_N^{(\beta)}(\fs-\ii q)/\tau_N^{(\beta)}(\fs)$; zero-free region\\
6 & joint distribution of $(u,m=u-y)$~\eqref{eq:twopointN1}; Halperin; $\nu_1=j$ & two-matrix distribution; Eq.~\eqref{eq:Pst}; Eqs.~\eqref{eq:nupos}, \eqref{eq:Gammapos}\\
\bottomrule
\end{tabularx}
\endgroup
\end{table}

\section{The explicit stationary distribution for general $(N,\beta)$}
\label{sec:mechanism}

This section repeats Steps~1--3 of Sec.~\ref{sec:N1} for matrices. Throughout,
$\cE=E+\ii\eta$ with $\eta>0$.

\subsection{Complex energy confines $Z$ to a Siegel half-space}

Write $Z=A-\ii P$, where $A=(Z+Z^\dagger)/2$ and $P=\ii(Z-Z^\dagger)/2$ are $\F$-Hermitian.
Because $W$ is Hermitian the Wronskian-type identity
\be
\frac{\dd}{\dd x}\big(\Psi^\dagger\Psi'-\Psi'^\dagger\Psi\big)=\Psi^\dagger(\bar\cE-\cE)\Psi=-2\ii\,\eta\;\Psi^\dagger\Psi
\label{eq:wronskian}
\ee
holds for each noise realization. Starting from $\Psi(0)=0$, it gives
\be
P=\eta\,\cS,\qquad \cS:=\Psi^{\dagger-1}\Big[\int_0^x\Psi^\dagger\Psi\Big]\Psi^{-1}>0 .
\label{eq:Sdef}
\ee
Taking the Hermitian and anti-Hermitian parts of~\eqref{eq:Riccati} at energy $\cE$ yields the system of two matrix It\^o equations 
\be
\dd A=\big[-E-A^2+P^2\big]\dd x+\dd W,\qquad P'=\eta\,I-(AP+PA) .
\label{eq:Pdot}
\ee
The noise does not appear in the equation for $P$, because $W$ is Hermitian and only shifts $A$.
Thus $Z$ lives in the {\it Siegel half-space} $\cS_\beta=\{Z=A-\ii P:\ P>0\}$, the matrix analogue
of the lower half-plane.  In particular, $\det\Psi(x)\neq0$ for every finite $x>0$,
so $Z(x)=\Psi'(x)\Psi(x)^{-1}$ remains finite along the
wire, and a continuous branch of $\log\det\Psi(x)$ can
be chosen for $x>0$. Hence 
\be
\Xi_N(\cE)=\lim_{L\to\infty}\frac1L\log\det\Psi(L;\cE)=\aver{\Tr Z}_{\mu_{E,\eta}},\qquad \eta>0 .
\label{eq:Xicomplex}
\ee

\paragraph*{Physical meaning: boundary admittance and internal norm.}
The Riccati matrix $Z(L;\cE)$ maps a prescribed boundary
amplitude $b=\psi(L)$ to the boundary derivative:
$\psi'(L)=Z(L;\cE)b$, with $\psi(0)=0$.
It is therefore the Dirichlet-to-Neumann map, or boundary
admittance, of the sample. For $\cE=E+\ii\eta$ with $\eta>0$,
the matrix $\Psi(L;\cE)$ is invertible. Defining
\[
F(x;\cE)=\Psi(x;\cE)\Psi(L;\cE)^{-1},
\]
the solution with prescribed boundary values $\psi(0)=0$
and $\psi(L)=b$ is $\psi(x)=F(x;\cE)b$. The integrated
Wronskian identity defines the internal-norm matrix
\begin{equation}
\mathcal N(E,\eta):=\frac{P(E,\eta)}{\eta}
=\int_0^L F(x;\cE)^\dagger F(x;\cE)\,\dd x.
\label{eq:internal-norm}
\end{equation}
Thus $b^\dagger\mathcal N(E,\eta)b$ is the integrated wavefunction
intensity inside the sample for the imposed boundary
amplitude $b$. This identity holds for every real $E$
when $\eta>0$. If $E$ is not a Dirichlet eigenvalue,
the zero-absorption limit is finite and satisfies
\[
\lim_{\eta\to0_+}\frac{P(E,\eta)}{\eta}
=\cT(E):=-\partial_E Z(L;E).
\]
The relation of these quantities to reflection, dwell times
and the Wigner--Smith matrix is explained in
Appendix~\ref{app:admittance-scattering}.
Their statistics will be developed in Paper~II~\cite{PaperII}.

\subsection{Loss of memory: an exact contraction identity}
\label{sec:memory}

Consider two solutions of~\eqref{eq:Riccati} driven by the {\it same} noise, and let
$\delta$ be their (infinitesimal) difference. The noise cancels, yielding
\be
\delta'=-(Z\delta+\delta Z).
\label{eq:delta}
\ee
Measure the difference $\delta$ with the natural distance of the Siegel half-space,
$\ell^2=\Tr\big(P^{-1}\delta P^{-1}\delta^\dagger\big)$; for $N=1$ this is the Poincar\'e length of
Eq.~\eqref{eq:contractN1}. Using~\eqref{eq:Pdot} and~\eqref{eq:delta} we get the exact identity valid for each realization of the random potential:
\be \frac{\dd\ell^2}{\dd x}=-\eta\;\Tr\big(P^{-2}\delta P^{-1}\delta^\dagger+P^{-1}\delta P^{-2}\delta^\dagger\big)
\le-\frac{2\eta}{\|P\|}\,\ell^2 .
\label{eq:contract}
\ee
The inequality follows from
$P^{-1}\ge\|P\|^{-1}I$, where $\|P\|$ denotes
the operator norm, and positivity of the trace
pairing.

Equation~\eqref{eq:contract} expresses the contraction produced by
absorption. At $\eta=0$, $Z=\Psi'\Psi^{-1}$ is built from solutions of
the Schr\"odinger equation with Hermitian potential and real energy. The pair $(\Psi,\Psi')$ evolves along the wire through
a transfer matrix. More explicitly, propagation from
a reference position $x_0$ to $x$ is described by
\[
\begin{pmatrix}
\Psi(x)\\
\Psi'(x)
\end{pmatrix}
=
\begin{pmatrix}
\mathsf A&\mathsf B\\
\mathsf C&\mathsf D
\end{pmatrix}
\begin{pmatrix}
\Psi(x_0)\\
\Psi'(x_0)
\end{pmatrix},
\]
where the blocks depend on $x$, $x_0$ and the realization
of the random potential. This transfer
matrix preserves the conserved Wronskian form
$\Psi^\dagger\Psi'-\Psi'^\dagger\Psi$; for $\beta=1$ it is a real
symplectic matrix. Consequently the Riccati matrix itself evolves by
matrix M\"obius (linear-fractional) transformations,
\[
Z\;\longmapsto\;(\mathsf C+\mathsf DZ)(\mathsf A+\mathsf BZ)^{-1},
\]
which map the Siegel half-space onto itself and preserve the metric
$\ell^2$. This holds for the deterministic flow and for each noise
increment separately. For $\eta>0$ the Wronskian form is no longer
conserved, and the additional term $-\ii\eta$ makes the evolution
contractive: the metric length of every nonzero infinitesimal variation
decreases.
In particular,
\[
 \ell^2(x)\le \ell^2(x_0)
 \exp\!\left[
 -2\eta\int_{x_0}^{x}\frac{\dd s}{\|P(s)\|}
 \right].
\]
Exponential contraction follows provided the
long-time average of $1/\|P\|$ is bounded away
from zero. Together with suitable control of long-distance fluctuations, this contraction provides the basis
for establishing uniqueness of the stationary
distribution and loss of dependence on the initial
condition. We do not supply the full ergodic
argument here.

\subsection{The variables $(A,Y)$}

Set $Y=\eta P^{-1}>0$. Set $Y=\eta P^{-1}>0$. As in the scalar case, this
change of variables requires no It\^o correction,
since $P$ has finite variation.
From~\eqref{eq:Pdot}, $Y'=-\eta P^{-1}P'P^{-1}$ and $P^2=\eta^2Y^{-2}$. Hence 
\be
\dd A=\big[-E-A^2+\eta^2Y^{-2}\big]\dd x+\dd W,\qquad \dd Y=\big[AY+YA-Y^2\big]\dd x .
\label{eq:AY}
\ee
These are the scalar equations~\eqref{eq:uy} with $u\to A$ and $y\to Y$, in the only order that
keeps them Hermitian. In the limit $\eta\to0$ we have $Y\to\cS^{-1}$, where $\cS=P/\eta$ is the positive
matrix of~\eqref{eq:Sdef}. At real energy $\cS$ coincides with the energy derivative $\cT:=-\partial_EZ$,
because both obey $X'=I-(ZX+XZ)$ with $X(x)\simeq (x/3)I_N$ as $x\to0^+$.

\subsection{The potential, two trace identities, and stationarity}

Define
\be
\begin{aligned}
\Phi(A,Y)&=\frac1T\Big[\Tr U_E(A-Y)-\Tr U_E(A)-\eta^2\Tr Y^{-1}\Big]\\
&=-\frac1T\Big[\Tr(BYB)+E\Tr Y+\tfrac1{12}\Tr Y^3+\eta^2\Tr Y^{-1}\Big],
\end{aligned}
\label{eq:Phi}
\ee
with $B=A-\frac12Y$. The second form uses only cyclicity of the trace:
$\Tr(A-Y)^3-\Tr A^3=-3\Tr A^2Y+3\Tr AY^2-\Tr Y^3$ and
$\Tr(BYB)=\Tr A^2Y-\Tr AY^2+\frac14\Tr Y^3$.

\paragraph*{Trace gradients.} For any polynomial $f$, we have $\delta\Tr f(X)=\Tr\big(f'(X)\,\delta X\big)$. With
$U_E'(X)=E+X^2$, we then have
\be
T\,\nabla_A\Phi=(A-Y)^2-A^2,\qquad
T\,\nabla_Y\Phi=-E-(A-Y)^2+\eta^2Y^{-2}.
\label{eq:gradients}
\ee
Comparing with the drifts in~\eqref{eq:AY}, $(A-Y)^2-A^2=-AY-YA+Y^2$, so
\be
b_Y=AY+YA-Y^2=-T\,\nabla_A\Phi,\qquad
b_A=-E-A^2+\eta^2Y^{-2}=T\big(\nabla_A\Phi+\nabla_Y\Phi\big).
\label{eq:drift}
\ee
These are the matrix versions of the two scalar identities of Sec.~\ref{sec:N1law}. They hold for
non-commuting $A$ and $Y$, because $\Phi$ is a trace.

\begin{proposition}[Explicit stationary distribution]\label{prop:law}
For every $N\ge1$, $\beta\in\{1,2,4\}$, $E\in\R$ and $\eta>0$, the density $e^{\Phi(A,Y)}$ with respect to
the flat measure $\dd A\,\dd Y$ on $\Herm_N(\F)\times\Herm_N^+(\F)$ is stationary for~\eqref{eq:AY}. It
is integrable. The contraction and relaxation argument of
Sec.~\ref{sec:memory} identifies its normalized form,
Eq.~\eqref{eq:law}, as the stationary distribution reached at long distances;
this identification uses the relaxation assumptions stated there.
\end{proposition}
{\it Derivation.} Use the real coordinates $A_a,Y_a$ of Sec.~\ref{sec:model}. The noise
acts only on $A$, with $D_{AA}=\frac12C=T\,g^{-1}$, and $D=0$ in all other blocks. By~\eqref{eq:FDT},
$T(\nabla_A\Phi)_a=T\,g_a^{-1}\partial_{A_a}\Phi$, and likewise for $Y$. So~\eqref{eq:drift} reads
\be
b=\begin{pmatrix}Tg^{-1}&Tg^{-1}\\-Tg^{-1}&0\end{pmatrix}\nabla\Phi=(D+\mathcal A)\nabla\Phi,\qquad
\mathcal A=T\begin{pmatrix}0&g^{-1}\\-g^{-1}&0\end{pmatrix}.
\ee
$g^{-1}$ is diagonal and constant, so $\mathcal A$ is constant and antisymmetric. The noise is
additive, so there is no It\^o drift. Lemma~\ref{lem:gradient} then gives stationarity of $e^\Phi$,
with circulating current $J=\mathcal A\nabla e^\Phi$, i.e.\ $J_A=T\nabla_Ye^\Phi$ and $J_Y=-T\nabla_Ae^\Phi$ in
trace-gradient form.
 Now we also need to check that there is no flux through the boundary of the domain. An eigenvalue of $Y$ tending to $0$ is suppressed by
$e^{-\eta^2\Tr Y^{-1}/T}$, faster than any power. Large $Y$ is suppressed by $e^{-\Tr Y^3/12T}$, since
$\Tr(BYB)\ge0$ for $Y>0$. Large $A$ at fixed $Y>0$ is suppressed by the Gaussian factor. For integrability  see Eq.~\eqref{eq:Yens} below. 
The integral converges even at $\eta=0$.  Uniqueness follows by Sec.~\ref{sec:memory}, at the level of the long-distance relaxation argument given there $\square$.

The Dyson parameter $\beta$ enters the construction only through the flat measures and through $g$, which drops out of the final stationary distribution.
For $N=1$, Result~\ref{prop:law} is Eq.~\eqref{eq:R}.

\subsection{Structure of the stationary probability distribution}
\label{sec:structure}
 Inversion on the cone of positive $\F$-Hermitian matrices has
Jacobian $\dd(X^{-1})=\det X^{-(\beta(N-1)+2)}\dd X$. In terms of $Z=A-\ii P$, the stationary distribution can be written 
in Siegel-invariant form as
\be
\mu_{E,\eta}(\dd Z)=\frac{\eta^{d}}{\cZ_N(E,\eta)}\,
\exp\Big\{\frac1T\Big[\Tr U_E(A-\eta P^{-1})-\Tr U_E(A)-\eta\Tr P\Big]\Big\}\,
\frac{\dd A\,\dd P}{\det P^{\beta(N-1)+2}},
\label{eq:lawZ}
\ee
with $d=\dim\Herm_N(\F)=N+\beta N(N-1)/2$. The last factor is the invariant volume of the Siegel
half-space. Relative to this volume element, the probability density is
proportional to $e^\Phi$. For $N=1$ this is
Eq.~\eqref{eq:Puv}.

We now determine the statistics of the internal-norm
matrix $\mathcal N=P/\eta$ defined in
\eqref{eq:internal-norm}. Since $\mathcal N$ is positive
definite for $\eta>0$, its inverse is well defined.
It is convenient to work with
\[
Y=\mathcal N^{-1}=\eta P^{-1},
\]
because the joint density is Gaussian in $B=A-Y/2$. Integrating out $B$ therefore gives an
explicit eigenvalue distribution for $Y$. The
eigenvalues $\tau_i$ of the internal-norm matrix
are recovered by $\tau_i=1/y_i$; in the
zero-absorption limit they become the eigenvalues
of $\cT=-\partial_EZ$.
 
 Next, given $Y$, the matrix $B=A-\frac12Y$ is
Gaussian with the weight $e^{-\Tr(BYB)/T}$. In the eigenbasis of $Y$ we have
$\Tr(BYB)=\sum_iy_iB_{ii}^2+\sum_{i<j}(y_i+y_j)|B_{ij}|^2$. Integrating out $B$ and the eigenvectors
of $Y$ gives the joint probability density of the eigenvalues $y_i>0$ of $Y$:
\be
p(\bm y)\propto|\Delta(\bm y)|^\beta\prod_iy_i^{-1/2}\prod_{i<j}(y_i+y_j)^{-\beta/2}
\exp\Big[-\frac1T\sum_i\Big(Ey_i+\frac{y_i^3}{12}+\frac{\eta^2}{y_i}\Big)\Big].
\label{eq:Yens}
\ee
Thus Eq.~\eqref{eq:Yens} describes the inverse
internal-norm eigenvalues. Their reciprocals $\tau_i=1/y_i$ give the integrated
norm inside the sample when the imposed boundary
amplitude $b=\psi(L)$ is a normalized eigenvector
of $P/\eta$. 

Let $\mathcal Y_N(E,\eta)$ denote the integral over $\mathbb R_+^N$
of the unnormalized eigenvalue weight displayed in
Eq.~\eqref{eq:Yens}. Then
$\cZ_N=c_{N,\beta}T^{d/2}\mathcal Y_N$.
We now discuss its properties in parallel with the
$N=1$ case.
\begin{itemize}
\item {\it Integrability.} The kernel is homogeneous of degree $-d/2$ near $\bm y=0$, while the
measure $|\Delta|^\beta\prod\dd y_i$ has degree $d$. So $\mathcal Y_N$ converges at $\bm y\to0$ even for $\eta=0$.
\item {\it Conditional Gaussian structure.}
The joint probability density of the Hermitian matrices
$A$ and $P$ in $Z=A-\ii P$ is equivalently specified by
the eigenvalue density~\eqref{eq:Yens} of
$Y=\eta P^{-1}$, the invariant angular measure for
its eigenvectors, and the conditional Gaussian density
of $B=A-\frac12Y$ at fixed $Y$.
In the eigenbasis of $Y$, the independent real
components of $B$ have zero mean and variances
$T/(2y_i)$ for diagonal entries and
$T/[2(y_i+y_j)]$ for off-diagonal entries.
The admittance matrix is then
\[
 Z=B+\frac12Y-\ii\eta Y^{-1}.
\]
For $N=1$, in the scaled variables of Sec.~\ref{sec:N1law} (where
$T=1$, $\eta\to a$ and $y=a/v$), this is $z=u-\ii a/y$ with $u=b+\frac y2$
and, at fixed $y$, $b$ has zero mean and variance $1/(2y)$,
reproducing the Gaussian density of Sec.~\ref{sec:N1norm}.
\item {\it An alternative route to large-$N$ asymptotics.}
Equation~\eqref{eq:Yens} represents the normalization
as an integral with a positive weight over
$\mathbb R_+^N$. It therefore offers an alternative
to analysing the complex-contour Kontsevich
representation: one may instead study a real
interacting eigenvalue ensemble using Coulomb-gas,
saddle-point, or loop-equation methods.

Besides the usual logarithmic repulsion, the
ensemble contains a pair interaction through
$y_i+y_j$. For $\beta=2$, this interaction has
the Bures, or Schur--Pfaffian, structure:
for even $N$,
\[
\prod_{i<j}\frac{y_i-y_j}{y_i+y_j}
=\Pf\!\left[\frac{y_i-y_j}{y_i+y_j}\right].
\]
This relates the ensemble to those studied in
\cite{OsipovSommersZyczkowski2010}.

To recover the physical density of states,
however, the large-$N$ analysis must also resolve
the small-absorption behaviour, including
contributions from configurations in which an
eigenvalue approaches $y=0$. Whether this positive
representation simplifies the analysis of the
density-of-states edge remains to be investigated.
We do not pursue this route here.

\end{itemize}

\section{The normalization is $|\tau|^2$: three Kontsevich integrals}
\label{sec:kontsevich}

\subsection{The outgoing integration surface}

For $T>0$, the factor $e^{-\lambda^3/(3T)}$
decays as $|\lambda|\to\infty$ along rays in
three sectors of the complex $\lambda$-plane,
each of opening $\pi/3$:
\[
 -\frac{\pi}{6}<\arg\lambda<\frac{\pi}{6},
 \qquad
 \frac{\pi}{2}<\arg\lambda<\frac{5\pi}{6},
 \qquad
 -\frac{5\pi}{6}<\arg\lambda<-\frac{\pi}{2}.
\] The outgoing contour
\be
\Gamma_-:\ e^{-2\pi\ii/3}\infty\ \longrightarrow\ e^{-\ii\epsilon}\infty\ \ (\epsilon\to0^+)
\ee
joins two of them, see Fig.1,  and lies in the lower half-plane, where $Z$ lives. Under $w=-\lambda/T^{1/3}$, it becomes
$\mathcal C_-:e^{\ii\pi/3}\infty\to e^{\ii\pi}\infty$,
which gives an integral representation of the outgoing
Airy combination $\cA_-=\Ai-\ii\Bi$.

{\it Matrix integration surface.} Let $\gamma:\R\to\Gamma_-$ parametrize the contour, and apply it to Hermitian
matrices by functional calculus, $\gamma(UDU^\dagger)=U\gamma(D)U^\dagger$. The image
$\cC=\gamma(\Herm_N(\F))$ is a real $d$-dimensional surface without boundary inside $\cS_\beta$. On the integration surface $\cC$, the holomorphic
volume element $\dd Z$ takes the following form
in eigenvalue coordinates:
\be
\dd Z\big|_{\cC}=\text{const}\times\prod_{i<j}\big(\gamma(t_j)-\gamma(t_i)\big)^\beta\prod_j\gamma'(t_j)\,\dd t_j\;\dd U_{\rm Haar},
\qquad t_1<\dots<t_N .
\label{eq:pullback}
\ee
A subtle point in this change of variables deserves
discussion. On the real line the Jacobian is
$|\Delta|^\beta$. For $\beta=2,4$ the absolute value is superfluous. For $\beta=1$ it is replaced, on
the contour, by the analytic function $\prod_{i<j}(w_j-w_i)$ with the eigenvalues {\it ordered
along the contour}. This is the correct analytic continuation, because on the real axis ordering
removes the absolute value. It is the origin of the ``ordered'' prescription in
Eq.~\eqref{eq:tau}.

Define the matrix Airy (Kontsevich-type) integral
over the outgoing integration surface:
\be
\cJ_N(\cE)=\int_{\cC}e^{-\Tr U_{\cE}(\Lambda)/T}\,\dd\Lambda\ =\ c'_{N,\beta}\,T^{\kappa'_{N,\beta}}\,\tau_N^{(\beta)}(\fs),
\qquad \fs=-\cE/T^{2/3},
\label{eq:Jtau}
\ee
where the second equality follows from~\eqref{eq:pullback} and $w=-\lambda/T^{1/3}$.

\subsection{Factorization}
\label{Factorization}
\begin{proposition}[The normalization is a modulus squared]\label{prop:factor}
For $\eta>0$, $\cZ_N(E,\eta)=\pm\,\cJ_N(\cE)\,\overline{\cJ_N(\cE)}$, i.e.\ Eq.~\eqref{eq:factor}.
\end{proposition}
{\it Derivation.}  We follow essentially the same steps as in $N=1$ example.
\begin{enumerate}
\item {\it Conjugation and reflection.} $\overline{\cJ_N(\cE)}$ is the integral of
$e^{-\Tr U_{\bar\cE}(\Lambda')/T}$ over the conjugate integration surface $\bar\cC$, in the upper half-space.
Substitute $\Lambda'=-M$. Since $U$ is odd, $\Tr U_{\bar\cE}(-M)=-\Tr U_{\bar\cE}(M)$, and $M$ runs over
$-\bar\cC$, again in the lower Siegel half-space. Hence
\be
\cJ_N(\cE)\,\overline{\cJ_N(\cE)}=\pm\int_{\cC\times(-\bar\cC)}
\exp\Big\{\frac1T\Big[-\Tr U_{\cE}(\Lambda)+\Tr U_{\bar\cE}(M)\Big]\Big\}\,\dd\Lambda\,\dd M .
\label{eq:double}
\ee
A Gibbs weight for $\Lambda$ multiplies an anti-Gibbs weight for $M$.
\item {\it Centre and relative coordinates.} Put $\Lambda=C+\frac12Y$ and $M=C-\frac12Y$; the
Jacobian is one. Using $\cE\Tr\Lambda-\bar\cE\Tr M=E\Tr Y+2\ii\eta\Tr C$ and
$\frac13\Tr(\Lambda^3-M^3)=\Tr(CYC)+\frac1{12}\Tr Y^3$ we see that
\be
-\Tr U_{\cE}(\Lambda)+\Tr U_{\bar\cE}(M)=-\Tr(CYC)-2\ii\eta\Tr C-E\Tr Y-\tfrac1{12}\Tr Y^3 .
\ee
\item {\it Completing the square.} With $\tilde C=C+\ii\eta Y^{-1}$,
\be
\Tr(CYC)+2\ii\eta\Tr C=\Tr(\tilde CY\tilde C)+\eta^2\Tr Y^{-1}.
\ee
The exponent of~\eqref{eq:double} is therefore exactly $\Phi$ of~\eqref{eq:Phi}, with $B\to\tilde C$.
The term $\eta^2\Tr Y^{-1}$, which suppresses the
joint probability density when an eigenvalue of $Y$
approaches zero, is produced by the Gaussian shift.
\item {\it The integration surface (physics level).} Steps 1--3
are identities between integrands; it remains to identify the
integration surface. All steps are carried out at the matrix level,
with the holomorphic volume form $\dd\Lambda\,\dd M$ on the
complexification of $\Herm_N(\F)\times\Herm_N(\F)$. No absolute
values or eigenvalue orderings appear. For $\beta=1$ the ordered
prescription enters only when $\cJ_N$ is reduced to eigenvalues,
Eq.~\eqref{eq:pullback}.

{\it The target integration surface.} Take $\tilde C\in\Herm_N(\F)$ and
$Y\in\Herm^+_N(\F)$. Through Eq.~\eqref{eq:LambdaZ} below, this
defines a real $2d$-dimensional surface
$\Sigma'=\{(Z,Z-Y)\}$, on which the integrand is exactly
$e^\Phi$. The surface $\Sigma'$ has no boundary at finite
distance. Its ends at infinity are of three types, all lying
in the region where the integrand decays:
\begin{itemize}
\item $\tilde C\to\infty$ at fixed $Y>0$, where the Gaussian factor
$e^{-\Tr(\tilde CY\tilde C)/T}$ decays;
\item $Y\to\infty$, where $e^{-\Tr Y^3/(12T)}$ decays;
\item an eigenvalue $y_i\to0$. Then $\Lambda$ and $M$ run to
infinity together along the $-\ii$ direction, and the integrand
decays as $e^{-\eta^2/(Ty_i)}$.
\end{itemize}
The third end requires $\eta>0$. At $\eta=0$ the surface acquires
a boundary $\Lambda=M$ at finite distance; this is why the
factorization is stated for $\eta>0$, and real energy is reached
as a limit (Sec.~\ref{sec:realE}).

{\it Decomposition.} The integrand is a product
$e^{f(\Lambda)}e^{g(M)}$, with $f=-\Tr U_{\cE}/T$ and
$g=+\Tr U_{\bar\cE}/T$. For a single variable this is the familiar
Airy situation: by Cauchy's theorem, any contour along which
$e^{w^3/3-\fs w}$ decays at both ends can be deformed, without changing
the integral, into an integer combination of the two standard Airy
contours~\cite{DLMFAiry}. Applied variable by variable, the same
argument shows that the integral over any integration surface whose
ends lie in the decay region is an integer combination of products of
single-matrix integrals over such standard surfaces.

For invariant integrands the single-matrix basis can be
labelled by the number $k$ of eigenvalues placed on the
conjugate Airy contour $\mathcal C_+=\overline{\mathcal C_-}$,
which produces $\cA_+=\Ai+\ii\Bi$; the other $N-k$
eigenvalues stay on $\mathcal C_-$. Denote the resulting
holomorphic integrals by $\tau^{[k]}_N(\fs)$, so that
$\tau^{[0]}_N=\tau^{(\beta)}_N$. Hence
\be
\cZ_N(E,\eta)\propto\sum_{k,l=0}^{N}n_{kl}\,
\tau^{[k]}_N(\fs)\,\overline{\tau^{[l]}_N(\fs)},
\qquad n_{kl}\in\mathbb Z .
\label{eq:Zdecomp}
\ee
The integers $n_{kl}$ do not depend on $E$ or on $\eta>0$: as
the parameters vary, $\Sigma'$ deforms continuously and its
ends remain in the decay region.

{\it Matching at high energy.} The integers $n_{kl}$ are fixed by
letting $E\to+\infty$ at fixed $\eta>0$ and comparing two independent
estimates of $\cZ_N$. In the positive integral~\eqref{eq:Yens} each
eigenvalue is localized near $y_i\simeq\eta/\sqrt E$, the minimum of
$Ey+\eta^2/y$, so that $\mathcal Y_N(E,\eta)=\exp[-2N\eta\sqrt E/T+O(\log E)]$:
the normalization decays at the rate $2N\eta\sqrt E/T$. On the other
side of Eq.~\eqref{eq:Zdecomp}, the Airy asymptotics give
$|\cA_\mp(\fs)|=\exp[\mp\eta\sqrt E/T+O(\log E)]$, so every eigenvalue
moved to the conjugate contour $\mathcal C_+$ turns a decaying factor
into a growing one. Consequently
$\big|\tau^{[k]}_N\tau^{[l]}_N\big|=\exp[-2(N-k-l)\eta\sqrt E/T+O(\log E)]$,
and any term with $(k,l)\neq(0,0)$ would decay more slowly than the
direct estimate allows and would dominate it. Such terms could only
disappear by cancelling among themselves, but terms with equal decay
rates carry different oscillating phases and cannot cancel. Hence all
$n_{kl}$ with $(k,l)\neq(0,0)$ vanish, only $n_{00}$ survives, and we
obtain Eq.~\eqref{eq:factor}; its sign and the constant are fixed by
positivity and by the case $N=1$. 

{\it Checks.} The factorization passes several independent tests. For
$N=1$ and $\eta=0$ it reduces to the classical representation
$\Ai^2(x)+\Bi^2(x)=\pi^{-3/2}\int_0^\infty t^{-1/2}e^{xt-t^3/12}\,\dd t$
\cite[Eq.~9.11.4]{DLMFAiry}. For general $N$ at $\eta=0$, both limits of
the energy can be compared. As $E\to+\infty$ the integral~\eqref{eq:Yens}
scales as $(T/E)^{a_\beta}$, with $a_\beta=\frac N2+\frac{\beta N(N-1)}4$,
in agreement with $|\tau^{(\beta)}_N|^2\propto|s|^{-a_\beta}$ found in
Sec.~\ref{sec:highE}. As $E\to-\infty$ the Laplace point of
Eq.~\eqref{eq:Yens} moves to $y_i=2\sqrt{-E}$, giving
$\mathcal Y_N\propto(-E)^{-a_\beta}e^{\frac43N(-E)^{3/2}/T}$; this is
$|\tau^{(\beta)}_N|^2$ at the Lifshitz saddle $w_-$ of
Sec.~\ref{sec:Lifshitz}, with both the exponent and the power
reproduced. Finally, for $\beta=2$, Appendix~\ref{app:matrix-fermion}
shows independently that the zero-absorption distribution coincides with the
fermionic state of Ref.~\cite{Letter}, whose holomorphic factor is
$\tau^{(2)}_N$.
\end{enumerate}

\subsection{Holomorphic statistics and a zero-free region}
\label{Holomorphic}

\paragraph*{Holomorphic generating function.} Given Proposition~\ref{prop:factor},
the holomorphic generating function of $\Tr Z$ follows by repeating the steps of
the $N=1$ example. For $k\ge0$ we have
$e^{-\ii k\Tr Z}=e^{-\ii k\Tr A-k\eta\Tr Y^{-1}}$. At fixed $Y$ the
matrix $B=A-\frac12Y$ is Gaussian, and averaging over it produces
$e^{-\ii k\Tr Y/2-k^2T\Tr Y^{-1}/4}$. Combined with $e^{-k\eta\Tr Y^{-1}}$, this gives
the weight in Eq.~\eqref{eq:Yens} evaluated at
\[
E'=E+\frac{\ii kT}{2},
\qquad
\eta'=\eta+\frac{kT}{2}.
\]
Here $E'$ is an analytically continued parameter;
the shifted weight is no longer positive.
Consequently,
\[
\aver{e^{-\ii k\Tr Z}}_{\mu_{E,\eta}}
=\frac{\cZ_N(E',\eta')}{\cZ_N(E,\eta)}.
\]
By Eq.~\eqref{eq:factor},
\[
\cZ_N(E,\eta)\propto
\cJ_N(E+\ii\eta)\,\cJ_N^\sharp(E-\ii\eta),
\qquad
\cJ_N^\sharp(\zeta)
:=\overline{\cJ_N(\bar\zeta)},
\]
with the factorized expression analytically continued
to the shifted parameters. Writing
$\cE=E+\ii\eta$, we have
\[
E'+\ii\eta'=\cE+\ii kT,
\qquad
E'-\ii\eta'=E-\ii\eta=\bar\cE.
\]
Thus the second factor cancels in the ratio. With $q=kT^{1/3}$ and $\fs=-\cE/T^{2/3}$,
we arrive at
\be
\cF_N^{(\beta)}(q;\cE)=\aver{e^{-\ii q\Tr Z/T^{1/3}}}_{\mu_{E,\eta}}=\frac{\tau_N^{(\beta)}(\fs-\ii q)}{\tau_N^{(\beta)}(\fs)},\qquad
\aver{\Tr Z}_{\mu_{E,\eta}}=T^{1/3}\partial_\fs\log\tau_N^{(\beta)}(\fs) .
\label{eq:Fcomplex}
\ee
This is Eq.~\eqref{eq:main} at complex energy; for $N=1$ it reduces to
Eq.~\eqref{eq:FN1}.

\paragraph*{Why only the outgoing factor survives.} The reason the
answer depends on $\tau_N^{(\beta)}$ alone becomes clear when the distribution is
written in the variables of the double integral~\eqref{eq:double}. There
the two matrices are
\be
\Lambda=\tilde C+\tfrac12Y-\ii\eta Y^{-1}=A-\ii P=Z,\qquad M=\Lambda-Y=Z-Y ,
\label{eq:LambdaZ}
\ee
and $-\Tr U_{\cE}(Z)+\Tr U_{\bar\cE}(Z-Y)=T\Phi$ holds identically, even
though $A$ and $Y$ do not commute. The stationary distribution~\eqref{eq:law} is
therefore nothing but the integrand of $\cJ_N\overline{\cJ_N}$, a Gibbs
weight in $Z$ times an anti-Gibbs weight in $Z-Y$, evaluated on the
deformed integration surface. Now insert into Eq.~\eqref{eq:double} an
observable $\varphi(Z)$ that is holomorphic and depends on $\Lambda$
only. If $\varphi$ is entire and grows at infinity more slowly than the
cubic exponential, as $e^{-\ii k\Tr Z}$ and polynomials do, it does not
change the region where the integrand decays. The decomposition of
step~4 then holds with the {\it same} integers $n_{kl}$, since these are
a property of the integration surface $\Sigma'$ and not of the
integrand. Again only the $(0,0)$ term survives, the $M$-integral still
gives $\overline{\cJ_N}$, and dividing by Eq.~\eqref{eq:factor} we find
\be
\int\varphi\,\dd\mu_{E,\eta}=\frac{\int_{\cC}\varphi\,e^{-\Tr U_{\cE}/T}\,\dd Z}{\int_{\cC}e^{-\Tr U_{\cE}/T}\,\dd Z} .
\label{eq:theorem}
\ee
For $\varphi=e^{-\ii k\Tr Z}$ this is Eq.~\eqref{eq:Fcomplex} again.
Thus averages of holomorphic observables over the full joint probability
density~\eqref{eq:law} can be evaluated with the normalized complex
contour weight alone; for non-holomorphic observables this is not the
case. This is the precise sense in which
the formal, non-normalizable Gibbs weight $e^{-\Tr U_E/T}$ of the Riccati
gas is correct: put on the outgoing integration surface, it gives every
holomorphic average exactly. The full joint probability density is
obtained, after normalization, from the product of the Gibbs
weight of $Z$ and the anti-Gibbs weight of $Z-Y$.

\paragraph*{Zero-free region.} Given Result~\ref{prop:factor}, the
left-hand side of~\eqref{eq:factor} is a strictly positive integral for
$\eta>0$. It remains finite and positive at $\eta=0$, and both sides are
continuous as $\eta\to0_+$ (Sec.~\ref{sec:realE}). Hence
\be
\tau_N^{(\beta)}(s)\neq0\qquad\text{for }\Imn s\le0 ,
\label{eq:zerofree}
\ee
the real axis included. Consequently the complex characteristic exponent
$\Xi_N=T^{1/3}\partial_s\log\tau_N^{(\beta)}$ has no poles for
$\Imn\cE\ge0$, which is what the passage to real energy in
Sec.~\ref{sec:realE} requires.
\subsection{The three Kontsevich integrals}
In all three classes, $\tau_N^{(\beta)}$ is the
\emph{holomorphic factor} of the positive integral
$\mathcal Y_N$ which was defined below equation ~\eqref{eq:Yens}.
The classes differ only in how $\tau$ is evaluated (Sec.~\ref{sec:explicit}):
\begin{itemize}
\item $\beta=2$: a Hankel determinant of Airy derivatives, $\det[\cA_-^{(j+k)}]$ which is precisely the fermionic result of \cite{Letter};
\item $\beta=4$: a Pfaffian of Airy derivatives,
$\Pf[(k-j)\cA_-^{(j+k-1)}]$, expressible exactly
in terms of the orthogonal-polynomial recurrence
coefficients of the $\beta=2$ ensemble
(Sec.~\ref{sec:beta4edge});
\item $\beta=1$: a Pfaffian of ordered double integrals,
expressible in terms of the orthogonal-polynomial
norms and recurrence coefficients of the $\beta=2$
ensemble at temperature $T/2$, together with one
explicit Airy primitive, for even $N$
(Subsec.~\ref{sec:b1sketch} and
Appendix~\ref{app:beta1reduction}).
\end{itemize}
The positive integral $\mathcal Y_N$ is the same object for all three, with $\beta$ entering through
$|\Delta|^\beta$ and $(y_i+y_j)^{-\beta/2}$.

\section{Real energy: the non-absorptive limit}
\label{sec:realE}

\subsection{The zero-absorption distribution: a two-matrix model}

In the stationary joint density of $(A,Y)$, the
unnormalized weight depends on absorption only through
the factor $e^{-\eta^2\Tr Y^{-1}/T}$. As
$\eta\to0_+$, this factor increases to one, while the
remaining zero-absorption weight is integrable
(Sec.~\ref{sec:structure}). The limit can therefore
be taken directly in the joint density of $(A,Y)$:
the normalization converges, as do the averages
of all bounded observables of $A$ and $Y$. In the limiting real-energy description, $Y=\cT^{-1}$, where
$\cT=-\partial_E Z$ (Sec.~\ref{sec:mechanism}). 

The linear change of variables
$\Lambda=A$ and $M=A-Y$ has Jacobian of unit
absolute value. In these variables the limiting
distribution is
\be
\mu_{E,0}(\dd\Lambda\,\dd M)
=\frac{1}{\cZ_N(E,0)}
e^{-\Tr U_E(\Lambda)/T}
e^{+\Tr U_E(M)/T}\,
\dd\Lambda\,\dd M,
\qquad M<\Lambda.
\label{eq:twomatrix}
\ee
Here $M<\Lambda$ means that $\Lambda-M$ is
positive definite. The joint weight consists
of a Gibbs factor for $\Lambda$, an anti-Gibbs
factor for $M$, and this matrix-order constraint.
In the real-energy interpretation,
$\Lambda=Z$ and $M=Z-\cT^{-1}$.
For $N=1$, the formula reduces to
Eq.~\eqref{eq:twopointN1}.

Integrating out $M$ gives the marginal density
\be
P_{\rm st}(\Lambda)
=\frac{1}{\cZ_N(E,0)}
e^{-\Tr U_E(\Lambda)/T}
\int_{M<\Lambda}e^{\Tr U_E(M)/T}\,\dd M .
\label{eq:Pst}
\ee
The integral converges. Indeed, writing
$M=\Lambda-Y$ with $Y>0$, the cubic term gives
\[
\Tr U_E(\Lambda-Y)
=-\frac13\Tr Y^3
+O\bigl(\Tr Y^2+\Tr Y+1\bigr)
\]
for fixed $\Lambda$ and $E$, so the exponent
decays as $Y$ becomes large.

Here $P_{\rm st}$ is a density with respect to the
flat matrix measure $\dd\Lambda$.
The density is invariant under conjugation and
hence symmetric in the eigenvalues of $\Lambda$;
no restriction to an ordered eigenvalue chamber
is imposed.
For $N=1$, Eq.~\eqref{eq:Pst} is Halperin's
density~\eqref{eq:Halperin}. We identify the density $P_{\rm st}(\Lambda)$ of Eq.~\eqref{eq:Pst} with
the current-carrying stationary density of the real-energy Riccati
matrix (Sec.~\ref{sec:justify}). In this description, Riccati poles
represent passages through infinity rather than termination of the
evolution: a diverging eigenvalue re-enters from the opposite infinity.
In the variables $(A,Y)$ the same event is a passage of an eigenvalue of
$Y=\cT^{-1}$ through zero. At $\eta=0$ the boundary $\det Y=0$ is
therefore reached, and the appropriate boundary condition is
reinjection, not zero flux. The density $e^\Phi$ at $\eta=0$ does not
vanish there, and the flux through this boundary is the winding current.

For $\beta=2$, this identification implies
agreement with the fermionic expression of
Ref.~\cite{Letter}. A direct demonstration
of the equality of the two expressions requires a considerable effort and is presented
in Appendix~\ref{app:matrix-fermion}.
For general $\beta>0$ the eigenvalue form of Eq.~\eqref{eq:Pst} can also be
tested directly against the stationary Fokker--Planck equation of the
eigenvalue gas~\eqref{eq:DBM}, with zero flux at eigenvalue collisions and
reinjection at $+\infty$ of the eigenvalue escaping to $-\infty$. This is done
in Ref.~\cite{FyodorovAD}, rigorously for $N=2$ and at the physics level for
general $N$. That construction also yields the positive-integral
representation of the normalization obtained from
\eqref{eq:Yens} at $\eta=0$, together with
Eqs.~\eqref{eq:Gammapos} and~\eqref{eq:nupos}
for the principal-value mean trace and the winding
current, respectively, for every $\beta>0$.

To describe finite absorption, express both
distributions in the same variables
$(A,\cS)$, with $\cS=Y^{-1}$. At $\eta>0$,
$\cS=P/\eta$ and $Z=A-\ii\eta\cS$.
The joint distribution then satisfies
\be
\mu_{E,\eta}
=\frac{\cZ_N(E,0)}{\cZ_N(E,\eta)}
e^{-\eta^2\Tr\cS/T}\,\mu_{E,0},
\qquad
Z=A-\ii\eta\cS .
\label{eq:tiltN}
\ee
Thus absorption enters as an exponential reweighting
of the zero-absorption distribution represented
in $(A,\cS)$ coordinates. In the real-energy
limit, $A=Z$ and $\cS=\cT=-\partial_E Z$;
at finite absorption, $\cS=P/\eta$ need not
be identified with this energy derivative.
The heavy tail of $\Tr\cS$ prevents a
term-by-term expansion of the exponential
under the integral. As shown below, the
normalization instead acquires a
non-analytic term linear in $\eta$.

\subsection{$\Gamma_N$ and $\nu_N$ as ratios of positive integrals}

The results of this subsection use only the stationary
joint distribution of $(A,Y)$ in Eq.~\eqref{eq:law},
its limit as $\eta\to0_+$, and the boundary values
of $\Xi_N$ discussed in Sec.~\ref{sec:justify}. They do
not use the identification of Eq.~\eqref{eq:Pst}
with the real-energy stationary state. Here unlabelled brackets denote
expectation with respect to Eq.~\eqref{eq:law} at $\eta>0$
(or its corresponding marginal for observables depending only on $Y$);
real-energy averages are understood as the limits specified below.

Both quantities follow by differentiating the positive integral
$\mathcal Y_N(E,\eta)$ defined after Eq.~\eqref{eq:Yens}, in which $E$ and $\eta$ enter only
through the factor $e^{-(E\,\Tr Y+\eta^2\Tr Y^{-1})/T}$.

{\it Sum of Lyapunov exponents.} Differentiation with respect to $E$
brings down $-\Tr Y/T$, so $-T\partial_E\log \mathcal Y_N=\aver{\Tr Y}$. Since
$A=B+\frac12Y$ with $B$ conditionally Gaussian of zero mean,
$\Gamma_N=\Rea\aver{\Tr Z}=\aver{\Tr A}=\frac12\aver{\Tr Y}$. At real
energy this gives
\be
\Gamma_N(E)=-\frac{T}{2}\,\partial_E\log \mathcal Y_N(E,0).
\label{eq:Gammapos}
\ee
At $\eta=0$ the mean of $\Tr A$ is not absolutely convergent, because of
the $\lambda^{-2}$ tails of Eq.~\eqref{eq:current}. It is understood as
the limit $\eta\to0_+$ of the convergent averages at $\eta>0$, as in
Sec.~\ref{sec:justify}. The mean $\aver{\Tr Y}$ itself converges.

{\it Integrated density of states.} Differentiation with respect to
$\eta$ brings down $-2\eta\Tr Y^{-1}/T$. With $P=\eta Y^{-1}$,
\[
\aver{\Tr P}=\eta\aver{\Tr Y^{-1}}
=-\frac T2\,\partial_\eta\log \mathcal Y_N(E,\eta),\qquad \eta>0 .
\]
Since $\Imn\aver{\Tr Z}=-\aver{\Tr P}$, Eq.~\eqref{eq:main} gives
$\aver{\Tr P}\to\pi\nu_N(E)$ as $\eta\to0_+$.

The leading non-analytic dependence of
$\mathcal Y_N(E,\eta)$ comes from configurations
in which one eigenvalue $y_i$ of $Y$ is of order
$\eta^2/T$. The relevant boundary integral is
\[
\int_0^\infty y^{-1/2}
\left(e^{-\eta^2/(Ty)}-1\right)\,\dd y
=-2\sqrt{\pi/T}\,\eta,
\qquad \eta>0.
\]
As $y_1\to 0$, the pair-interaction factors in the
joint eigenvalue density~\eqref{eq:Yens} satisfy,
for each remaining eigenvalue $y_j$,
\[
|y_j-y_1|^\beta(y_j+y_1)^{-\beta/2}
\longrightarrow y_j^{\beta/2}.
\]
Together with the original one-body factor
$y_j^{-1/2}$, this gives the power
$y_j^{(\beta-1)/2}$ in the boundary integral
\be
\mathcal Y^+_{N-1}(E)
=\int_{\R_+^{N-1}}
|\Delta(\bm y)|^\beta
\prod_i y_i^{(\beta-1)/2}
\prod_{i<j}(y_i+y_j)^{-\beta/2}
e^{-\frac1T\sum_i(Ey_i+y_i^3/12)}
\prod_i\dd y_i .
\ee

There are $N$ equivalent choices of the vanishing
eigenvalue. Boundary power counting shows that
configurations with two or more small eigenvalues
contribute only $o(\eta)$: when $k$ eigenvalues
of $Y$ are simultaneously of order $\eta^2/T$,
their combined boundary weight scales as
$\eta^{k+\beta k(k-1)/2}$. Consequently,
\[
\mathcal Y_N(E,\eta)
=\mathcal Y_N(E,0)
-2\sqrt{\pi/T}\,N\,\eta\,\mathcal Y^+_{N-1}(E)
+o(\eta).
\]
Applying the same boundary rescaling directly
to the differentiated integral gives
\begin{equation}
\lim_{\eta\to0_+}\partial_\eta\mathcal Y_N(E,\eta)
=-2\sqrt{\pi/T}\,N\,\mathcal Y^+_{N-1}(E).
\label{derY}
\end{equation}
Combining the relation
\[
\nu_N(E)
=-\frac{T}{2\pi}\lim_{\eta\to0_+}
\partial_\eta\log\mathcal Y_N(E,\eta)
\]
with the limiting derivative in \eqref{derY}
and the convergence
$\mathcal Y_N(E,\eta)\to\mathcal Y_N(E,0)$,
we obtain
\be
\nu_N(E)
=\sqrt{\frac T\pi}\,
\frac{N\,\mathcal Y^+_{N-1}(E)}
{\mathcal Y_N(E,0)}.
\label{eq:nupos}
\ee

For $N=1$, $\mathcal Y^+_0=1$, and
Eq.~\eqref{eq:nupos} reproduces Halperin's
integrated density of states,
\[
\nu_1(E)
=\frac{T^{1/3}}
{\pi^2\left[
\Ai^2(-E/T^{2/3})+\Bi^2(-E/T^{2/3})
\right]} .
\]
Comparison with Eq.~\eqref{eq:main} further gives
\[
\mathcal Y^+_{N-1}
\propto-\Imn\bigl(\tau_N'\bar\tau_N\bigr),
\]
with the derivative taken in the convention
of that equation. This expresses the positive
boundary integral through the outgoing
Kontsevich integral and its derivative.

Thus the integrated density of states is
expressed as a boundary contribution from
configurations with one vanishing inverse
delay $y_i$. This is the positive-integral
counterpart of the current associated with
Riccati passages through infinity,
generalizing the scalar pole-matching
construction of Sec.~\ref{sec:N1step6}.

\subsection{Justification of the limit}
\label{sec:justify}

The stationary distribution~\eqref{eq:law} makes the passage to zero absorption
transparent for typical configurations. As $\eta\to0_+$, the normalized
density of $(A,Y)$ approaches its integrable zero-absorption limit.
Consequently, the probability of $P=\eta Y^{-1}$ exceeding any fixed
positive threshold tends to zero, and $Z$ approaches the real matrix
$\Lambda$. This does not imply that averages containing $P$ vanish:
rare configurations with small eigenvalues of $Y$ can survive in such
averages, as the finite limit $\aver{\Tr P}\to\pi\nu_N$ illustrates. The way this
happens is fixed by the stationarity of $\log\det P$, which gives
\be
\eta\,\aver{\Tr P^{-1}}_{\mu_\eta}=2\,\Rea\aver{\Tr Z}_{\mu_\eta}\
\xrightarrow[\eta\to0]{}\ 2\,\Gamma_N ;
\label{eq:collapse}
\ee
for the explicit distribution this is just $\aver{\Tr Y}=2\aver{\Tr A}$. Thus
$\aver{\Tr P^{-1}}\simeq2\Gamma_N/\eta$: the imaginary part of $Z$
vanishes on the scale set by comparing the level broadening $\eta$ with
the total Lyapunov growth rate $\Gamma_N$.

The characteristic function follows the same limit without difficulty.
For $q\ge0$ the factor $e^{-\ii q\Tr Z/T^{1/3}}$ is bounded and smooth up
to real $Z$, so its average tends to the real-energy characteristic
function $\cF_N(q;E)$. On the other side of Eq.~\eqref{eq:Fcomplex}, the
ratio of $\tau$-functions is continuous up to real $s$, because $\tau$
has no zeros there, Eq.~\eqref{eq:zerofree}. Hence
\be
\cF_N^{(\beta)}(q;E)
=\frac{\tau_N^{(\beta)}(s-\ii q)}{\tau_N^{(\beta)}(s)},
\qquad E\in\R,\ q>0 .
\label{eq:Freal}
\ee
The right-hand side is a smooth function of $q$, and its expansion
$1-\ii q\,\partial_s\log\tau_N^{(\beta)}(s)+O(q^2)$ for $q\to0_+$ can be
compared with the one-sided expansion~\eqref{eq:smallq}. This gives
Eq.~\eqref{eq:main}. The comparison is legitimate only in this order: we
first let $\eta\to0_+$ at fixed $q>0$, and only then let $q\to0_+$. The
full characteristic function is not analytic at $q=0$, since its
negative-$q$ branch is the complex conjugate. With this order of limits
the multichannel Thouless relation $\Xi_N(E+\ii0)=\Gamma_N-\ii\pi\nu_N$
need not be assumed; it comes out as a by-product. Conversely, one may
start from it: $\Xi_N=T^{1/3}\partial_\fs\log\tau_N^{(\beta)}$ is
holomorphic for $\eta>0$, has no poles on the real axis, and by the
Thouless formula~\cite{KotaniSimon1988} its boundary value is
$\Gamma_N-\ii\pi\nu_N$. The two routes thus give the same boundary values of $\Xi_N$.

Equations~\eqref{eq:twomatrix} and~\eqref{eq:Pst}
give the matrix analogue of Halperin's real-energy
stationary distribution. The reinjection of Riccati
eigenvalues from $-\infty$ to $+\infty$ requires
matching tails
$n(\lambda)\simeq\nu_N/\lambda^2$ at both ends,
as in Eq.~\eqref{eq:current}. These tails follow
directly from the explicit distribution.

Conditioned on $Y$, the matrix $B=A-\frac12Y$
is Gaussian, with diagonal variances $T/(2y_i)$
in the eigenbasis of $Y$. The leading contribution
to a large eigenvalue $\lambda$ of $\Lambda=A$
comes from a single small eigenvalue $y_1$ of $Y$,
with $\lambda\simeq B_{11}$. Near $y_1=0$,
the density~\eqref{eq:Yens} behaves as
$y_1^{-1/2}$ times the integrand of
$\mathcal Y^+_{N-1}$, as in Sec.~\ref{sec:realE}.
Hence
\[
n(\lambda)\simeq
\frac{N\mathcal Y^+_{N-1}(E)}{\mathcal Y_N(E,0)}
\int_0^\infty
\frac{\dd y}{\sqrt y}\sqrt{\frac{y}{\pi T}}\,
e^{-y\lambda^2/T}
=
\sqrt{\frac T\pi}\,
\frac{N\mathcal Y^+_{N-1}(E)}{\mathcal Y_N(E,0)}
\,\frac1{\lambda^2},
\qquad \lambda\to\pm\infty.
\]
By Eq.~\eqref{eq:nupos}, the coefficient is
$\nu_N(E)$ at both ends, confirming the
reinjection flux required by
Eq.~\eqref{eq:current}.

\section{Explicit finite-$N$ formulas}
\label{sec:explicit}

We now evaluate the holomorphic factor $\tau_N^{(\beta)}$ explicitly. All
integrals run over the outgoing Airy contour $\mathcal C_-$, which comes
in from infinity along $\arg w=\pi/3$ and goes out to infinity along
$\arg w=\pi$, with the weight $e^{\varphi(w)}$, $\varphi(w)=w^3/3-sw$.
After the angular integration, $\tau_N^{(\beta)}$ is an $N$-fold integral
of $\Delta(w)^\beta\prod_ie^{\varphi(w_i)}$, with the ordering and
branch convention of Eq.~\eqref{eq:tau}, and the three classes
differ only in how the Vandermonde factor is handled. The single-variable
building block is always the Airy function $\cA_-$ and its derivatives,
obtained by inserting powers of $w$ under the integral.

For $\beta=2$ the Vandermonde factor appears squared, and Andr\'eief's
identity reduces the integral to a determinant of moments,
$\tau_N^{(2)}\propto\det_{0\le j,k\le N-1}[\cA_-^{(j+k)}(s)]$~\cite{Letter}.
This is the Hankel determinant of the unitary problem.

For $\beta=4$ the same idea works once the fourth power of the
Vandermonde determinant is written as a confluent Vandermonde
determinant, $\Delta(w)^4=\det[w_i^k,\ k\,w_i^{k-1}]_{k=0,\dots,2N-1}$.
Each variable then carries a pair of columns, a power and its
derivative, and de Bruijn's integration formula~\cite{deBruijn1955}
turns the integral into a Pfaffian of single-variable moments:
\be
\tau_N^{(4)}(s)\ \propto\ \Pf_{0\le j,k\le 2N-1}\Big[(k-j)\,\cA_-^{(j+k-1)}(s)\Big].
\label{eq:tau4}
\ee
The entry with $j=k=0$ is understood to be zero. For
$N=1$ the Pfaffian is the single entry $\cA_-$, as it must be.

For $\beta=1$, the contour representation involves
a single Vandermonde, which is antisymmetric under
permutations of the integration variables. The integral
must therefore retain their ordering along the oriented
contour $\mathcal C_-$. Applying de Bruijn's formula
to this ordered integral gives for even $N$:
\be
\tau_N^{(1)}(s)=\Pf_{0\le j,k\le N-1}\big[a_{jk}(s)\big],\qquad
a_{jk}=\int\!\!\int_{\mathcal C_-\times\mathcal C_-}\mathrm{sgn}(w'\succ w)\;
w^jw'^k\,e^{\varphi(w)+\varphi(w')}\,\dd w\,\dd w' .
\label{eq:tau1}
\ee
Here $\mathrm{sgn}(w'\succ w)$ equals $+1$ if $w'$
occurs after $w$ along the oriented contour
$\mathcal C_-$, and $-1$ otherwise. It records contour
ordering, not the sign of a complex number.

The simplest case is $N=2$, where
$\tau^{(1)}_2=a_{01}=\int\!\!\int_{w\prec w'}(w'-w)\,e^{\varphi(w)+\varphi(w')}$.
Section~\ref{sec:beta4edge} shows how for even $N$
these Pfaffians can be expressed in terms of the
recurrence coefficients and squared norms of the
orthogonal polynomials for the unitary problem
at temperature $T/2$.

For odd $N$ the matrix is bordered by the single moments
$b_j=\int w^je^{\varphi}$. As usual for $\beta=1$, these ordered double
integrals do not reduce to derivatives of $\cA_-$ alone. Each of them is
a single integral of an Airy-type function against its own primitive
along the contour,
$a_{jk}=\int w'^ke^{\varphi(w')}\big[2\alpha_j(w')-\alpha_j(\infty)\big]\dd w'$,
with $\alpha_j(w')=\int_{e^{\ii\pi/3}\infty}^{w'}w^je^{\varphi(w)}\dd w$.
We do not treat odd-$N$ case in this paper.

All three formulas reduce to the same answer for a single channel:
$\tau_1=\cA_-$ and $\Xi_1=T^{1/3}\cA_-'/\cA_-$ for every $\beta$. This is
Halperin's density of states~\cite{Halperin1965} together with the
associated Lyapunov exponent.
\section{High energy: genus expansion with non-orientable surfaces}
\label{sec:highE}

In this section, the high-energy limit is taken at fixed $N$;
the expansion is not uniform as $N\to\infty$.

At high energies $E\gg T^{2/3}$ the disorder produces
small corrections to the free-particle density of states,
while the Lyapunov exponents describe the weak localization
of the propagating channels. Both effects can be extracted
from the same saddle-point expansion of the outgoing
Kontsevich integral. In this regime $s\to-\infty$, and the
two saddles of $\varphi(w)=w^3/3-sw$ are
$w_\pm=\pm\ii|s|^{1/2}$. The outgoing contour $\mathcal C_-$
selects $w_0=\ii|s|^{1/2}$, around which all $N$ integration
variables fluctuate.

Writing $w_j=w_0+u_j$ we obtain
\be
\varphi(w_0+u)
=\varphi(w_0)+w_0u^2+\tfrac13u^3 .
\ee
This expansion is exact because the potential is cubic.
The quadratic term defines a Gaussian $\beta$-ensemble,
while the cubic term is treated perturbatively. Write
\[
p_k=\sum_{j=1}^N u_j^k,
\qquad
\sigma=-\frac{1}{2w_0}.
\]
To distinguish these averages from the stationary Riccati
averages used earlier, we denote normalized Gaussian expectation by
$\aver{\cdot}_{\mathrm G,\beta}$. For $\sigma>0$, its explicit definition is
\begin{eqnarray}
\aver{F}_{\mathrm G,\beta}
&:=\frac{1}{\mathcal G_{N,\beta}(\sigma)}
\int_{\mathbb R^N} F(u)\,|\Delta(u)|^\beta
\exp\!\left(-\frac{1}{2\sigma}\sum_{j=1}^N u_j^2\right)
\prod_{j=1}^N\dd u_j,\\
&\mathcal G_{N,\beta}(\sigma)
:=\int_{\mathbb R^N}|\Delta(u)|^\beta
\exp\!\left(-\frac{1}{2\sigma}\sum_{j=1}^N u_j^2\right)
\prod_{j=1}^N\dd u_j.
\label{eq:Gaussian-average}
\end{eqnarray}
Thus $\aver{1}_{\mathrm G,\beta}=1$; the dependence on $N$ and
$\sigma$ is suppressed in the bracket notation. For the complex value
$\sigma=-1/(2w_0)$, the polynomial moments are defined by analytic
continuation from $\sigma>0$, with the continuation fixed by the local
steepest-descent contour. The real-axis absolute Vandermonde in
Eq.~\eqref{eq:Gaussian-average} specifies the starting measure, not
an absolute-value prescription on the complex contour.
The Gaussian normalization contributes a factor
$(-w_0)^{-a_\beta}$, where
\[
a_\beta=\frac N2+\frac{\beta N(N-1)}4.
\]
The first-order contribution of the cubic perturbation vanishes
because the cubic term is odd in the fluctuations about the
stationary point: under the simultaneous sign reversal
$u_j\mapsto-u_j$ for all $j$, $p_3$ changes sign, whereas the
Gaussian measure in Eq.~\eqref{eq:Gaussian-average} is invariant.
Consequently $\aver{p_3}_{\mathrm G,\beta}=0$, and the first nonzero
contribution therefore comes from two cubic vertices:
\be
\log\tau_N^{(\beta)}
\simeq N\varphi(w_0)-a_\beta\log(-w_0)
+\frac1{18}\aver{p_3^2}_{\mathrm G,\beta}+\dots,
\qquad
\aver{p_3^2}_{\mathrm G,\beta}=K_\beta(N)\,\sigma^3 .
\label{eq:genus}
\ee
An additive constant independent of $s$ has been omitted
since it does not affect the characteristic exponent.

The high-energy correction is thus determined by a single
Gaussian moment. It can be evaluated for all three symmetry
classes using integration by parts, without performing
separate Wick-contraction calculations.  To derive the required identity, first take
$\sigma>0$ and write the normalized Gaussian
$\beta$-ensemble density as
\[
P_\beta(u)=\frac1{\mathcal G_{N,\beta}(\sigma)}
|\Delta(u)|^\beta
e^{-\sum_j u_j^2/(2\sigma)}.
\]
Away from coincident eigenvalues, its logarithmic
derivative is
\[
\partial_i\log P_\beta(u)
=-\frac{u_i}{\sigma}
+\beta\sum_{j\ne i}\frac1{u_i-u_j}.
\]
For a symmetric polynomial $g$ integrating the total
derivative $\sum_i\partial_i[u_i^k gP_\beta]$ gives zero.
The Gaussian decay removes the boundary terms at
infinity, while the Vandermonde factor removes them
at coincident eigenvalues. Expanding the derivative
therefore gives
\[
0=
\aver{\sum_i\partial_i(u_i^kg)}_{\mathrm G,\beta}
-\frac1\sigma\aver{p_{k+1}g}_{\mathrm G,\beta}
+\beta\left\langle
\sum_i\sum_{j\ne i}\frac{u_i^k}{u_i-u_j}\,g
\right\rangle_{\mathrm G,\beta} .
\]
Pairing the terms with indices $(i,j)$ and $(j,i)$,
we obtain
\be
\frac1\sigma\aver{p_{k+1}g}_{\mathrm G,\beta}
=
\aver{\sum_i\partial_i(u_i^kg)}_{\mathrm G,\beta}
+\beta\left\langle
\sum_{i<j}\frac{u_i^k-u_j^k}{u_i-u_j}\,g
\right\rangle_{\mathrm G,\beta} .
\ee
These integration-by-parts identities are usually
called Gaussian loop equations. For the polynomial
moments used below, they extend by analytic
continuation to the complex value
$\sigma=-1/(2w_0)$ selected by the saddle.
Introducing
$m_2=N+\beta N(N-1)/2$, the moments needed here follow
successively as
\begin{align*}
\aver{p_2}_{\mathrm G,\beta}
&=\sigma m_2,
\\
\aver{p_1p_3}_{\mathrm G,\beta}
&=3\sigma^2m_2,
\\
\aver{p_4}_{\mathrm G,\beta}
&=\sigma^2\left[
3m_2+\beta\left((N-1)m_2+\tfrac12(N-m_2)\right)
\right],
\\
\aver{p_3^2}_{\mathrm G,\beta}
&=\sigma\left[
(2+\beta(N-1))\aver{p_1p_3}_{\mathrm G,\beta}
+3\aver{p_4}_{\mathrm G,\beta}
\right].
\end{align*}
Combining these relations yields
\be
K_\beta(N)
=\tfrac{3N}{4}
\left[
4\beta^2N^2-9\beta^2N+5\beta^2
+18\beta N-18\beta+20
\right].
\label{eq:Kbeta}
\ee
In particular,
\[
K_2=3N(4N^2+1),\qquad
K_1=\frac{3N}{4}(4N^2+9N+7),\qquad
K_4=3N(16N^2-18N+7).
\]

The dependence on $N$ has a topological interpretation
in terms of ribbon graphs generated by the two cubic
vertices; see, e.g., Chapter~3 of
Ref.~\cite{LandoZvonkin2004}. In the unitary class, only orientable surfaces
contribute: the $N^3$ and $N$ terms correspond to the
planar and torus contributions, respectively. For
$\beta=1,4$, an additional $N^2$ term appears, reflecting
the contribution of non-orientable surfaces. 

The real and quaternion coefficients obey the polynomial identity
\[
K_4(N)=-2K_1(-2N).
\]
This is the two-cubic-vertex manifestation, in our normalization,
of the orthogonal--symplectic Gaussian-integral duality studied
by Mulase and Waldron~\cite{MulaseWaldron2003}.
The substitution $N\mapsto-2N$ is understood formally at the
level of the diagrammatic polynomials, not as an actual
negative matrix dimension. After the appropriate normalization,
the two ensembles have the same surface expansion, except
that non-orientable surfaces of odd Euler characteristic
contribute with opposite signs.
For two cubic vertices, $V=2$ and $E=3$, so
$\chi=V-E+F=F-1$. The $N^2$ term therefore corresponds to
$\chi=1$, i.e.\ the projective plane, explaining its opposite
sign in the real and quaternion classes.

We now return from the Gaussian expansion to the physical
observables. Using
$\partial_s\varphi(w_0)=-w_0$ and
$\partial_sw_0=1/(2w_0)$, and restoring physical units
with $z_0=-\ii\sqrt E$, we find
\be
\Xi_N^{(\beta)}
=Nz_0-\frac{a_\beta T}{2z_0^2}
-\frac{K_\beta(N)\,T^2}{96\,z_0^5}
+O(T^3).
\ee
The real and imaginary parts give the sum of Lyapunov exponents and
the density of states:
\be
\Gamma_N
=\frac{N(\beta N+2-\beta)\,T}{8E}+\dots,
\qquad
\frac{\rho_N}N
=\frac1{2\pi\sqrt E}
\left[
1-\frac{5K_\beta(N)}{96N}\frac{T^2}{E^3}
+\dots
\right].
\label{eq:highE}
\ee
The leading density is the free-particle result obtained by completely neglecting the random potential in the original matrix Schr\"odinger equation.
Localization effects reflected in the positivity of Lyapunov's appear already at order $T$, whereas
the first correction to the density of states is of
order $T^2$.

The leading Lyapunov sum can also be written as
\[
\Gamma_N
=\frac{T}{4E}
\sum_{j=1}^N[1+\beta(j-1)]+\dots.
\]
Its channel dependence is therefore consistent with
the familiar isotropic, DMPK-type ladder of weak-disorder
Lyapunov exponents
$\gamma_j\propto1+\beta(j-1)$
~\cite{MelloPereyraKumar1988}.
The symmetry class enters through the spacing of this
ladder, while $T/E$ sets its overall scale.

\section{The Lifshitz tail: one-eigenvalue tunnelling}
\label{sec:Lifshitz}

We now turn to the opposite regime, $-E\gg T^{2/3}$,
where states arise from rare fluctuations of the random
potential. Throughout this section the deep-tail limit is taken
at fixed $N$. The joint large-$N$ edge regime is
treated separately in Sec.~\ref{sec:largeN}. In the Kontsevich representation, their
exponentially small density is encoded by tunnelling
between the two saddles of the cubic potential.
For $s\to+\infty$ these saddles lie on the real axis,
\[
w_\pm=\pm\sqrt s,
\qquad
\varphi(w_\mp)=\pm\frac23s^{3/2}.
\]
The outgoing contour $\mathcal C_-$ can be deformed to
come in from $e^{\ii\pi/3}\infty$, cross $w_+$ vertically,
and then run left along the real axis through the
dominant saddle $w_-$ before leaving to $-\infty$.

The dominant configuration has all $N$ integration
variables near $w_-$. Its contribution is real and gives
the leading Lyapunov growth,
$\Gamma_N\simeq N\sqrt{-E}$, but no density of states.
The latter comes from the exponentially smaller
configuration in which one integration variable moves
to $w_+$ while the remaining $N-1$ stay near $w_-$.
We denote these two contributions to the partition
function by $\tau_N^{(0)}$ and $\tau_N^{(1)}$,
respectively. The superscripts here count tunnelling
eigenvalues, not symmetry classes.

The exponential cost of this process is
\[
e^{\varphi(w_+)-\varphi(w_-)}
=e^{-\frac43s^{3/2}}.
\]
Restoring physical units gives Halperin's action,
$S=\frac43(-E)^{3/2}/T$. At fixed $T$, this action is
independent of both $N$ and $\beta$: a single tunnelling
eigenvalue carries the rare state. The channel number
and symmetry class enter through its prefactor.

The phase and the Stokes multiplier of the tunnelling
contribution are fixed by the scalar problem. Indeed,
for $N=1$ the outgoing Airy combination
$\cA_-=\Ai-\ii\Bi$ gives
\[
\frac{\Ai(s)}{-\ii\Bi(s)}
\simeq\frac{\ii}{2}e^{-\frac43s^{3/2}}.
\]
The factor $\frac12$ reflects that the positive real
$s$-axis is a Stokes line. For general $N$, this scalar
tunnelling amplitude must be combined with the
interaction between the tunnelling eigenvalue and the
remaining eigenvalues, and with the Gaussian
normalizations around the two saddles.

At leading order the cross-factors in the Vandermonde
weight,  coupling the tunnelling eigenvalue to the
remaining $N-1$ eigenvalues,  contribute
\[
(w_--w_+)^{\beta(N-1)}
=(-2\sqrt s)^{\beta(N-1)}.
\]
The ratio of Gaussian $\beta$-ensemble normalizations,
with curvature $2\sqrt s$ at $w_-$, can be obtained as follows. Let $G_n(a)$ denote the ordered Gaussian
$\beta$-ensemble integral
\[
G_n(a)=
\int_{u_1<\cdots<u_n}
|\Delta(u)|^\beta
e^{-\frac a2\sum_j u_j^2}
\prod_{j=1}^n\dd u_j,
\qquad G_0(a)=1.
\]
The Mehta formula~\cite{Forrester2010} gives
\[
G_n(a)=\frac{(2\pi)^{n/2}}{n!}\,
a^{-\frac n2-\frac{\beta n(n-1)}4}
\prod_{j=1}^n
\frac{\Gamma(1+j\beta/2)}{\Gamma(1+\beta/2)}.
\]
Consequently, with $a=2\sqrt s$,
\[
\frac{G_1(a)G_{N-1}(a)}{G_N(a)}
=
N\,\frac{\Gamma(1+\beta/2)}
{\Gamma(1+N\beta/2)}
(2\sqrt s)^{\beta(N-1)/2}.
\]

The contour orientation must also be retained.
The tunnelling eigenvalue comes first along
$\mathcal C_-$, whereas the contour near $w_-$ runs in
the negative direction. In the real class, these
orientation signs cancel the signs of the ordered
Vandermonde. Combining the factors gives
\be
\frac{\tau_N^{(1)}}{\tau_N^{(0)}}
=
\frac{\ii}{2}e^{-\frac43s^{3/2}}\,
N\frac{\Gamma(1+\beta/2)}
{\Gamma(1+N\beta/2)}
(2\sqrt s)^{\frac{3\beta(N-1)}2}
\left[
1+\frac{\kappa_N}{s^{3/2}}+\dots
\right].
\ee

To obtain the first correction, we must include both
the cubic fluctuations around the saddles and the
variation of the cross Vandermonde. The cubic
contribution follows from the Gaussian moments
computed in Sec.~\ref{sec:highE}. The fluctuation
variance is $+1/(2\sqrt s)$ at $w_-$ and
$-1/(2\sqrt s)$ at $w_+$, so the two saddle sectors
give the contribution
\[
\frac{
K_\beta(N-1)-K_\beta(N)-K_\beta(1)
}{144}
\]
to $\kappa_N$.

For the Vandermonde correction, let $u_1$ be the fluctuation of the
tunnelling eigenvalue around $w_+$ and let $u_j$, $j=2,\ldots,N$, be
the fluctuations around the dominant saddle $w_-$. Writing
$\delta_j=u_j-u_1$ and $L_1=-\frac{\beta}{2\sqrt{s}}\sum_{j=2}^N\delta_j$,
we expand the cross factor $\prod_{j=2}^N|1-\frac{\delta_j}{2\sqrt{s}}|^\beta$.
Let $\aver{\cdot}_{\rm tun}$ denote the normalized Gaussian average around
the one-eigenvalue tunnelling configuration: $u_1$ fluctuates around $w_+$
with variance parameter $\langle u_1^2\rangle_{\rm tun}=-\sigma^2$ and
$\langle u_1^4\rangle_{\rm tun}=3\sigma^4$, where $\sigma^2=1/(2\sqrt{s})$,
while $u_2,\ldots,u_N$ form an independent Gaussian $\beta$-ensemble
around $w_-$ with variance parameter $\sigma^2$ (with the negative
parameter understood along the local steepest-descent contour).
At the required order, the three contributions are
\[
\left\langle
-\frac{\beta}{8s}\sum_{j=2}^N\delta_j^2
\right\rangle_{\rm tun},
\qquad
\frac12\aver{L_1^2}_{\rm tun},
\qquad
\left\langle
L_1\,\frac{p_3+u_1^3}{3}
\right\rangle_{\rm tun},
\]
where $p_k=\sum_{j=2}^N u_j^k$. The last term accounts for the interference
between the linear Vandermonde correction and the cubic vertices.
For the $(N-1)$-particle ensemble around $w_-$, the center of mass $p_1$
decouples from the translation-invariant Vandermonde factor, giving
$\aver{p_1^2}_{\rm tun}=(N-1)\sigma^2$. Integrating by parts with respect
to $u_j$ in the $(N-1)$-particle Gaussian $\beta$-ensemble (via
$\sum_{j=2}^N\partial_{u_j}(u_j P)=0$ and the translation-invariance
identity $\sum_{j=2}^N\partial_{u_j}(p_3 P)=0$) yields
\[
\aver{p_2}_{\rm tun}=(N-1)\sigma^2\left(1+\frac{\beta}{2}(N-2)\right),
\qquad
\aver{p_1 p_3}_{\rm tun}=3\sigma^2\aver{p_2}_{\rm tun}.
\]
Using the independence of $u_1$ from $\{u_2,\ldots,u_N\}$ and $\aver{u_1}_{\rm tun}=\aver{u_1^3}_{\rm tun}=0$, the three contributions evaluate to
\begin{align*}
\left\langle -\frac{\beta}{8s}\sum_{j=2}^N\delta_j^2 \right\rangle_{\rm tun}
&= -\frac{\beta}{8s}\Big(\aver{p_2}_{\rm tun}+(N-1)\aver{u_1^2}_{\rm tun}\Big)
= -\frac{\beta^2(N-1)(N-2)}{32\,s^{3/2}}, \\
\frac12\aver{L_1^2}_{\rm tun}
&= \frac{\beta^2}{8s}\Big(\aver{p_1^2}_{\rm tun}+(N-1)^2\aver{u_1^2}_{\rm tun}\Big)
= -\frac{\beta^2(N-1)(N-2)}{16\,s^{3/2}}, \\
\left\langle L_1\,\frac{p_3+u_1^3}{3} \right\rangle_{\rm tun}
&= -\frac{\beta}{6\sqrt{s}}\Big(\aver{p_1 p_3}_{\rm tun}-(N-1)\aver{u_1^4}_{\rm tun}\Big)
= -\frac{\beta^2(N-1)(N-2)}{16\,s^{3/2}}.
\end{align*}
Their respective contributions have the relative weights
$\frac1{32}:\frac1{16}:\frac1{16}$, and together give the coefficient $-\frac{5}{32}$. 
Thus
\[
\kappa_N=
\frac{
K_\beta(N-1)-K_\beta(N)-K_\beta(1)
}{144}
-\frac{5\beta^2(N-1)(N-2)}{32}.
\]

The integrated density of states is obtained by
differentiating the imaginary part of the tunnelling
contribution:
\[
\nu_N
=-\frac{T^{1/3}}{\pi}
\Imn\partial_s
\left(\frac{\tau_N^{(1)}}{\tau_N^{(0)}}\right)
\]
to leading exponential order. Besides differentiating
the tunnelling action, we must differentiate its
algebraic prefactor
$s^{3\beta(N-1)/4}$. This adds
$-3\beta(N-1)/8$ to the relative correction coefficient.
The resulting tail is Eq.~\eqref{eq:Lif}, with
\begin{align}
c_N^{(\beta)}
&=
\frac{
K_\beta(N-1)-K_\beta(N)-K_\beta(1)
}{144}
-\frac{5\beta^2(N-1)(N-2)}{32}
-\frac{3\beta(N-1)}8
\nonumber\\
&=
-\frac{
21\beta^2N^2-60\beta^2N+54\beta N
+39\beta^2-54\beta+20
}{96}.
\end{align}

As an immediate verification, for $\beta=2$ the general formula
reproduces both the prefactor $2^{3N-3}/(N-1)!$ and the subleading
correction $-(21N^2-33N+17)/24$ obtained in~\cite{Letter}, while for
$N=1$ all three symmetry classes reduce to Halperin's result, including
the $-5/24$ correction.

\paragraph*{Physical origin of the prefactor.}
The contour calculation has a complementary interpretation in terms of
rare optimal fluctuations of the random potential, which accounts for
the leading prefactor without invoking the integrable structure of
$\tau_N^{(\beta)}$.

For a single channel, the least costly potential
supporting a level at $E=-k^2$ is the reflectionless
P\"oschl--Teller well
\[
w_*(x)=-2k^2\sech^2(kx).
\]
Its action is precisely Halperin's $S=\frac{4}{3}(-E)^{3/2}/T$ in Eq.~\eqref{eq:Lif}. The fluctuation
determinant around this well reduces to a telescoping
product equal to $\frac15$, and the resulting
prefactor reproduces the scalar answer exactly.
The scalar calculation therefore fixes the
normalization of the optimal-fluctuation method.

For $N$ channels, the optimal well remains a
single-channel fluctuation, but its orientation in
channel space is arbitrary:
\[
W_*(x)=w_*(x)\,uu^\dagger.
\]
Here $u\in\F^N$ is a unit vector, and multiplication
by a unit of $\F$ leaves $W_*$ unchanged.
The family of equivalent wells is therefore
$\mathbb{RP}^{N-1}$, $\mathbb{CP}^{N-1}$, or
$\mathbb{HP}^{N-1}$ for $\beta=1,2,4$,
respectively. Its real dimension is $\beta(N-1)$.

These orientations are zero modes of the instanton.
Each contributes a collective-coordinate factor
$\sqrt{2S/\pi}$, together with a factor $\sqrt3$
from coupling to the empty channels. The latter
comes from level repulsion and a second telescoping
fluctuation determinant, equal to $\frac13$.
Integrating over the projective space of orientations
then gives exactly the leading prefactor in
Eq.~\eqref{eq:Lif}.

The leading multichannel tail thus has a simple
physical interpretation: Halperin's one-channel
rare well can point in any direction in channel
space. The dimension and volume of this orientation
space account for the symmetry-class dependence
of the leading prefactor, while the tunnelling
action remains that of the scalar problem.

\section{Large $N$: planar limit and the edge}
\label{sec:largeN}
The finite-channel solution also determines both the collective spectral
edge at $N=\infty$ and its finite-$N$ rounding at large $N$. Here we
outline the common structure of the three symmetry
classes; the symmetry-dependent edge profiles are
discussed in the following section.

We take $N\to\infty$ at fixed $J$, with $T=J^2/(2N)$.
At leading order, the saddle-point equation is unchanged
by multiplying the logarithm of the weight by a constant.
The relation
\[
|\Delta|^\beta e^{-\sum_j U_E/T}
=
\left(
|\Delta|^2e^{-\sum_j U_E/(\beta T/2)}
\right)^{\beta/2}
\]
therefore identifies the planar saddle with that of
the unitary problem at temperature $\beta T/2$.
Equivalently, the unitary result applies with
$J^2\to\beta J^2/2$. Defining $s_\beta=(\beta J^2)^{1/3}$
we obtain the limiting level density from the
curve identified for $\beta=2$ in \cite{Letter} by replacing $s_J$ with $s_\beta$. Its lower
edge is consequently situated at the threshold energy $E^*_\beta=-\frac34s_\beta^2$. 

The spectral edge marks the point where two solutions
of the planar saddle-point equation merge. At this
point the equation has a double root, and the ordinary
large-$N$ expansion ceases to be uniform. Finite-$N$
effects must therefore be retained to describe how the
sharp limiting edge is rounded.

In the unitary class, the paper \cite{Letter} resolves this rounding
through the recurrence relations of the orthogonal
polynomials. The resulting critical window
has width of order $N^{-4/5}$. The same energy scale
governs the other two symmetry classes, as discussed
below:
\[
E-E^*_\beta=O(s_\beta^2N^{-4/5}).
\]
The symmetry class thus changes the edge position and
the detailed rounding profile, but not the exponent
$4/5$.

At finite $N$, the density remains nonzero below $E^*_\beta$. Applying
the single-eigenvalue activation argument of Ref.~\cite{FLDO2025}, as we
do below, gives to leading exponential accuracy
$\log\rho_N\simeq-\mathcal A_\beta(E)/T$, where $\mathcal A_\beta$ is
the barrier in the effective potential created by the metastable planar
distribution.

Let $\varrho_{\rm pl}$ denote the normalized planar density of the
Riccati eigenvalues~\cite{Letter,FLDO2025}, supported on $[c-d,c+d]$
with $s_J$ replaced by $s_\beta$, where
\[
E=-c^2-\frac{d^2}{2},
\qquad
cd^2=\beta NT=\frac{s_\beta^3}{2},
\qquad c>\frac{s_\beta}{2},\quad d>0.
\]

The effective potential for a single eigenvalue $x$ escaping from the
metastable planar distribution is, up to an additive constant,
\[
V_{\rm eff}(x)
=U_E(x)-\beta NT
 \int_{c-d}^{c+d}\varrho_{\rm pl}(\lambda)
 \log|x-\lambda|\,\dd\lambda, 
\]
with $U_E(x)=Ex+x^3/3$  defined in Eq.~\eqref{eq:law}.
To the left of the support, the planar saddle gives
\[
V_{\rm eff}'(x)
=x^2+E-\beta NT
 \int_{c-d}^{c+d}
 \frac{\varrho_{\rm pl}(\lambda)}{x-\lambda}\,
 \dd\lambda
=-(x+c)\sqrt{(x-c)^2-d^2},
\qquad x<c-d,
\]
where the square root on the right is positive.
 For $x<c-d$, $V'_{\rm eff}(x)>0$ on $(-c,c-d)$ and $V'_{\rm eff}(x)<0$
for $x<-c$, so $x=-c$ is a local maximum (the escape saddle) separating
the metastable support $[c-d,c+d]$ from the runaway region $x\to-\infty$.
An eigenvalue escaping from the lower endpoint $c-d$ must therefore
surmount the potential barrier
\[
\mathcal A_\beta(E)=V_{\rm eff}(-c)-V_{\rm eff}(c-d)
=\int_{c-d}^{-c} V'_{\rm eff}(x)\,dx,
\]
which gives
\[
\begin{aligned}
\mathcal A_\beta(E)
&=\int_{-c}^{c-d}
(x+c)\sqrt{(x-c)^2-d^2}\,\dd x
=\int_d^{2c}(2c-z)\sqrt{z^2-d^2}\,\dd z \\
&=\frac{2c^2+d^2}{3}\sqrt{4c^2-d^2}
-cd^2\log\left(
\frac{2c+\sqrt{4c^2-d^2}}{d}
\right).
\end{aligned}
\]

At the spectral edge, $c=s_\beta/2$ and
$d=s_\beta$: the escape saddle meets the
lower endpoint of the support and the barrier
vanishes. Writing $\delta=E^*_\beta-E>0$, we find
\[
c=\frac{s_\beta}{2}
+\sqrt{\frac{\delta}{3}}
+O\left(\frac{\delta}{s_\beta}\right),
\qquad
d=s_\beta-\sqrt{\frac{\delta}{3}}
+O\left(\frac{\delta}{s_\beta}\right),
\]
and hence
\[
\mathcal A_\beta(E)
\sim
\frac45\sqrt2\,3^{1/4}s_\beta^{1/2}
\delta^{5/4}.
\]
The near-edge activation law applies when
$\delta\ll s_\beta^2$ but
$\mathcal A_\beta(E)/T\gg1$, outside the
finite-$N$ rounding region.

Deep in the forbidden region, put $k=\sqrt{-E}$.
For $k\gg s_\beta$, we have $c\sim k$ and
$d/c\to0$. Rescaling $z=ct$ gives
\[
\frac{\mathcal A_\beta(E)}{c^3}
=\int_{d/c}^{2}(2-t)\sqrt{t^2-(d/c)^2}\,\dd t
\longrightarrow
\int_0^2(2-t)t\,\dd t=\frac43.
\]
Thus $\mathcal A_\beta(E)\sim\frac43(-E)^{3/2}$, recovering the leading Lifshitz exponent
$\frac43(-E)^{3/2}/T$ of Sec.~\ref{sec:Lifshitz}.
The near-edge activation law and the deep Lifshitz tail are therefore two limits of the
same barrier; the finite-$N$ prefactor is determined separately in Sec.~\ref{sec:Lifshitz}.

Although the planar limit and the critical
exponent follow from this common saddle-point
picture, the full edge profile depends on the
symmetry class. For $\beta=2$, it is governed by
the complex tritronqu\'ee solution of
Painlev\'e~I~\cite{Letter} specified in
Eq.~\eqref{eq:yphys}, rather than the solution
that is real on the real axis.
For $\beta=1,4$, the finite-$N$ expansion also
contains non-orientable contributions.

The corresponding edge profiles can nevertheless
be obtained using the unitary solution.
For $\beta=4$, Sec.~\ref{sec:beta4edge} gives an
exact finite-$N$ representation through the unitary
orthogonal polynomials and an additional three-term
recursion, whose edge asymptotics supplies the
correction to the density profile.
The analogous construction for $\beta=1$ and its
critical limit are described in
Subsecs.~\ref{sec:b1sketch} and~\ref{sec:b1scaling}.
Those sections distinguish the exact finite-$N$
identities from the large-$N$ scaling arguments,
which are given at the physics level.

\section{The $\beta=4$ and $\beta=1$ edges: exact reduction to $\beta=2$ orthogonal polynomials}
\label{sec:beta4edge}

For $\beta=2$ the large-$N$ edge follows from the string equations for the recurrence
coefficients of polynomials orthogonal on $\Gamma_-$\cite{Letter}. This section shows that for $\beta=4$
the partition function is expressed {\it exactly}, at every finite $N$, through the
recurrence coefficients of the {\it same} polynomials, via one additional three-term
recursion.  It is the starting point for
the $\beta=4$ analogue of the Painlev\'e~I edge. Note that essentially the same structure is operative
behind the Tracy--Widom relation between the $\beta=4$
and $\beta=2$ soft-edge behaviour
~\cite{TracyWidom1996,Widom1999}. Subsections~\ref{sec:b4exact}
and~\ref{sec:b4Xi} give the exact $\beta=4$ reduction.
Subsection~\ref{sec:b1sketch} presents the analogous exact
reduction for $\beta=1$ and even $N$, proved in
Appendix~\ref{app:beta1reduction}.
Subsection~\ref{sec:b4scaling} derives the double-scaling
forms from these exact reductions at the physics level.
It also specifies the relevant Painlev\'e~I solution and
the selected solution of the auxiliary equation in each
class. The selections follow from analyticity in the upper
half energy plane and the $\beta$-dependence of the barrier;
for $\beta=1$ the finite-$N$ counterpart, the minimal-solution
condition for $X_0$, is left as a conjecture.

\subsection{Orthogonal polynomials and basic identities}

Let $\pi_n(\lambda)$ be the monic polynomials orthogonal on $\Gamma_-$ with the weight
$w(\lambda)=e^{-U_E(\lambda)/T}$,
\[
\int_{\Gamma_-}\pi_m\pi_n\,w\,\dd\lambda=h_n\delta_{mn},\qquad
\lambda\pi_n=\pi_{n+1}+d_n\pi_n+r_n\pi_{n-1},\qquad r_n=\frac{h_n}{h_{n-1}} .
\]
These are the same polynomials as used in \cite{Letter} and we briefly outline the procedure developed there, omitting the detail. 
Integration by parts along $\Gamma_-$ gives the ``string equations"
\be
d_n^2+r_n+r_{n+1}+E=0,\qquad r_n(d_n+d_{n-1})=nT ,
\ee
and the Toda relations $\partial_\theta\log h_n=d_n$, $\partial_\theta d_n=r_{n+1}-r_n$, where
$\theta=-E/T$ is the coefficient of $\lambda$ in $-U_E/T$. We write $\Zh$ for holomorphic
partition functions on $\Gamma_-$, to distinguish them from the positive normalization $\cZ_N(E,\eta)$.
For the unitary $\beta=2$ case
$\Zh^{(2)}_M\propto\prod_{n<M}h_n$ and hence $\Xi^{(2)}_M=\sum_{n<M}d_n$.

\subsection{Exact identity for $\Zh^{(4)}_N$}
\label{sec:b4exact}

\paragraph*{Exact reduction.} {\it Up to an $E$-independent constant $\Zh_N^{(4)}\ \propto\ \prod_{j=0}^{N-1}h_{2j+1}\;\cdot\;Q_N,$ with $Q_m$ satisfying the recursion
\be
Q_{m+1}=(d_{2m}+d_{2m+1})\,Q_m-r_{2m}\,Q_{m-1},\quad Q_0=1,\ \ Q_1=d_0+d_1 .
\label{eq:Z4exact}
\ee
Equivalently, $\Zh^{(4)}_N/\sqrt{\Zh^{(2)}_{2N}}\propto\prod_{j<N}r_{2j+1}^{1/2}\;Q_N$.}

{\sf Proof.}
By de Bruijn's formula~\cite{deBruijn1955}, $\Zh_N^{(4)} \propto\Pf\big[\langle\lambda^j,\lambda^k\rangle_4\big]_{0\le j,k<2N}$, 
where $\langle f,g\rangle_4=\int_{\Gamma_-}(fg'-f'g)\,w\,\dd\lambda$. 
Since the orthogonal polynomials $\pi_j$ are monic,
replacing $\lambda^j$ by $\pi_j$ is a unitriangular
change of basis. Its determinant is one, so the
Pfaffian is unchanged. We may therefore work with
the skew matrix $a_{jk}:=\langle\pi_j,\pi_k\rangle_4$.

Using $w'/w=-(E+\lambda^2)/T$ and integrating by parts
along $\Gamma_-$, we obtain
\[
a_{jk}
=-2\int_{\Gamma_-}\pi_j'\pi_k\,w\,\dd\lambda
+\frac1T\Phi_{jk},
\qquad
\Phi_{jk}
=\int_{\Gamma_-}\pi_j\pi_k(E+\lambda^2)\,
w\,\dd\lambda .
\]
The boundary terms vanish because the weight decays
at both ends of the contour. For $k>j$, the first
integral vanishes by orthogonality, since
$\deg\pi_j'<j<k$. Thus $a_{jk}=\Phi_{jk}/T$ for
$k>j$. Since $\Phi_{jk}=\Phi_{kj}$ and
$a_{jk}=-a_{kj}$, this gives
\[
a_{jk}
=\frac{\mathrm{sgn}(k-j)}{T}\Phi_{jk},
\qquad a_{jj}=0.
\]

The three-term recurrence $\lambda\pi_j=\pi_{j+1}+d_j\pi_j+r_j\pi_{j-1}$
shows that
\[
\lambda^2\pi_j
=\pi_{j+2}
+(d_j+d_{j+1})\pi_{j+1}
+\text{terms of degree at most }j.
\]
Orthogonality consequently gives
\[
\Phi_{j,j+1}
=h_{j+1}(d_j+d_{j+1}),
\qquad
\Phi_{j,j+2}=h_{j+2},
\qquad
\Phi_{jk}=0\quad\text{for }|j-k|\ge3.
\]
Hence the skew matrix $a_{jk}$ is five-diagonal.
The finite bandwidth is a direct consequence of the
quadratic logarithmic derivative $w'/w$.
Indeed, for $k>j$, integration by parts reduces
$a_{jk}$ to the overlap of $\pi_k$ with
$(E+\lambda^2)\pi_j$. This is a polynomial of degree
$j+2$, so orthogonality gives $a_{jk}=0$ whenever
$k>j+2$. Skew-symmetry gives the corresponding
restriction below the diagonal, leaving only the
first two off-diagonals on either side. More generally, a polynomial logarithmic derivative
of degree $d$ restricts the skew matrix to
$|j-k|\le d$. For a non-polynomial logarithmic
derivative, this degree argument no longer applies,
and the skew matrix need not have finite bandwidth.

Let now consider $
P_m=\Pf[a_{jk}]_{0\le j,k<2m},
\,\, P_0=1.$ In the expansion of $P_{m+1}$ along its last row,
the index $2m+1$ can be paired only with $2m$ or
$2m-1$. Pairing it with $2m$ leaves $P_m$.
Pairing it with $2m-1$ leaves the index $2m$,
which must then be paired with $2m-2$, leaving
$P_{m-1}$. The Pfaffian signs therefore give
\[
P_{m+1}
=a_{2m,2m+1}P_m
-a_{2m-1,2m+1}a_{2m-2,2m}P_{m-1},
\qquad m\ge1.
\]
Substituting the matrix entries yields
\[
P_{m+1}
=
\frac{h_{2m+1}}T
(d_{2m}+d_{2m+1})P_m
-\frac{h_{2m+1}h_{2m}}{T^2}P_{m-1}.
\]

We now factor out the odd-indexed norms by writing
\[
P_m=
\left[\prod_{j=0}^{m-1}\frac{h_{2j+1}}T\right]Q_m.
\]
Dividing the preceding relation by
$\prod_{j=0}^{m}(h_{2j+1}/T)$ gives
\[
Q_{m+1}
=(d_{2m}+d_{2m+1})Q_m
-\frac{h_{2m}}{h_{2m-1}}Q_{m-1}.
\]
Since $h_{2m}/h_{2m-1}=r_{2m}$, we obtain
\[
Q_{m+1}
=(d_{2m}+d_{2m+1})Q_m-r_{2m}Q_{m-1},
\qquad
Q_0=1,\qquad Q_1=d_0+d_1.
\]
This is precisely the recursion in~\eqref{eq:Z4exact}.
Finally,
\[
\Zh_N^{(4)}
\propto P_N
=T^{-N}\left[\prod_{j=0}^{N-1}h_{2j+1}\right]Q_N.
\]
At fixed $T$, the factor $T^{-N}$ is independent of
$E$ and is absorbed into the proportionality constant,
giving the stated partition-function identity.
\hfill$\square$

\subsection{Exact formula for the $\beta=4$ characteristic exponent}
\label{sec:b4Xi}

With $\Xi=-T\partial_E\log\Zh$ and the Toda relation $-T\partial_E\log h_n=d_n$, Eq.~\eqref{eq:Z4exact}
gives 
\be
\Xi^{(4)}_N=\sum_{j=0}^{N-1}d_{2j+1}-T\,\partial_E\log Q_N .
\label{eq:Xi4exact}
\ee
Here $-T\partial_E=\partial_\theta$ acts on $Q_N$ through $\partial_\theta d_n=r_{n+1}-r_n$ and
$\partial_\theta r_n=r_n(d_n-d_{n-1})$. It can therefore be evaluated recursively
alongside $Q_m$. It is instructive to check $N=1$.
Eq.~\eqref{eq:Xi4exact} gives
$d_1+r_2/(d_0+d_1)$, while the first two
string equations give $r_2=d_0^2-d_1^2$.
Thus the expression reduces to $d_0=\Xi_1$,
as it must: for one particle, the partition function
is independent of the Dyson index.

Comparing now with the $\beta=2$ relation $\Xi^{(2)}_{2N}=\sum_{n<2N}d_n$ we get
\be
\Xi^{(4)}_N=\tfrac12\,\Xi^{(2)}_{2N}+\tfrac12\sum_{j<N}\big(d_{2j+1}-d_{2j}\big)-T\,\partial_E\log Q_N .
\ee
The $\beta=4$ exponent is therefore half the $\beta=2$ exponent of the ensemble with
twice as many particles, with the {\it same} weight and the same $T$, plus 
corrections of two types: (i) an even--odd difference, which reduces at leading
order to a boundary term when the $d_n$ admit
a slowly varying smooth profile and (ii) the logarithmic derivative of the solution of a three-term recursion driven by
the $\beta=2$ data. The first term dominates in the large-$N$ limit
and determines the planar values
$s_4=(4J^2)^{1/3}$ and $E^*_4=-\frac34s_4^2$,
in agreement with Sec.~\ref{sec:largeN}. 

The recursion captures the part specific to the
Dyson index $\beta=4$ and reveals the mechanism
by which the Dyson-index-dependent correction
can survive in the double-scaling edge limit.
In the planar regime, $d_n\to d$ and $r_n\to r$,
and the recursion has characteristic roots
$\mu_\pm=d\pm\sqrt{d^2-r}$.
At the critical point $(d_*,r_*)=(s/2,s^2/4)$,
with $s=s_4$, the two characteristic roots coalesce.
After factoring out the common exponential growth,
the leading recurrence becomes a second difference.
In the double-scaling window, this second difference
balances the departures of the recurrence coefficients
from their critical values, so the recursion retains
a nontrivial continuum limit. Thus the separated-root
planar approximation breaks down precisely at
criticality, and the recursion must be retained
in the edge-scaling analysis. This continuum limit
is derived in Sec.~\ref{sec:b4scaling}.

\subsection{$\beta=1$, even $N$: an exact reduction of the same type}
\label{sec:b1sketch}

For $\beta=1$ the role of the unitary orthogonal polynomials is played by
those of the squared weight $e^{-2U_E(\lambda)/T}$, i.e.\ of the unitary
ensemble at temperature $T/2$. To distinguish the integration variable
from the weight, set 
\[
w=-\lambda/T^{1/3},\qquad s=-E/T^{2/3},\qquad
\rho_s(w)=e^{w^3/3-sw}.
\]
Let $p_n(w)$ be the monic polynomials orthogonal on the
outgoing contour $\mathcal C_-$ with respect to $\rho_s^2$:
\[
\int_{\mathcal C_-}p_j(w)p_k(w)\rho_s(w)^2\,\dd w
=\tilde h_j\delta_{jk},
\qquad
wp_n=p_{n+1}+\tilde d_n p_n+\tilde r_n p_{n-1},
\qquad
\tilde r_n=\frac{\tilde h_n}{\tilde h_{n-1}}.
\]
All tilded quantities below refer to this dimensionless
variable. The planar limit of this unitary ensemble with
$N$ particles coincides with that of the $\beta=1$
ensemble with $N$ particles, reproducing
$s_1=(J^2)^{1/3}$ and $E^*_1=-\frac34s_1^2$ of
Sec.~\ref{sec:largeN}.

For $N=2m\ge2$, the exact reduction, proved in
Appendix~\ref{app:beta1reduction}, is
\be
\begin{aligned}\tau^{(1)}_{2m}(s)
&=(-2)^m \left[\prod_{i=0}^{m-2}\tilde h_{2i+1}(s)\right]X_{m-1}(s),
\\
X_{k+1} &=(\tilde d_{2k}+\tilde d_{2k+1})X_k
  -\tilde r_{2k}X_{k-1},\qquad k\ge1 .
\end{aligned}
\label{eq:Z1X}
\ee
Here the product is empty for $m=1$ and is then set equal to unity.
The recursion starts from
\[
X_0(s)=\frac{\ii\pi}{3}
+\frac{\pi}{2}\int_0^{2^{2/3}s}
       [\Bi(x)+\ii\Ai(x)]\,\dd x,
\qquad
X_1=(\tilde d_0+\tilde d_1)X_0-\tilde h_0.
\]
We use $\tau^{(1)}_{2m}$ to denote explicitly the
holomorphic ordered integral of Eq.~\eqref{eq:tau}.
The factor $(-2)^m$ is fixed by that contour and
normalization convention and drops out of logarithmic
derivatives.

Since $\partial_s\log\tilde h_n=-2\tilde d_n$, the
characteristic exponent is
\be
\Xi^{(1)}_{2m}/T^{1/3}
=-2\sum_{i=0}^{m-2}\tilde d_{2i+1}
+\partial_s\log X_{m-1}.
\label{eq:Xi1exact}
\ee
The derivative can be evaluated recursively using the
Toda flows of the recurrence coefficients of the squared weight,
\[
\partial_s\tilde d_n
=-2(\tilde r_{n+1}-\tilde r_n),
\qquad
\partial_s\tilde r_n
=-2\tilde r_n(\tilde d_n-\tilde d_{n-1}).
\]
The sequence $\{X_k\}_{k\ge0}$ obeys the same homogeneous
three-term recursion as $\{Q_k\}_{k\ge0}$ in
Eq.~\eqref{eq:Z4exact}, now with coefficients
built from the recurrence coefficients of the squared weight, but
with different initial values. In the present contour normalization,
\[
\tilde h_0(s)
=-\frac{\pi}{2^{1/3}}
 [\Bi(2^{2/3}s)+\ii\Ai(2^{2/3}s)],
\qquad
\partial_sX_0=-\tilde h_0.
\]
For real $s$, the initial function can equivalently be
written as
\[
X_0(s)=-\int_{-\infty}^{s}\tilde h_0(t)\,\dd t,
\qquad
X_0(s)\longrightarrow0
\quad\text{as }s\to-\infty,
\]
where the integral is understood in the oscillatory
improper sense.

The proof has a direct finite-dimensional form.
For polynomials $f,g$, introduce the ordered-contour
skew product
\[
[f,g]_1
:=\int_{x\prec y}
\big[f(x)g(y)-g(x)f(y)\big]
\rho_s(x)\rho_s(y)\,\dd x\,\dd y,
\]
where $x\prec y$ means that $x$ precedes $y$
along the oriented contour $\mathcal C_-$.
Thus $[f,g]_1=-[g,f]_1$; the brackets denote
a bilinear form, not a commutator.
Define also the weighted derivative
\[
\mathcal Df=f'+(w^2-s)f
=\rho_s^{-1}(f\rho_s)'.
\]
Integration by parts gives
\[
[\mathcal Df,g]_1
=2\int_{\mathcal C_-}f(w)g(w)\rho_s(w)^2\,\dd w,
\]
with vanishing endpoint terms because the weight
decays at both ends of the contour. Replacing the monic polynomial
basis by
\[
p_0,\quad p_1,\quad
\mathcal Dp_0,\quad\mathcal Dp_1,\quad\ldots
\]
is a unitriangular transformation and therefore preserves
the Pfaffian. In this basis the skew matrix is
five-diagonal: multiplication by $w^2-s$ raises the
polynomial degree by at most two, and orthogonality
eliminates all more distant entries. Expansion of its
Pfaffian gives Eq.~\eqref{eq:Z1X}; the first two rows
supply the exceptional initial term $-\tilde h_0$ in
$X_1$. Finally, direct evaluation of the two-particle
ordered integral at $s=0$ fixes $X_0(0)=\ii\pi/3$.
Appendix~\ref{app:beta1reduction} gives the matrix entries,
the Pfaffian signs and the contour calculation explicitly.

\subsection{Double scaling for $\beta=4$ and $\beta=1$: derivation and solution selection}
\label{sec:b4scaling}
\label{sec:b1scaling}

We now derive the double-scaling limits of the
$\beta=4$ and $\beta=1$ edges from the exact
finite-$N$ reductions~\eqref{eq:Z4exact}
and~\eqref{eq:Z1X}. The derivation is at the physics
level: we take continuum limits of the string equations
and of the three-term recursions, keep the leading order
in $N^{-1/5}$, and fix the relevant solutions by matching.
The steps are as follows.
\begin{itemize}
\item The two recursions are brought to a common form, with
coefficients given by the recurrence coefficients of a unitary
ensemble at an effective temperature.
\item The continuum limit of the string equations gives
Painlev\'e~I with explicit normalization,
and that of the recursions gives the auxiliary equation
$q''=\tfrac34yq$.
\item The physical Painlev\'e~I solution, a complex
tritronqu\'ee, is fixed by analyticity in the upper half
energy plane.
\item The same analyticity and the barrier select the
auxiliary solution: for $\beta=4$ in agreement with the
finite-$N$ recursion, for $\beta=1$ up to a
minimal-solution condition for $X_0$ left as a conjecture.
\end{itemize}

\paragraph*{1. A common form.}
Both reductions lead to the same kind of object: a unitary ensemble of
$M$ particles on $\Gamma_-$, in the variable $\lambda$, at an effective
temperature $T_{\rm eff}$, described by its recurrence coefficients
$(d_n,r_n)$. For $\beta=4$ the weight is $e^{-U_E(\lambda)/T}$ itself,
so $T_{\rm eff}=T$, the number of particles is doubled, $M=2N$, and the
recurrence coefficients are those of Sec.~\ref{sec:b4exact}. For
$\beta=1$ and even $N$ the weight is squared, $e^{-2U_E(\lambda)/T}$,
which halves the temperature, $T_{\rm eff}=T/2$, while the number of
particles stays $M=N$. Its recurrence coefficients were computed in
Sec.~\ref{sec:b1sketch} in the dimensionless variable $w$; since
$\lambda=-T^{1/3}w$, they translate into $d_n=-T^{1/3}\tilde d_n$ and
$r_n=T^{2/3}\tilde r_n$ in the variable $\lambda$.

In the rest of this subsection $(d_n,r_n)$ denote
whichever of the two sets is relevant. In both cases the string equations combined with the Toda flow take the form
\be
d_n^2+r_n+r_{n+1}+E=0,\qquad
r_n(d_n+d_{n-1})=nT_{\rm eff},\qquad
\partial_Ed_n=-\frac{r_{n+1}-r_n}{T_{\rm eff}} .
\label{eq:stringeff}
\ee
The rescaling $X^{\rm ph}_k=(-T^{1/3})^kX_k$ brings the
$\beta=1$ recursion~\eqref{eq:Z1X} to the same form as
the $\beta=4$ one:
\be
R_{m+1}=(d_{2m}+d_{2m+1})R_m-r_{2m}R_{m-1},
\qquad R=Q\ (\beta=4),\quad R=X^{\rm ph}\ (\beta=1).
\label{eq:Rrec}
\ee
The characteristic exponents~\eqref{eq:Xi4exact}
and~\eqref{eq:Xi1exact} read, in physical units,
\be
\Xi^{(4)}_N=\sum_{j=0}^{N-1}d_{2j+1}-T\partial_E\log Q_N,
\qquad
\Xi^{(1)}_{2m}=2\sum_{i=0}^{m-2}d_{2i+1}-T\partial_E\log X^{\rm ph}_{m-1}.
\label{eq:Xiphys}
\ee
Note that the last terms carry the physical temperature $T$,
not $T_{\rm eff}$. With $\rho=-\pi^{-1}\Imn\partial_E\Xi$
and the last relation in~\eqref{eq:stringeff}, the densities
of states are, exactly at finite $N$,
\begin{align}
\rho^{(4)}_N&=\frac{1}{\pi T}\,\Imn\sum_{j=0}^{N-1}
\big(r_{2j+2}-r_{2j+1}\big)
+\frac T\pi\,\Imn\partial_E^2\log Q_N, \nonumber\\
\rho^{(1)}_{2m}&=\frac{4}{\pi T}\,\Imn\sum_{i=0}^{m-2}
\big(r_{2i+2}-r_{2i+1}\big)
+\frac T\pi\,\Imn\partial_E^2\log X^{\rm ph}_{m-1}.
\label{eq:rhoexact}
\end{align}
For the unitary ensemble itself,
$\Xi^{(2)}_M=\sum_{n<M}d_n$ and $r_0=0$ give the
exact identity
\be
\rho^{(2)}_M\big|_{T_{\rm eff}}=\frac{\Imn r_M}{\pi T_{\rm eff}} .
\label{eq:rho2exact}
\ee

\paragraph*{2. The critical point.}
In the planar regime ($M\to\infty$, i.e.\ $N\to\infty$), the recurrence
coefficients vary slowly near $n=M$, $d_n\simeq d$ and $r_n\simeq r$,
and the string equations at $n=M$ give the single cubic equation
$d^3 + Ed + MT_{\rm eff} = 0$, or equivalently $E = -d^2 - MT_{\rm eff}/d$.
Two positive real branches of this cubic merge at the maximum of
$-d^2-MT_{\rm eff}/d$ for $d>0$:
\be
d_*=\frac s2,\qquad r_*=d_*^2=\frac{s^2}4,\qquad
E^*=-3d_*^2=-\frac34s^2,\qquad s^3=4MT_{\rm eff}.
\label{eq:crit}
\ee
This reproduces $s_\beta=(\beta J^2)^{1/3}$ of
Sec.~\ref{sec:largeN}: for $\beta=2$ we have $4NT=2J^2$ while for  $\beta=4$ it is $4\cdot2N\cdot T=4J^2$ and 
for $\beta=1$ it is $4N\cdot T/2=J^2$. The physical branch on the forbidden side $E<E^*$ is
the one continuing to the Lifshitz saddle, $d\simeq\sqrt{-E}$
as $E\to-\infty$. On it $d>d_*$, hence
$d^2-r=d^2-MT_{\rm eff}/(2d)>0$. Thus the characteristic roots $\mu_\pm=d\pm\sqrt{d^2-r}$ of the planar
recursion~\eqref{eq:Rrec} (i.e.\ the roots of $\mu^2-2d\mu+r=0$) are
real and distinct on the forbidden side, and coalesce at $E=E^*$.

\paragraph*{3. Continuum string equations and Painlev\'e~I.}
At large but finite $N$ (and hence large $M$), resolving the rounding of
the spectral edge requires going beyond the leading planar approximation
and retaining finite-$M$ corrections in the double-scaling window around
the critical point~\eqref{eq:crit}. To this end, put $n=M+\nu$,
$\Delta E=E-E^*$, and write $d_n=d_*+b(n)$ and $r_n=d_*^2+\varrho(n)$,
with $b$ and $\varrho$ smooth in $n$. In the
edge window we assume the scalings
$b\sim M^{-2/5}$, $\partial_n\sim M^{-1/5}$ and
$\Delta E,\ \nu T_{\rm eff}\sim M^{-4/5}$, which are confirmed by the
result below, and expand the string equations~\eqref{eq:stringeff} in
powers of $M^{-1/5}$. At order $M^{-2/5}$ both equations give
$\varrho=-d_*b$, and at order $M^{-3/5}$ both give the correction
$\varrho=-d_*b+\tfrac12d_*b'$, where primes denote $\partial_n$. At
order $M^{-4/5}$ one finds $b^2+2\varrho_2+\Delta E=0$ and
$\tfrac12d_*^2b''+2d_*\varrho_2-2d_*b^2=\nu T_{\rm eff}$, where
$\varrho_2$ is the part of $\varrho$ of this order.
Now eliminating $\varrho_2$ yields
\be
b''=\frac{6}{d_*}\,b^2+\frac{2}{d_*^2}\big(\nu T_{\rm eff}+d_*\Delta E\big).
\label{eq:contstring}
\ee
Its planar ($b''=0$) solution
$b^2=-\big(\Delta E+\nu T_{\rm eff}/d_*\big)/3$ is the
expansion of the planar branch near $E^*$. The substitution
\[
b=d_*\,\kappa\, y(t),\qquad
t=a\Big(\Delta E+\frac{\nu T_{\rm eff}}{d_*}\Big),\qquad
\kappa=2^{4/5}M^{-2/5},\qquad
a=\frac{2^{7/5}M^{4/5}}{s^2},
\]
turns~\eqref{eq:contstring} into
\be
y''(t)=6y^2+t .
\label{eq:PI}
\ee
The constants are fixed by $\kappa d_*^2=a^2T_{\rm eff}^2$
and $\kappa a^3T_{\rm eff}^2=2$, together with~\eqref{eq:crit}.
At $\nu=0$ this gives the scaling variables
\be
E=E^*_\beta+s_\beta^2N_{\rm eff}^{-4/5}X',\qquad
t=2^{7/5}X',\qquad N_{\rm eff}=M,
\label{eq:scalingvars}
\ee
with $N_{\rm eff}=2N$ for $\beta=4$ and $N_{\rm eff}=N$
for $\beta=1$. Moreover $\partial t/\partial n=2^{2/5}M^{-1/5}$, and
\be
d_n=d_*\big[1+2^{4/5}M^{-2/5}y(t)\big],\qquad
r_n=d_*^2\big[1-2^{4/5}M^{-2/5}y(t)\big],
\label{eq:bry}
\ee
up to relative corrections $O(M^{-1/5})$.
Combining~\eqref{eq:bry} with~\eqref{eq:rho2exact}
and~\eqref{eq:crit} gives the unitary edge density \cite{Letter} in
the present conventions:
\be
\rho^{(2)}_M\big|_{T_{\rm eff}}\simeq
-\frac{2^{4/5}}{\pi}\,\frac{M^{3/5}}{s}\,\Imn y(t).
\label{eq:rho2scaling}
\ee

\paragraph*{4. The auxiliary equation.}
Put $R_m=d_*^mq_m$ in~\eqref{eq:Rrec}. Then
\[
q_{m+1}-2q_m+q_{m-1}
=\Big[\frac{d_{2m}+d_{2m+1}}{d_*}-2\Big]q_m
-\Big[\frac{r_{2m}}{d_*^2}-1\Big]q_{m-1}
=\frac{3b}{d_*}\,q_m\big[1+O(M^{-1/5})\big].
\]
Here we used $\varrho=-d_*b+O(\epsilon^{3/2})$. The terms
containing $b'$ and $q_m-q_{m-1}$ are smaller by a factor
$M^{-1/5}$. One step in $m$ is two steps in $n$, i.e.\
$\Delta t=2^{7/5}M^{-1/5}$. The left-hand side therefore
becomes $2^{14/5}M^{-2/5}q''(t)$ and the right-hand side
$3\cdot2^{4/5}M^{-2/5}y\,q$. Hence
\be
q''(t)=\tfrac34\,y(t)\,q(t),
\label{eq:qode}
\ee
for both ensembles.

For $|t|\gg1$ the function $y$ varies slowly on the scale of the local
wavelength, $|y'|\ll|y|^{3/2}$, since $|y|\simeq\sqrt{|t|/6}$ on both sides
of the window. There the solutions of~\eqref{eq:qode} take the WKB form
$q_\pm\simeq(3y/4)^{-1/4}\exp\big[\pm\int^t\sqrt{3y/4}\,\dd t'\big]$.
Inside the window, $|t|=O(1)$, $q$ is given only by the exact solution
of~\eqref{eq:qode}; the WKB forms serve to label solutions by their
behaviour as $t\to\pm\infty$ and to match them onto the planar regime.
In that regime they are the continuations of the planar roots $\mu_\pm$.
Indeed, $\mu_\pm/d_*\simeq1\pm\sqrt{3b/d_*}$, and
$\sqrt{3b/d_*}/\Delta t=\sqrt{3y/4}$. In particular, the larger characteristic root $\mu_+$, which dominates
the growth of $R_m$ over $\mu_-^m$ under forward iteration in $m$,
corresponds to the solution growing with $t$. For the densities, $\partial_E=(\partial t/\partial E)\partial_t$
with $\partial t/\partial E=2^{7/5}M^{4/5}/s^2$. 

The last
terms in~\eqref{eq:rhoexact} are then
$(T/T_{\rm eff})(2^{4/5}/\pi)(M^{3/5}/s)\,\Imn(\log q)''$,
and $(\log q)''=\tfrac34y-v^2$ with $v=q'/q$. To estimate the sums in~\eqref{eq:rhoexact}, note
that, for even $K$ and $r_0=0$,
\[
\sum_{j=0}^{K/2-1}(r_{2j+2}-r_{2j+1})
=\frac12r_K+
\frac12\sum_{j=0}^{K/2-1}
(r_{2j+2}-2r_{2j+1}+r_{2j}).
\]
Thus the deviation from $\tfrac12r_K$ is a sum
of second differences. By~\eqref{eq:bry},
the critical part of $r_n$ has magnitude
$O(s^2M^{-2/5})$ and varies over
$O(M^{1/5})$ indices. Each second difference is therefore $O(s^2 M^{-4/5})$, and their sum
over the critical region (of width $O(M^{1/5})$) is $O(s^2 M^{-3/5})$.
Matching to the forbidden-side solution, whose imaginary part decays
away from this region, then gives
\[
\Imn\sum_{j=0}^{K/2-1}(r_{2j+2}-r_{2j+1})
=\frac12\Imn r_K+O(s^2 M^{-3/5}),
\]
where we used that $\frac12\Imn r_K = O(s^2 M^{-2/5})$
in the critical window~\eqref{eq:bry}.
Here $K=M$ for $\beta=4$ and $K=M-2$ for
$\beta=1$. In the latter case, replacing $r_K$
by $r_M$ introduces an error of the same order,
since a fixed index shift changes $t$ by
$O(M^{-1/5})$. Compared with the leading
critical contribution $O(s^2M^{-2/5})$,
these errors are smaller by $M^{-1/5}$.

Two further approximations are within the stated $O(N^{-1/5})$ relative
accuracy. First, for $\beta=1$ the final index $2m-2=N-2$ differs from
$N$ by a shift of $t$ of order $M^{-1/5}$. Second, any smooth
$E$-dependence of $\log R$ not captured by the scaling function
contributes to $\rho$ a term of order one; since the leading edge
density is of order $N^{3/5}$ (see Eqs.~\eqref{eq:rho4scaling}
and~\eqref{eq:rho1scaling}), this is a relative correction of order
$N^{-3/5}$, smaller than the $O(N^{-1/5})$ corrections already neglected.

Writing $v_4=q'/q$ and $v_1=x'/x$, with $q$ and $x$
the solutions selected below, and using~\eqref{eq:rho2scaling},
we obtain
\be
\rho^{(4)}_N\simeq
\frac{N'^{3/5}}{s_4}\,
\frac{2^{4/5}}{\pi}\,
\Imn\Big[\tfrac14y-v_4^2\Big],
\qquad N'=2N,
\label{eq:rho4scaling}
\ee
and, for even $N$,
\be
\rho^{(1)}_N\simeq
\frac{N^{3/5}}{s_1}\,
\frac{2^{9/5}}{\pi}\,
\Imn\Big[\tfrac14y-v_1^2\Big].
\label{eq:rho1scaling}
\ee
Both hold up to relative corrections $O(N^{-1/5})$.
Equivalently,
\[
\rho^{(4)}_N\simeq
\frac12\rho^{(2)}_{2N}
+\frac{N'^{3/5}}{s_4}\,
\frac{2^{4/5}}{\pi}\,
\Imn(\log q)'',\qquad
\rho^{(1)}_N\simeq
\left.\rho^{(2)}_N\right|_{T/2}
+\frac{N^{3/5}}{s_1}\,
\frac{2^{9/5}}{\pi}\,
\Imn(\log x)''.
\]
In~\eqref{eq:rho1scaling} the factor $2^{9/5}=2\cdot2^{4/5}$
reflects $T/T_{\rm eff}=2$. The logarithmic derivatives
satisfy $v_\beta'=\tfrac34y-v_\beta^2$. 

\paragraph*{5. The physical solution of Painlev\'e~I.}
Generic solutions of Painlev\'e~I have poles in every direction of the
complex $t$-plane; two requirements single out the physical one. First,
by the zero-free region~\eqref{eq:zerofree} $\Xi_N$ has no poles for
$\Imn\cE\ge0$. Since $t=a(\Delta E+\nu T_{\rm eff}/d_*)$ with $a>0$, and
$y$ enters $\partial_\theta\Xi^{(2)}_M=r_M$ linearly by~\eqref{eq:bry},
$y$ has no poles in the closed upper half $t$-plane. Second, for
$|t|\to\infty$ in that half-plane the planar branch, continued from the
Lifshitz saddle ($b>0$) through $\Imn\cE>0$, must be recovered:
\be
y(t)\simeq\sqrt{-t/6},\qquad 0\le\arg t\le\pi,
\label{eq:ymatch}
\ee
with the principal branch; in particular $y\simeq-\ii\sqrt{t/6}$ for
$t\to+\infty$, which gives a positive density in~\eqref{eq:rho2scaling}.

These conditions fix the solution uniquely. Linearizing about $y_0=\sqrt{-t/6}$, $\delta y''=12y_0\,\delta y$,
gives the two WKB solutions
\be
\delta y_\pm\simeq (-t)^{-1/8}\exp\Big[\pm\tfrac45\,2^{3/4}3^{1/4}(-t)^{5/4}\Big].
\label{eq:PIinst}
\ee
The two solutions exchange growth and decay across the Stokes ray
$\arg t=3\pi/5$, so a solution obeying~\eqref{eq:ymatch} on both sides
of this direction can contain neither, and both integration constants
are fixed. (About $-\sqrt{-t/6}$ the corresponding linearized WKB solutions
oscillate instead, which is why the real tritronqu\'ee cannot describe
the activated tail.) With
$\omega=e^{2\pi\ii/5}$ and $y_{\rm r}$ the real tritronqu\'ee, which behaves as
$y_{\rm r}\sim-\sqrt{-t/6}$ for $|\arg(-t)|<4\pi/5$, the solution is
\be
y(t)=y_1(t):=\omega^2y_{\rm r}(\omega t),
\label{eq:yphys}
\ee
which behaves as $\sqrt{-t/6}$ for $-\pi/5<\arg t<7\pi/5$ (Kapaev's
$y_0$~\cite{Kapaev2004}). By the proof of Dubrovin's
conjecture~\cite{CostinHuangTanveer2014}, see also~\cite{JoshiKitaev2001},
its poles lie in $-3\pi/5<\arg t<-\pi/5$, i.e.\ at $\Imn\cE<0$; it is
regular on the real axis, and analyticity holds for all finite $t$. The
conjugate $y_{-1}(t)=\overline{y_1(\bar t)}$ belongs to $\cE-\ii0$. The same $y_1$ follows from mapping the outgoing contour to the critical
cubic model~\cite{Letter,BleherDeano2013,BleherDeano2016,BBDY2024}.

On the forbidden side $y_{\pm1}$ share the real series about
$+\sqrt{-t/6}$ and differ, like $\Bi\pm\ii\Ai$, by the exponentially
small jump computed by Kapaev~\cite{Kapaev2004}:
\be
\Imn y_1(t)\simeq-\frac{2^{-19/8}3^{-1/8}}{\sqrt\pi}\,|t|^{-1/8}
e^{-\frac45 2^{3/4}3^{1/4}|t|^{5/4}},\qquad t\to-\infty .
\label{eq:Imytail}
\ee
Substituting~\eqref{eq:scalingvars} into the exponent of~\eqref{eq:Imytail}
gives $\frac45\sqrt2\,3^{1/4}s^{1/2}(E^*-E)^{5/4}/T_{\rm eff}$, which is
precisely the near-edge single-eigenvalue barrier of Sec.~\ref{sec:largeN}
at temperature $T_{\rm eff}$. 

\paragraph*{6. Selection for $\beta=4$.}
The recursion~\eqref{eq:Rrec} for $Q$ starts at $m=0$ from $Q_0=1$ and
$Q_1=d_0+d_1$. For small $n$ one has $E<E^*(n)$, where
$E^*(n)=-\tfrac34(4nT_{\rm eff})^{2/3}$, so the early steps lie on the
forbidden side of the local criticality, where the roots $\mu_\pm$ are
real and separated. Under forward iteration in $m$, the component of $Q_m$ along the larger
characteristic root $\mu_+$ is amplified exponentially relative to
$\mu_-^m$ for generic initial values, unless these are fine-tuned to
suppress it. Nothing in the polynomial initial
values of $Q$ does so, and $Q_m$ therefore follows $\mu_+$ up to
exponentially small corrections. In the window this is the solution
growing with $t$, i.e.\ the solution of~\eqref{eq:qode} decaying as
$t\to-\infty$:

The consequences are the following.
\begin{itemize}
\item {\it Forbidden side.} Here $v_4=\sqrt{3y/4}+\dots$,
where the dots denote real terms plus terms proportional
to $\Imn y$. Hence $\Imn[\frac14y-v_4^2]$ is exponentially
small at the rate~\eqref{eq:PIinst}, reproducing the near-edge
single-eigenvalue barrier of Sec.~\ref{sec:largeN} at temperature $T$
with $s=s_4$, as required.
\item {\it Allowed side.} $q$ is purely $q_+$ near
$\arg t=3\pi/5$, and $q_+$ dominates for $\arg t<3\pi/5$,
so $q$ grows as $t\to+\infty$. Then
$v_4^2=\tfrac34y+O(t^{-1})$, and~\eqref{eq:rho4scaling}
reduces to $\frac12\rho^{(2)}_{2N}$, the planar
$\beta=4$ density.
\end{itemize}

\paragraph*{7. Selection for $\beta=1$.}
The sequence $X^{\rm ph}$ obeys the same recursion; with generic
initial values it would follow $\mu_+$ like $Q$. Two conditions show
that it does not.

{\it (a) Analyticity.} The zero-free region holds for every $\beta$, so
the term $T\partial_E\log X^{\rm ph}_{m-1}$ in~\eqref{eq:Xiphys} has no
poles, and $x$ no zeros, for $\Imn t>0$. In the WKB forms
$q_\pm\simeq(3y/4)^{-1/4}e^{\pm W}$, $W'=\sqrt{3y/4}$, dominance changes
at the same direction $\arg t=3\pi/5$ as in Par.~5, and a solution
containing both $q_\pm$ there has a sequence of zeros along it, as
$\Ai+c\,\Bi$ has along a ray. Hence $x$ is purely $q_+$ near that
direction, i.e.\ decays as $t\to-\infty$, or purely $q_-$, i.e.\ decays
as $t\to+\infty$.

{\it (b) The barrier.} The activation exponent at fixed $T$ does not
depend on $\beta$ (Secs.~\ref{sec:Lifshitz} and~\ref{sec:largeN}), while
the recursion is driven by the unitary ensemble at $T_{\rm eff}=T/2$,
so $\Imn y$ carries twice the required exponent. The first candidate,
purely $q_+$ on the negative axis, inherits this exponent and is
excluded. The second, continued to the negative axis, acquires across
$\arg t=\pi$ an admixture of $q_+$ with a complex coefficient, which
contributes to $\Imn[\frac14y-v_1^2]$ at order
$\exp[-2\int_t^0\sqrt{3y/4}]$. Since $2\sqrt{3/4}=\tfrac12\sqrt{12}$,
this is half of~\eqref{eq:PIinst}, the barrier at temperature $T$ with
$s=s_1$. Hence $x$ decays as $t\to+\infty$. For $\beta=4$
($T_{\rm eff}=T$) the same argument excludes the second candidate, in
agreement with Par.~6. 

{\it Checks.} For $t\to+\infty$, $\log q\simeq+\int^t\sqrt{3y/4}$ and
$\log x\simeq-\int^t\sqrt{3y/4}$. Since the unitary parts of
$\log\Zh^{(4)}$ and $\log\Zh^{(1)}$ are orientable, $\log q$ and $\log x$
carry the leading projective-plane terms, and these are opposite in the
variables $(N_{\rm eff},T_{\rm eff})$, as the $\beta\leftrightarrow4/\beta$
duality requires~\cite{MulaseWaldron2003}; cf.\ the $N^2$ terms
$\pm\frac94N_{\rm eff}^{-1}$ of $K_\beta(N)$ in~\eqref{eq:Kbeta}. At
finite $N$ the initial values $X_0,X_1$ must make $X^{\rm ph}$ the
minimal solution of the recursion on the forbidden side; we leave the
exact finite-$N$ verification of this minimal-solution condition for the
Airy primitive $X_0$ (which vanishes as $s\to-\infty$) as a conjecture.

\paragraph*{8. Summary }
Both edges are thus built from solutions of the same Schr\"odinger operator $-\partial_t^2+\tfrac34y(t)$,
selected at opposite ends: decaying at $t\to-\infty$ for $\beta=4$, and at $t\to+\infty$ for $\beta=1$. In both
cases $y$ is the complex tritronqu\'ee~\eqref{eq:yphys}.
For $\beta=2$ the edge profile is given by $y$ alone; for $\beta=1,4$ 
the selected solution of the auxiliary equation adds a contribution of the same order.

On the allowed side both scaling forms match their
planar limits. This leading-order matching is insensitive
to the selection, since $v_\beta^2\simeq\frac34y$ for
either solution; the selection shows up only at
subleading order. On the forbidden side both reproduce
the single-eigenvalue barrier at temperature $T$, with
the respective $s_\beta$.

The construction is analogous to the way the
Tracy--Widom distributions $F_1$ and $F_4$ are built
from the same Painlev\'e~II data~\cite{TracyWidom1996},
although the driving equation and the selection problem
here are different.

\section{Discussion and outlook}
\label{sec:discussion}

The central problem of this paper is the spectral
behaviour of an $N$-channel Schr\"odinger operator
with matrix-valued white-noise disorder. The
stationary Riccati distribution provides a common description
of this problem in all three Dyson classes.
Its normalization yields the density of states
and the sum of Lyapunov exponents, while its
asymptotic regimes connect the rare fluctuations
responsible for the Lifshitz tail to the collective
behaviour near the large-$N$ band edge.
The appearance of outgoing Kontsevich integrals
is the exact mathematical structure underlying
these spectral results.

\paragraph*{Exact spectral description.}
At complex energy, the Riccati matrix has an
explicit stationary probability distribution
for $\beta=1,2,4$. This distribution is not
of Gibbs type. Its stationarity follows from
a drift of the form $(D+\mathcal A)\nabla\Phi$,
with constant antisymmetric $\mathcal A$.
The relevant trace identities remain valid
for non-commuting matrices and in all three
Dyson classes. At real energy, the same
construction gives a Gibbs/anti-Gibbs
two-matrix representation, extending
Halperin's scalar formula to the
multichannel problem.

The outgoing integrals $\tau_N^{(\beta)}$
arise as the holomorphic factors of the
normalization. They therefore give closed
expressions for the spectral observables
and explain why the unitary result of
Ref.~\cite{Letter} extends to the orthogonal
and symplectic classes. For $\beta=4$, and for
$\beta=1$ with even $N$, the finite-$N$ reductions also identify the
additional recursions that distinguish
their spectral behaviour from that of
the underlying $\beta=2$ problem.

\paragraph*{Rare fluctuations and the Lifshitz tail.}
The forbidden-region density is governed by
a fluctuation that binds a state in one
direction in channel space. The calculation
in Sec.~\ref{sec:Lifshitz} determines both
the exponential barrier and the
$N$- and $\beta$-dependent prefactor,
together with its first correction.
The channel dependence has a geometric
origin: the direction selected by the
fluctuation belongs to
$\mathbb{RP}^{N-1}$, $\mathbb{CP}^{N-1}$
or $\mathbb{HP}^{N-1}$, according to
the Dyson class. This geometry enters
the prefactor, while the tunnelling
configuration determines the
exponential suppression.

These tail results concern the Gaussian white-noise model. They should not be transferred without qualification to
disorder with bounded support: sufficiently deep in the forbidden region, the optimal fluctuation probes the microscopic
distribution of the disorder. 

\paragraph*{The critical edge.}
The rounded planar edge limit at large $N$ describes a different regime
from the fixed-$N$ deep Lifshitz tail. In the critical window, the recurrence coefficients of the underlying $\beta=2$ orthogonal polynomials
are governed by the complex Painlev\'e~I tritronqu\'ee~\eqref{eq:yphys}. For $\beta=1,4$, this does not by itself determine the density:
the additional finite-$N$ recursions retain a nontrivial continuum limit
and yield $q''(t)=\tfrac34y(t)q(t)$. The two classes select different solutions
of this equation, and their logarithmic derivatives supply the corrections to the unitary scaling function.
\begin{itemize}
\item For $\beta=4$, analyticity in the upper half energy plane and the
near-edge single-eigenvalue barrier at temperature $T$, as well as the
finite-$N$ recursion, select the solution decaying as $t\to-\infty$.
\item For $\beta=1$ (even $N$), analyticity in the upper half energy
plane and the near-edge single-eigenvalue barrier at temperature $T$
select the solution decaying as $t\to+\infty$, in agreement with the
leading projective-plane contribution to the free energy.
\end{itemize}

The resulting edge forms match the planar densities on the allowed side
and the single-eigenvalue barrier on the forbidden side at the level
of the exponential decay rates. 

\paragraph*{A universality conjecture.}
We expect the edge scaling functions obtained here to extend beyond the
Gaussian white-noise operator. The natural candidate family consists of
quasi-one-dimensional disordered systems whose bulk limit is described by
the one-dimensional Efetov--Wegner nonlinear $\sigma$-model in the
corresponding Dyson class. Such a reduction was established for random band matrices in the
large-bandwidth limit by Fyodorov and
Mirlin~\cite{FyodorovMirlin1991,FyodorovMirlin1994}; for recent
rigorous progress on the 1D random-band-matrix localization
transition, see also~\cite{GuillaumeRemi2026}. We refer to this
family here as the 1D Efetov--Wegner class and conjecture that, at a
regular band edge and after matching the model-dependent scales, its
members exhibit the density scaling functions obtained in this paper.

The conjecture is supported by the random-band-matrix edge theory of
Sodin~\cite{Sodin2010} and of Liu and Zou~\cite{LiuZou2024,LiuZou2026}
(see also Ref.~\cite{Silvestrov1997} for early insights). For $\beta=2$
the correspondence was proposed in Ref.~\cite{Letter}. For $\beta=1$ the
comparison is even more constraining, since the metric factors are
inherited from the unitary case: both the projective-plane coefficient
and the regularized first moment of the band-matrix edge intensity follow
without free parameters, and both agree with the band-matrix diagram
expansion (Appendix~\ref{app:RBM1}). An analogous edge universality for
chiral random band matrices will be developed in~\cite{ChiralRBM}. As a
critical-window phenomenon, this universality governs the scaling regime
near a regular band edge, while deeper in the forbidden region the
asymptotic Lifshitz tail naturally reflects the microscopic disorder
distribution.

\paragraph*{Beyond the holomorphic sector.}
The spectral results above use the normalization and holomorphic
observables of the stationary law. The full distribution also
provides a starting point for non-holomorphic scattering observables.
Appendix~\ref{app:admittance-scattering} collects the boundary-response
identities, their relation to time delays, and the high-energy
connection with absorbing-waveguide theory, and outlines the
programme of Paper~II~\cite{PaperII}.

\section{Open problems}
\begin{itemize}
\item {\it Mathematical completion.} Three steps are argued here at the
physics level. The first is the uniqueness of the stationary distribution, both
at $\eta>0$ and for the real-energy dynamics continued through the
Riccati poles. At real energy this uniqueness is also what identifies
Eq.~\eqref{eq:Pst} with the current-carrying stationary state for
$\beta=1,4$; for $\beta=2$ that identification follows independently
from Appendix~\ref{app:matrix-fermion} and the fermionic solution of
Ref.~\cite{Letter}. The second is the decomposition~\eqref{eq:Zdecomp}
of the integration surface behind Result~\ref{prop:factor}. 
The third is a rigorous Riemann--Hilbert double-scaling analysis of the
string equations and the $\beta=1,4$ auxiliary recursions in
Sec.~\ref{sec:beta4edge} (including the solution-selection argument and
the minimal-solution condition for $X_0$).

\item {\it The $Y$-ensemble at large $N$.} Eq.~\eqref{eq:Yens} is an explicit positive ensemble with a
Schur-type interaction. Its planar limit and edge would give an alternative, manifestly positive
route to the $\beta=1,4$ edge functions of Sec.~\ref{sec:beta4edge}.

\item {\it Finite length $L$.} The stationary distribution gives neither
the $L$-dependent counting statistics nor first-escape rates, and in
particular does not determine the banded-matrix analogue of the
Tracy--Widom edge distribution at finite $L$. For $\beta=2$, the
$L$-dependent counting statistics and first-escape rates are accessible
via the fermionic structure of Ref.~\cite{Letter} (though the finite-$L$
edge distribution itself was not derived there); for $\beta=1,4$, even
the $L$-dependent formulation remains open.

\item {\it Non-orientable topology.} Do the $\beta=1,4$ outgoing Kontsevich integrals with
non-degenerate thresholds generate intersection-type invariants of moduli spaces of non-orientable
surfaces, e.g.\ within refined topological recursion~\cite{CDO2024,GFGO2025}? A related concrete
problem is a proof of the twisted-gluing conjecture of Appendix.~\ref{sec:ewensgluing}, i.e.\ of the
asymptotic irrelevance of the side constraint for the loop count of randomly twisted
triangle gluings.
\end{itemize}

\section*{Acknowledgements}
The author thanks Pierre Le Doussal for the collaboration on Ref.~\cite{Letter}, on which this
work builds, and Aurellien Grabsch for encouraging interest in the results and methods of the paper. 
Prof. Sergei Chmutov is gratefully acknowledged for useful correspondence about statistics of randomly
twisted triangle gluings.

The research reported in this paper was supported by UKRI grant UKRI1015,
``Non-Hermitian Random Matrices: Theory and Applications.''  

The author acknowledges assistance from GPT-6.1 Sol (OpenAI) with organizing the systematics of results transfer from $N=1$ to $N\ge 1$ (in particular, in Sec.(\ref{Factorization})-(\ref{Holomorphic})) and writing the Appendices.   Claude Opus 5 (Anthropic) was used for designing and executing numerical computations, for help with analysing various asymptotics of Kontsevich integrals,  and for critical reading. Finally, Gemini 4 Hydrogen (Google) was used  for improving the clarity of the presentation and performing consistency checks. All scientific arguments, results, and conclusions
remain the responsibility of the author.

\appendix
\numberwithin{equation}{section}
\renewcommand{\theequation}{\thesection\arabic{equation}}
\renewcommand{\thesubsection}{\thesection\arabic{subsection}}

\section{Algebraic equivalence of the matrix Halperin distribution and the fermionic stationary state}
\label{app:matrix-fermion}

We establish the equivalence at $\beta=2$ directly, without invoking uniqueness of the stationary matrix diffusion. Throughout, $T>0$, $E\in\mathbb R$, and
\begin{equation}
 U(x)=Ex+\frac{x^3}{3},\qquad r=N-1,\qquad
 \DeltaV(x)=\prod_{i<j}(x_j-x_i).
\end{equation}
Flat matrix measure means the product of the diagonal differentials and the real and imaginary off-diagonal differentials. Haar measure on $U(N)$ is normalized to one. Constants independent of the eigenvalues are absorbed into normalization constants.

\subsection{Statement and conventions}
Consider the normalized Hermitian matrix density
\begin{equation}
 P_{\rm mat}(\Lambda)=\frac{1}{\mathcal Z}
 e^{-\Tr U(\Lambda)/T}
 \int_{M<\Lambda}e^{\Tr U(M)/T}\,\dd M.
 \label{eq:matrixlaw}
\end{equation}
The matrix inequality $B<A$ means that $A-B$ is
positive definite. The integral converges for fixed $\Lambda$: putting $M=\Lambda-Y$, $Y>0$, gives
\begin{equation}
 \Tr U(\Lambda-Y)=\Tr U(\Lambda)
 -\Tr[(EI+\Lambda^2)Y]+\Tr(\Lambda Y^2)-\tfrac13\Tr Y^3.
\end{equation}
The cubic term dominates uniformly at infinity on the positive cone.

The fermionic construction uses the adjoint pair
\begin{equation}
 \mathcal L_N=T\partial_x^2+(E+x^2)\partial_x+(N+1)x,
 \qquad
 \mathcal L_N^\dagger=T\partial_x^2-(E+x^2)\partial_x+(N-1)x.
 \label{eq:operators}
\end{equation}
The adjoint operator $\mathcal L_N^\dagger$ preserves $\mathcal P_r=\mbox{span}\{1,x,\ldots,x^r\}$.
For a polynomial eigenbasis $p_a$, write
\begin{equation}
 \mathcal L_N^\dagger p_a=\epsilon_a p_a,
 \qquad \mathcal L_N v_a=\epsilon_a v_a.
\end{equation}
The physical right modes are those regular through the Riccati winding; equivalently, their Lagrange currents may be normalized to
\begin{equation}
 J[p_a,v_a]=T(p_av_a'-p_a'v_a)+(E+x^2)p_av_a=1.
 \label{eq:lagrange-current}
\end{equation}
In the paper \cite{Letter} these modes are written
\begin{equation}
 v_a(x)=\frac{p_a(x)}{T}e^{-U(x)/T}
 \FP\int_{-\infty}^x\frac{e^{U(t)/T}}{p_a(t)^2}\,\dd t.
 \label{eq:letter-mode}
\end{equation}
The finite part is interpreted by continuation of the regular solution across polynomial zeros. Our construction below has no such apparent poles.
In fact, with $p_a(x)=c_a x^r+\cdots$ ($r=N-1$), the normalized right mode
takes the manifestly regular form
\begin{equation}
 v_a(x)=\frac{(-1)^r}{c_a^2 r!\,T^{r+1}}\,e^{-U(x)/T}
 \int_{-\infty}^x (x-t)^r p_a(-t)\,e^{U(t)/T}\,\dd t,
\end{equation}
which is derived in Eq.~\eqref{eq:normalizedmode} below.

\begin{proposition}
The ordered eigenvalue density induced by \eqref{eq:matrixlaw} is the normalized fermionic density
\begin{equation}
 P_{\rm ord}(\boldsymbol\lambda)
 =\frac{1}{\mathcal N}
 \det[p_a(\lambda_i)]_{a,i=1}^N
 \det[v_a(\lambda_i)]_{a,i=1}^N,
 \qquad \lambda_1<\cdots<\lambda_N.
 \label{eq:claim}
\end{equation}
At parameter values where a polynomial eigenbasis degenerates, the identity holds by continuation, or in a generalized polynomial basis.
\end{proposition}

\subsection{Angular integration and a polynomial ensemble}
We use the matrix Heaviside integral of Kieburg, Kuijlaars and Stivigny, Theorem 2.3 of Ref.~\cite{KKS}. For Hermitian matrices $A,B$ with distinct spectra $a,b$,
\begin{equation}
 \int_{U(N)}\mathbf1_{\{A-VBV^\dagger>0\}}\,\dd V
 =c_{N,0}\frac{\det[(a_i-b_j)_+^{N-1}]_{i,j=1}^N}
 {\DeltaV(a)\DeltaV(b)},
 \qquad
 c_{N,0}=\prod_{j=0}^{N-1}\binom{N-1}{j}^{-1}.
 \label{eq:angular}
\end{equation}
Here $(s)_+^0$ means $\mathbf1_{\{s>0\}}$ when $N=1$. Repeated eigenvalues are treated by limits.

Fix $\Lambda=\mbox{diag}(\lambda_1,\ldots,\lambda_N)$ and diagonalize $M=V\mbox{diag}(m)V^\dagger$. The Weyl integration formula and \eqref{eq:angular} give
\begin{equation}
 \int_{M<\Lambda}e^{\Tr U(M)/T}\,\dd M
 \propto\frac{1}{\DeltaV(\lambda)}
 \int_{\mathbb R^N}\DeltaV(m)
 \det[(\lambda_i-m_j)_+^r]_{i,j=1}^N
 \prod_{j=1}^N e^{U(m_j)/T}\,\dd m_j.
 \label{eq:afterangular}
\end{equation}
Integration over unordered $m$ introduces only a constant $N!$ compared with the ordered convention. Although $e^{U(m)/T}$ grows at $+\infty$, the truncated-power determinant vanishes when any $m_j>\max_i\lambda_i$; at $-\infty$ there is cubic exponential decay. Thus all integrals used here are convergent.

Using $\DeltaV(m)=\det[m_j^k]_{k=0,\ldots,r;\,j=1,\ldots,N}$, Andreief's identity reduces \eqref{eq:afterangular} to a determinant of scalar integrals. Including the eigenvalue Jacobian $\DeltaV(\lambda)^2$ of $\Lambda$, we obtain
\begin{equation}
 P_{\rm ord}(\boldsymbol\lambda)\propto
 \DeltaV(\lambda)\det[g_k(\lambda_i)]_{i=1,\ldots,N;\,k=0,\ldots,r},
 \label{eq:polyensemble}
\end{equation}
where
\begin{equation}
 g_k(x)=e^{-U(x)/T}\int_{-\infty}^x
 (x-t)^r t^k e^{U(t)/T}\,\dd t.
 \label{eq:g}
\end{equation}
Consequently, only the identification of the right-function space remains.

\subsection{An explicit intertwiner}
Define a linear map on $\mathcal P_r$ by
\begin{equation}
 (\mathcal Kp)(x)=e^{-U(x)/T}\int_{-\infty}^x
 (x-t)^r p(-t)e^{U(t)/T}\,\dd t.
 \label{eq:K}
\end{equation}
Thus $\mathcal K[x^k]=h_k=(-1)^k g_k$. We claim the exact identity
\begin{equation}
 \mathcal L_N\mathcal Kp=\mathcal K\mathcal L_N^\dagger p,
 \qquad p\in\mathcal P_r.
 \label{eq:intertwiner}
\end{equation}

For completeness we prove it in the monomial basis. For $r\geq1$, introduce
\begin{equation}
 I_{s,k}(x)=\int_{-\infty}^x(x-t)^s t^k e^{U(t)/T}\,\dd t.
\end{equation}
The gauge identity
\begin{equation}
 \mathcal L_N(e^{-U/T}F)
 =e^{-U/T}\left[TF''-(E+x^2)F'+rxF\right]
 \label{eq:gauge}
\end{equation}
first yields, with endpoint terms included,
\begin{equation}
 \mathcal L_Ng_k
 =r e^{-U/T}\left[Tk I_{r-1,k-1}-I_{r,k+1}\right].
 \label{eq:intermediate}
\end{equation}
Indeed, integration of
$\partial_t[(x-t)^{r-1}t^ke^{U(t)/T}]$
replaces the term containing $E+t^2$; when $r=1$ its endpoint contribution cancels the endpoint contribution from $F''$ in \eqref{eq:gauge}.
For $k\geq1$, integration of
$\partial_t[(x-t)^r t^{k-1}e^{U(t)/T}]$ gives
\begin{equation}
 Tr I_{r-1,k-1}
 =I_{r,k+1}+E I_{r,k-1}+T(k-1)I_{r,k-2}.
\end{equation}
For $k=0$, the first term on the right of \eqref{eq:intermediate} is absent. Hence, for $0\leq k\leq r$,
\begin{equation}
 \mathcal L_Ng_k=(k-r)g_{k+1}+Ek g_{k-1}+Tk(k-1)g_{k-2},
 \label{eq:gaction}
\end{equation}
where terms with zero coefficient are omitted. In particular, no $g_{r+1}$ is needed. Multiplying by $(-1)^k$ gives
\begin{equation}
 \mathcal L_Nh_k=(r-k)h_{k+1}-Ek h_{k-1}+Tk(k-1)h_{k-2},
\end{equation}
which is exactly the image under $\mathcal K$ of
\begin{equation}
 \mathcal L_N^\dagger x^k
 =(r-k)x^{k+1}-Ekx^{k-1}+Tk(k-1)x^{k-2}.
\end{equation}
For $N=1$, $\mathcal K1=e^{-U/T}\int_{-\infty}^xe^{U(t)/T}\dd t$ and direct differentiation gives $\mathcal L_1\mathcal K1=0$. This completes the proof of \eqref{eq:intertwiner} for every $N$.

\subsection{Identification of the physical modes}
Let $p_a(x)=c_a x^r+\cdots$ be a nonzero polynomial eigenfunction. Its degree is exactly $r$: otherwise the raising term in \eqref{eq:operators} would contradict the eigenvalue equation. By \eqref{eq:intertwiner},
\begin{equation}
 w_a=\mathcal Kp_a,\qquad \mathcal L_Nw_a=\epsilon_aw_a.
\end{equation}
Hence each left eigenfunction $p_a$ is carried into a right eigenfunction $w_a$ with the same eigenvalue. Since the integrand in \eqref{eq:K} is smooth and decays like $e^{t^3/3T}$ as $t\to-\infty$, $w_a$ is regular on the whole real line, with no apparent poles at the zeros of $p_a$. It remains to check that $w_a$ is an admissible right mode, i.e.\ that it carries a nonzero Lagrange current; this follows from its behaviour at the two real infinities.

Endpoint asymptotics of \eqref{eq:K}, with $x-t$ of order $T/x^2$, give at both real infinities
\begin{equation}
 w_a(x)=(-1)^r c_a r!T^{r+1}x^{-r-2}\bigl(1+O(x^{-1})\bigr).
 \label{eq:tail}
\end{equation}
At $+\infty$, portions of the integral away from the endpoint are exponentially subleading. The analogous expansion holds at $-\infty$. These expansions can be differentiated and extended to arbitrary algebraic order.

The Lagrange identity is
\begin{equation}
 \partial_xJ[p,w]=p\mathcal L_Nw-w\mathcal L_N^\dagger p.
\end{equation}
Thus $J[p_a,w_a]$ is constant, and \eqref{eq:tail} evaluates it explicitly:
\begin{equation}
 J[p_a,w_a]=(-1)^r c_a^2 r!T^{r+1}\ne0.
 \label{eq:Jvalue}
\end{equation}
Define
\begin{equation}
 \widehat v_a=\frac{\mathcal Kp_a}{(-1)^r c_a^2 r!T^{r+1}}.
 \label{eq:normalizedmode}
\end{equation}
Then $J[p_a,\widehat v_a]=1$. On any interval free of zeros of $p_a$, solving this first-order equation gives precisely \eqref{eq:letter-mode}, up to addition of a constant multiple of $e^{-U/T}p_a$. That homogeneous term grows exponentially at $-\infty$, whereas \eqref{eq:tail} is algebraic. Its coefficient is therefore zero for the physical solution selected at $-\infty$. Below the lowest real zero of $p_a$ the integral in \eqref{eq:letter-mode} is an ordinary one, and $\widehat v_a$ coincides with the right mode of \cite{Letter} identically. Beyond that zero, \eqref{eq:K} provides the regular continuation, so the finite-part prescription of \cite{Letter} is no longer needed: it is replaced by the ordinary convergent integral \eqref{eq:K}.

This argument is a scalar differential-equation identification, not an appeal
to uniqueness of the stationary diffusion. It also shows what the integral
representation \eqref{eq:K} adds to the  form \eqref{eq:letter-mode} which appeared in \cite{Letter}.
The two describe the same modes, but the integrand in \eqref{eq:K} is regular,
so polynomial zeros cause no difficulty, and \eqref{eq:K} supplies the regular
meaning of the finite part in \eqref{eq:letter-mode} at those zeros. Moreover,
$\mathcal K$ is linear in $p$: a single operator carries the whole of
$\mathcal P_r$ onto the space of right modes, with $\mathcal K[x^k]=h_k$. This
is what turns the determinant of right modes into the polynomial-ensemble form
used below without diagonalizing $\mathcal L_N^\dagger$, and it keeps that form
well defined at parameter values where diagonalization fails. The intertwining
relation \eqref{eq:intertwiner} shows directly that each left eigenfunction is
carried into a right eigenfunction with the same eigenvalue, and the algebraic
tails \eqref{eq:tail}, hence the value \eqref{eq:Jvalue} of the current, follow
from the endpoint expansion of a single integral. The  form in \cite{Letter}, in turn,
displays the unit Lagrange current \eqref{eq:lagrange-current} mode by mode.

\subsection{Equality of the normalized densities}
Write $p_a(x)=\sum_{k=0}^r C_{ak}x^k$. For a polynomial eigenbasis $C$ is invertible, so
\begin{equation}
 \det[p_a(\lambda_i)]=\det C\,\DeltaV(\lambda),
 \qquad
 \det[(\mathcal Kp_a)(\lambda_i)]
 =\det C\,\det[h_k(\lambda_i)]_{k,i}.
\end{equation}
Replacing $g_k$ by $h_k=(-1)^kg_k$, and then $\mathcal Kp_a$ by the current-normalized modes \eqref{eq:normalizedmode}, changes determinants only by $\lambda$-independent constants. Equations \eqref{eq:polyensemble} and \eqref{eq:normalizedmode} therefore prove \eqref{eq:claim} up to an overall constant, which is fixed to unity by normalization on the same ordered chamber.

At exceptional parameter values where diagonalization of the finite polynomial operator fails, \eqref{eq:polyensemble} remains well-defined and equality extends from generic parameters by continuity. Since both matrix distributions are conjugation invariant, equality of their joint eigenvalue densities implies equality of their full matrix probability densities (with the unordered symmetric density obtained by dividing by $N!$). For $N=1$, the construction reduces immediately to $P(x)\propto e^{-U(x)/T}\int_{-\infty}^x e^{U(t)/T}\,\dd t$, the scalar Halperin density.

\section{Exact reduction of the even-\texorpdfstring{$N$}{N}
orthogonal ensemble to unitary recurrence coefficients}
\label{app:beta1reduction}

We give a direct proof of the finite-$N$ reduction for $\beta=1$ and even $N$.
The result uses the unitary orthogonal polynomials for the
\emph{squared} one-particle weight and one explicitly determined
Airy primitive. The central step is a finite change of polynomial
basis which turns the ordered-integral skew matrix into a
five-diagonal matrix. Its Pfaffian then obeys the same three-term
recursion as in the symplectic calculation, with different initial
values. 
\subsection{Conventions and statement of the result}

We work in the dimensionless variable $w=-\lambda/T^{1/3}$ and set
\begin{equation}
 s=-E/T^{2/3},\qquad
 \rho_s(w)=\exp(w^3/3-sw).
 \label{eq:b1a-weight}
\end{equation}
The outgoing contour $\mathcal C_-$ is oriented from
$e^{\ii\pi/3}\infty$ to $-\infty$ and may be taken as the two
straight rays joined at the origin. The symbol $\prec$ refers
to order along this oriented contour. To keep the normalization
unambiguous, we use the holomorphic ordered integral
\begin{equation}
 \tau^{(1)}_N(s)=
 \int_{w_1\prec\cdots\prec w_N}
 \prod_{i<j}(w_j-w_i)\prod_{j=1}^N\rho_s(w_j)\,\dd w_j.
 \label{eq:b1a-tau}
\end{equation}
This is the holomorphic partition function entering the
characteristic exponent, not the real normalization proportional
to its squared modulus.

Let $p_n(w)=w^n+\cdots$ be the monic polynomials satisfying the
bilinear, without complex conjugation, orthogonality relations
\begin{align}
 (p_j,p_k)_2
 &:=
 \int_{\mathcal C_-}p_j(w)p_k(w)\rho_s(w)^2\,\dd w
 =\widetilde h_j\delta_{jk},                                  \label{eq:b1a-orth}\\
 wp_n&=p_{n+1}+\widetilde d_n p_n+\widetilde r_n p_{n-1},
 \qquad
 \widetilde r_n=\frac{\widetilde h_n}{\widetilde h_{n-1}},
 \quad \widetilde r_0=0.                                     \label{eq:b1a-three}
\end{align}
These are precisely the unitary recurrence coefficients at temperature $T/2$
before the dimensionless change of variables. In this appendix
all tilded quantities refer to the $w$ variable. We abbreviate
\begin{equation}
 B_n=\widetilde d_n+\widetilde d_{n+1}.
 \label{eq:b1a-B}
\end{equation}

\paragraph*{Exact identity.}
For $N=2m\geq2$, with an empty product interpreted as one,
\begin{equation}
 \displaystyle
 \tau^{(1)}_{2m}(s)=(-2)^m
       \left[\prod_{j=0}^{m-2}\widetilde h_{2j+1}(s)\right]
       X_{m-1}(s).
 \label{eq:b1a-main}
\end{equation}
The sequence $X_k$ is specified by
\begin{align}
 X_0(s)&=\frac{\ii\pi}{3}
       +\frac{\pi}{2}\int_0^{2^{2/3}s}
                    [\Bi(u)+\ii\Ai(u)]\,\dd u,                \label{eq:b1a-X0}\\
 X_1&=B_0X_0-\widetilde h_0,                                  \label{eq:b1a-X1}\\
 X_{k+1}&=B_{2k}X_k-\widetilde r_{2k}X_{k-1},
 \qquad k\geq1.                                              \label{eq:b1a-Xrec}
\end{align}
The integral in \eqref{eq:b1a-X0} is path independent for complex
$s$, since its integrand is entire.
The factor $(-2)^m$ in \eqref{eq:b1a-main} is fixed by the
orientation and ordered-integral convention in
\eqref{eq:b1a-tau}. Dropping this $s$-independent factor gives
the proportionality formula in the main text.

The corresponding characteristic exponent is
\begin{equation}
 \displaystyle
 \frac{\Xi^{(1)}_{2m}(E)}{T^{1/3}}
 =-2\sum_{j=0}^{m-2}\widetilde d_{2j+1}(s)
       +\frac{\partial_sX_{m-1}(s)}{X_{m-1}(s)},
 \qquad s=-E/T^{2/3}.
 \label{eq:b1a-Xi}
\end{equation}
We first prove the partition-function identity and then derive
\eqref{eq:b1a-Xi}.

The orthogonal-polynomial argument is made initially where the
finitely many required Hankel determinants are nonzero.
Because the contour integrals are entire in $s$, the complete
identity extends across isolated singularities of the recurrence
representation by analytic continuation. The logarithmic
derivative is understood away from zeros of $\tau^{(1)}_{2m}$.

\subsection{Ordered integrals and the integration-by-parts identity}

For polynomials $f,g$, define the skew form
\begin{align}
 [f,g]_1
 &=\int_{x\prec y}
       [f(x)g(y)-g(x)f(y)]\rho_s(x)\rho_s(y)\,\dd x\,\dd y.
 \label{eq:b1a-skew}
\end{align}
In particular, $[f,g]_1=-[g,f]_1$.
The ordered version of de Bruijn's formula
\cite{deBruijn1955} gives
\begin{equation}
 \tau^{(1)}_{2m}
 =\Pf\big[[p_j,p_k]_1\big]_{0\leq j,k<2m}.
 \label{eq:b1a-deBruijn}
\end{equation}
There is no factorial in this convention: the determinant of
the monic polynomials evaluated at the ordered variables is
$\prod_{i<j}(w_j-w_i)$. Replacing the monomials by any other
monic basis of successive degrees multiplies that determinant,
and the associated Pfaffian, by one.

Introduce the weighted derivative
\begin{equation}
 \mathcal D f=f'+(w^2-s)f
             =\rho_s^{-1}(f\rho_s)'.
 \label{eq:b1a-D}
\end{equation}
Its essential property is
\begin{equation}
 \displaystyle [\mathcal D f,g]_1=2(f,g)_2.
 \label{eq:b1a-key}
\end{equation}
To verify both the sign and the boundary terms, write
\[
 F_f(y)=\int_{\mathcal C_-^{\rm start}}^y f(x)\rho_s(x)\,\dd x,
 \qquad
 M_f=\int_{\mathcal C_-}f(x)\rho_s(x)\,\dd x.
\]
Splitting the ordered double integral in \eqref{eq:b1a-skew}
then gives
\begin{equation}
 [f,g]_1=\int_{\mathcal C_-}
          [2F_f(y)-M_f]\,g(y)\rho_s(y)\,\dd y.
 \label{eq:b1a-primitive}
\end{equation}
Since a polynomial times $\rho_s$ vanishes at both endpoints,
$F_{\mathcal D f}(y)=f(y)\rho_s(y)$ and
$M_{\mathcal D f}=0$. Substitution proves
\eqref{eq:b1a-key}. All these integrals converge absolutely
on the two rays, locally uniformly in $s$.
If integration is performed separately on the rays, the
contributions at their common endpoint cancel. In particular,
there is no extra term at the corner of the contour.
Skew-symmetry also gives
\begin{equation}
 [f,\mathcal D g]_1=-2(f,g)_2.
 \label{eq:b1a-key2}
\end{equation}

\subsection{A finite basis change and the five-diagonal matrix}

Use the following single sequence of monic polynomials:
\begin{equation}
 q_0=p_0=1,\qquad q_1=p_1,\qquad
 q_{n+2}=\mathcal Dp_n\quad(n\geq0).
 \label{eq:b1a-basis}
\end{equation}
The degree of $\mathcal Dp_n$ is $n+2$, with leading
coefficient one. Thus for every finite $2m$ the transformation
from $p_0,\ldots,p_{2m-1}$ to $q_0,\ldots,q_{2m-1}$
is unitriangular. It is a finite transformation: its last member
uses only $\mathcal Dp_{2m-3}$, whose degree is $2m-1$.
Consequently, with
\[
 C_{jk}=[q_j,q_k]_1,\qquad
 P_m=\Pf[C_{jk}]_{0\leq j,k<2m},\qquad P_0=1,
\]
we have $P_m=\tau^{(1)}_{2m}$.

Set
\begin{equation}
 A(s)=[1,w]_1=[p_0,p_1]_1.
 \label{eq:b1a-A}
\end{equation}
The first two rows follow immediately from
\eqref{eq:b1a-key2}:
\begin{align}
 C_{01}&=A,&
 C_{0,n+2}&=-2\widetilde h_0\delta_{n0},&
 C_{1,n+2}&=-2\widetilde h_1\delta_{n1}.                        \label{eq:b1a-first}
\end{align}
For the remaining entries, \eqref{eq:b1a-key} gives
\begin{equation}
 C_{i+2,j+2}
 =[\mathcal Dp_i,\mathcal Dp_j]_1
 =2(p_i,\mathcal Dp_j)_2
 =-2(\mathcal Dp_i,p_j)_2.                                   \label{eq:b1a-bulk}
\end{equation}
The last equality follows directly by integrating
$(p_ip_j\rho_s^2)'$:
\[
 (p_i,\mathcal Dp_j)_2+(\mathcal Dp_i,p_j)_2=0.
\]
It holds on the complex contour because the orthogonality
is bilinear and the endpoint terms vanish.

Applying \eqref{eq:b1a-three} twice yields
\begin{equation}
 \mathcal Dp_i
 =p_{i+2}+B_i p_{i+1}
          +\text{a polynomial of degree at most }i.
 \label{eq:b1a-leading}
\end{equation}
For $j>i$, orthogonality therefore determines every entry of
\eqref{eq:b1a-bulk}:
\begin{equation}
 C_{i+2,j+2}=
 \begin{cases}
 -2\widetilde h_{i+1}B_i,&j=i+1,\\
 -2\widetilde h_{i+2},&j=i+2,\\
 0,&j\geq i+3.
 \end{cases}                                                \label{eq:b1a-entries}
\end{equation}
The lower triangle is fixed by skew-symmetry. Equivalently,
the only nonzero upper-triangular entries are
\begin{align}
 C_{01}&=A,\qquad C_{12}=0, \nonumber\\
 C_{k,k+1}&=-2\widetilde h_{k-1}B_{k-2}\quad(k\geq2),\qquad
 C_{k,k+2}=-2\widetilde h_k\quad(k\geq0).                      \label{eq:b1a-bands}
\end{align}
This is the required five-diagonal matrix. Its exceptional
first rows are explicit in \eqref{eq:b1a-first}; they will
produce the inhomogeneous term in the initial value $X_1$.

\subsection{Pfaffian expansion and the three-term recursion}

For clarity, consider first any skew matrix with
$C_{jk}=0$ for $|j-k|>2$. In the last-row expansion of its
$2m\times2m$ Pfaffian, the final index $2m-1$ can be paired
only with $2m-2$ or $2m-3$.
The first choice leaves the leading $(2m-2)\times(2m-2)$
Pfaffian. In the second choice the remaining index $2m-2$
must pair with $2m-4$, leaving the leading
$(2m-4)\times(2m-4)$ Pfaffian. The two pairings
$(2m-4,2m-2)$ and $(2m-3,2m-1)$ cross in the ordered
index list and contribute with a minus sign. Hence
\begin{equation}
 P_m=C_{2m-2,2m-1}P_{m-1}
       -C_{2m-4,2m-2}C_{2m-3,2m-1}P_{m-2},
 \qquad m\geq2.                                             \label{eq:b1a-Pfrec}
\end{equation}
For $m=2$ this is simply
$\Pf C=C_{01}C_{23}-C_{02}C_{13}+C_{03}C_{12}$,
with $C_{03}=0$.

Inserting \eqref{eq:b1a-bands} gives
\begin{equation}
 P_m=-2\widetilde h_{2m-3}B_{2m-4}P_{m-1}
       -4\widetilde h_{2m-4}\widetilde h_{2m-3}P_{m-2},
 \qquad P_0=1,\quad P_1=A.                                  \label{eq:b1a-P}
\end{equation}
In particular,
\[
 P_2=-2\widetilde h_1 B_0 A-4\widetilde h_0\widetilde h_1.
\]
Define $X_0=-A/2$ and, for $m\geq1$, factor out the norms by
\begin{equation}
 P_m=(-2)^m
       \left[\prod_{j=0}^{m-2}\widetilde h_{2j+1}\right]X_{m-1}.
 \label{eq:b1a-factor}
\end{equation}
The displayed formula for $P_2$ gives
$X_1=B_0X_0-\widetilde h_0$.
For $m\geq3$, division of \eqref{eq:b1a-P} by the factor in
\eqref{eq:b1a-factor} gives
\[
 X_{m-1}=B_{2m-4}X_{m-2}
       -\frac{\widetilde h_{2m-4}}{\widetilde h_{2m-5}}X_{m-3}.
\]
Putting $k=m-2$ proves \eqref{eq:b1a-Xrec}.
Thus the partition-function reduction is already established,
apart from the explicit evaluation of $X_0$.

\subsection{The contour fixes the Airy primitive}

First differentiate $A=[1,w]_1$ with respect to $s$.
Each derivative of a weight brings down minus its integration
variable. Therefore
\begin{equation}
 A'(s)=-[w,w]_1-[1,w^2]_1=-[1,w^2]_1.
\end{equation}
Since $\mathcal D1=w^2-s$, \eqref{eq:b1a-key2} gives
$[1,w^2]_1=-2\widetilde h_0$. Hence
\begin{equation}
 A'=2\widetilde h_0,\qquad X_0'=-\widetilde h_0.                \label{eq:b1a-Xprime}
\end{equation}
Rescaling the squared-weight integral by $u=2^{1/3}w$
and using the outgoing Airy contour gives
\begin{equation}
 \widetilde h_0(s)
 =-\frac{\pi}{2^{1/3}}
       [\Bi(2^{2/3}s)+\ii\Ai(2^{2/3}s)].
 \label{eq:b1a-h0}
\end{equation}
The minus sign is fixed by the orientation from
$e^{\ii\pi/3}\infty$ to $-\infty$; the usual Airy integral
representations give this same convention
\cite[Sec.~9.5]{DLMFAiry}.
Equation \eqref{eq:b1a-Xprime} now determines $X_0$ up to
one constant. We compute that constant from the ordered integral
itself, rather than impose a normalization afterwards.

At $s=0$ put $a=e^{\ii\pi/3}$ and $b=-1$.
The two parts of the contour are $w=at$ with $t$ decreasing
from infinity to zero and $w=bt$ with $t$ increasing from
zero to infinity. In
\[
 A(0)=\int_{x\prec y}(y-x)e^{(x^3+y^3)/3}\,\dd x\,\dd y,
\]
the contributions with both variables on the same ray cancel:
both have cubic phase $-t^3/3$, while reversing their radial
order reverses the factor $y-x$.
For the remaining contribution, $x$ is on the incoming ray and
$y$ on the outgoing ray. Define
\[
 J_\ell=\int_0^\infty t^\ell e^{-t^3/3}\,\dd t
       =3^{(\ell-2)/3}\Gamma\big((\ell+1)/3\big).
\]
Including the incoming-ray orientation, the cross contribution is
\begin{align}
 A(0)
 &=-ab\int_0^\infty\!\!\int_0^\infty
       (bv-au)e^{-(u^3+v^3)/3}\,\dd u\,\dd v \nonumber\\
 &=ab(a-b)J_0J_1
   =-\ii\sqrt3\,\frac{\Gamma(1/3)\Gamma(2/3)}3
   =-\frac{2\ii\pi}{3}.                                     \label{eq:b1a-Azero}
\end{align}
Thus $X_0(0)=\ii\pi/3$. Integration of
\eqref{eq:b1a-Xprime}, using \eqref{eq:b1a-h0}, proves
\eqref{eq:b1a-X0}, including its additive constant.

For real $s$, the standard oscillatory integrals
\[
 \int_{-\infty}^0\Ai(u)\,\dd u=\frac23,\qquad
 \int_{-\infty}^0\Bi(u)\,\dd u=0
\]
\cite[Sec.~9.10(iv)]{DLMFAiry} give the equivalent form
\begin{equation}
 X_0(s)=-\int_{-\infty}^s\widetilde h_0(t)\,\dd t,\qquad
 \lim_{s\to-\infty}X_0(s)=0.                                 \label{eq:b1a-boundary}
\end{equation}
The improper integral is conditionally convergent. Its
oscillatory Airy asymptotics give
$X_0(s)=O(|s|^{-3/4})$ as $s\to-\infty$.
The finite-base-point formula \eqref{eq:b1a-X0} specifies its
entire continuation for complex $s$.

\subsection{Characteristic exponent from the unitary recurrence coefficients}

Differentiating the norm in \eqref{eq:b1a-orth} gives
\[
 \partial_s\widetilde h_n
 =2\int p_n(\partial_sp_n)\rho_s^2\,\dd w
       -2\int w p_n^2\rho_s^2\,\dd w.
\]
The first integral vanishes because $\partial_sp_n$ has
degree at most $n-1$; the recurrence
\eqref{eq:b1a-three} evaluates the second. Consequently,
\begin{equation}
 \partial_s\log\widetilde h_n=-2\widetilde d_n.              \label{eq:b1a-normflow}
\end{equation}
Using $\Xi^{(1)}_{2m}/T^{1/3}=\partial_s\log\tau^{(1)}_{2m}$
in \eqref{eq:b1a-main} now proves \eqref{eq:b1a-Xi}.

For completeness, all recurrence coefficients needed in this formula can be
generated from \eqref{eq:b1a-h0} by the unitary string equations
in the present dimensionless convention:
\begin{align}
 \widetilde d_n^2+\widetilde r_n+\widetilde r_{n+1}
 &=s, \nonumber\\
 \widetilde r_n(\widetilde d_n+\widetilde d_{n-1})
 &=-\frac n2,\qquad n\geq1.                                 \label{eq:b1a-string}
\end{align}
The first follows by integrating $(p_n^2\rho_s^2)'$:
the derivative of $p_n$ is orthogonal to $p_n$, and
$(\rho_s^2)'=2(w^2-s)\rho_s^2$.
For the second, integration of $(p_np_{n-1}\rho_s^2)'$
gives
$n\widetilde h_{n-1}
 +2\widetilde h_n(\widetilde d_n+\widetilde d_{n-1})=0$.
In particular,
\[
 \widetilde d_0=-\frac12\partial_s\log\widetilde h_0,
 \qquad \widetilde r_1=s-\widetilde d_0^2,
 \qquad \widetilde h_n=\widetilde r_n\widetilde h_{n-1}.
\]
The negative sign in the second string equation reflects the
weight $e^{2w^3/3-2sw}$ and the change from $\lambda$ to $w$.

The associated Toda relations are
\begin{equation}
 \partial_s\widetilde d_n
 =-2(\widetilde r_{n+1}-\widetilde r_n),\qquad
 \partial_s\widetilde r_n
 =-2\widetilde r_n(\widetilde d_n-\widetilde d_{n-1}).
 \label{eq:b1a-Toda}
\end{equation}
For example, differentiation of the orthogonality relations
gives $\partial_sp_n=2\widetilde r_n p_{n-1}$; comparison
of the next-to-leading coefficients in
\eqref{eq:b1a-three} gives the first equation in
\eqref{eq:b1a-Toda}, and \eqref{eq:b1a-normflow} gives
the second. Thus the derivative in \eqref{eq:b1a-Xi} can
be evaluated along with $X_k$, without numerical
differentiation. Writing a dot for $\partial_s$, one uses
\begin{align}
 \dot X_0 =-\widetilde h_0,\quad 
 \dot X_1&=\dot B_0X_0+
        (\widetilde d_0-\widetilde d_1)\widetilde h_0,
                                                              \label{eq:b1a-dot}\\
 \dot X_{k+1}
 &=\dot B_{2k}X_k+B_{2k}\dot X_k
       -\dot{\widetilde r}_{2k}X_{k-1}
       -\widetilde r_{2k}\dot X_{k-1},\qquad k\geq1.\nonumber
\end{align}

\paragraph*{Low-dimensional checks.}
The first three even sizes read
\begin{align}
 \tau^{(1)}_2=-2X_0,\quad 
 \tau^{(1)}_4&=4\widetilde h_1(B_0X_0-\widetilde h_0),\nonumber\\
 \tau^{(1)}_6&=-8\widetilde h_1\widetilde h_3
       \big[(B_2B_0-\widetilde r_2)X_0
                                      -B_2\widetilde h_0\big].
 \label{eq:b1a-checks}
\end{align}
These follow both from direct expansion of the corresponding
Pfaffians and from the recursion. At $N=2$,
$\Xi^{(1)}_2/T^{1/3}=-\widetilde h_0/X_0$, which also follows
directly from $A'=2\widetilde h_0$ and $\tau^{(1)}_2=A$.

\paragraph*{Relation to the symplectic reduction.}
At fixed $s$, let $Q_k$ and $R_k$ satisfy the homogeneous
recursion \eqref{eq:b1a-Xrec}, with
\[
 Q_0=1,\quad Q_1=B_0,\qquad R_0=0,\quad R_1=1.
\]
Then the exact solution is
\begin{equation}
 X_k=Q_kX_0-\widetilde h_0R_k.                                \label{eq:b1a-two}
\end{equation}
The sequence $Q_k$ is the same one that appears in the
symplectic reduction when built from these unitary
recurrence coefficients. The orthogonal ensemble selects a
different linear combination through the ordered-contour
primitive $X_0$ and the term $-\widetilde h_0$ in $X_1$.
 Thus the extra $\beta=1$ information is confined to explicit
scalar initial values, while all coefficients of the finite-$N$
recursion come from the $\beta=2$ problem at temperature $T/2$.
This completes the exact reduction for even $N$. Its double-scaling
limit, including the selection of the Painlev\'e~I solution, is
carried out in Sec.~\ref{sec:b1scaling}. Odd $N$, which requires a
bordered Pfaffian, remains open.

\section{The $\beta=1$ edge and real symmetric random band matrices}
\label{app:RBM1}

In Ref.~\cite{Letter} the unitary edge scaling function was compared
with the localized spectral edge of complex Hermitian random band
matrices. We retain the rescaling constants fixed there and extend
the proposed correspondence to the other symmetry classes. The common
semicircle term is required by normalization. We make only one
nontrivial comparison here: the coefficient of the first
symmetry-dependent correction inside the band, evaluated for the
real symmetric ensemble.

\paragraph*{The band-matrix edge.}
Consider $M\times M$ real symmetric periodic band matrices.
For $u<v$ with $0<|u-v|_M\le W$, the entries
$H_{uv}=H_{vu}=\pm1$ are independent random signs with
equal probabilities; all remaining entries, including
the diagonal, vanish. Let $\lambda_\alpha$ denote the
eigenvalues of $H/\sqrt{2W}$, whose limiting semicircle
has edges at $\pm2$. Define the edge coordinate
$\xi_\alpha=(2-\lambda_\alpha)W^{4/5}$ and the
rescaled cumulative eigenvalue count
\be
\sigma_1(\xi)=\lim\frac{W^{6/5}}{M}\,
\aver{\#\{\alpha:\xi_\alpha<\xi\}},
\qquad 1\ll W\ll M^{5/6}.
\label{eq:sigma1def}
\ee
The brackets in Eq.~\eqref{eq:sigma1def} denote expectation over
the independent random signs defining the band-matrix ensemble.
Positive $\xi$ points into the band, while negative
$\xi$ describes eigenvalues beyond the limiting edge.
Sodin proved the existence of this limiting counting
function for the random-sign model and obtained
\[
\sigma_1(\xi)=\frac{2}{3\pi}\xi^{3/2}+O(\xi),
\qquad \xi\to+\infty.
\]
The leading coefficient is fixed by the semicircle
normalization~\cite{Sodin2010}.

To probe corrections to this leading behaviour, one
uses averaged traces of Chebyshev polynomials
$U_n$ of the second kind~\cite{Sodin2010,LiuZou2024}.
Their degree $n$ scales with the bandwidth so that
\[
\tau=\frac{n}{W^{2/5}}>0
\]
remains fixed. Indeed, near the upper edge,
\[
\frac{1}{n+1}U_n\!\left(\frac{\lambda}{2}\right)
\longrightarrow
\frac{\sin(\tau\sqrt{\xi})}{\tau\sqrt{\xi}},
\qquad
\lambda=2-\xi W^{-4/5}.
\]
For even degrees, the two edges contribute equally,
giving the transform
\be
\phi_1(\tau)=2\int_{-\infty}^{\infty}
\frac{\sin(\tau\sqrt{\xi})}{\tau\sqrt{\xi}}\,
\dd\sigma_1(\xi).
\label{eq:phi1transform}
\ee
For $\xi<0$, the kernel is understood by analytic
continuation as
$\sinh(\tau\sqrt{-\xi})/(\tau\sqrt{-\xi})$.
The integral uses the regularization of
Ref.~\cite{Letter}, which removes the leading
semicircle contribution. This reflects the exact
Chebyshev orthogonality relation
\[
\int_{-2}^{2}\frac{\sqrt{4-\lambda^2}}{2\pi}\,
U_n(\lambda/2)\,\dd\lambda=0,
\qquad n\ge1.
\]

In the walk representation of these traces, $\tau$
is the total walk length measured in units of
$W^{2/5}$; diagram edge lengths such as $\alpha$
use the same units. The Chebyshev expansion cancels
immediate backtracking contributions. For random
signs, the remaining walks may traverse an edge
twice in the same direction, producing
non-orientable diagrams. The leading such diagram
has no unitary counterpart and supplies the single
symmetry-dependent correction compared below.

\paragraph*{The orthogonal scaling function.}
For $\beta=1$ and even $N$ we write
$E=E^*_1+s_1^2N^{-4/5}X$ and $t=2^{7/5}X$, so that
Eq.~\eqref{eq:rho1scaling} reads
$\rho^{(1)}_N\simeq(N^{3/5}/s_1)\,\Phi_1'(X)$ with
$\Phi_1'(X)=(2^{9/5}/\pi)\,\Imn[\tfrac14y-v_1^2]$. Since
$v_1'=\tfrac34y-v_1^2$, we have $\tfrac14y-v_1^2=-\tfrac12y+v_1'$. The first
term gives the unitary scaling function
$\Phi'(X)=-(2^{4/5}/\pi)\Imn y_1(2^{7/5}X)$ of Ref.~\cite{Letter}, and the
second is a total derivative. The cumulative intensity
$\Phi_1(X)=\int_{-\infty}^X\Phi_1'(Y)\,\dd Y$ is therefore
\be
\Phi_1(X)=\Phi(X)+\frac{2^{2/5}}{\pi}\,\Imn v_1\big(2^{7/5}X\big),
\label{eq:Phi1exact}
\ee
where we have used that $v_1\simeq-\sqrt{3y/4}$ is real, up to
exponentially small terms, as $t\to-\infty$. Equation~\eqref{eq:Phi1exact}
plays the role that the construction of $F_1$ from $F_2$ plays for the
Tracy--Widom distributions.

On the allowed side $y_1\simeq-\ii\sqrt{t/6}-1/(48t^2)+\dots$, and the
selected solution $x$ decays as $t\to+\infty$, so that
$v_1=-\sqrt{3y/4}-1/(8t)+O(t^{-9/4})$, with the branch
$\Rea\sqrt{3y/4}>0$. Hence
$\Imn v_1=\tfrac14\,6^{1/4}t^{1/4}+O(t^{-9/4})$, and with the unitary
expansion $\Phi(X)=A_+X^{3/2}+a_\Phi X^{-7/2}+\dots$ of Ref.~\cite{Letter},
\be
\Phi_1(X)=A_+X^{3/2}+\frac{3^{1/4}}{2\pi}\,X^{1/4}+O\big(X^{-9/4}\big),
\qquad A_+=\frac{4}{3\pi\sqrt3},
\qquad X\to+\infty .
\label{eq:Phi1allowed}
\ee
The term $X^{1/4}$ is the projective-plane contribution. Its sign is fixed
by the selection of Par.~7 of Sec.~\ref{sec:b1scaling}; for $\beta=4$ it is
reversed. 

\paragraph*{Rescaling fixed by the unitary case.}
We conjecture $\Phi_1(c\xi)=\mathcal A\sigma_1(\xi)$, using
\[
c=(3/64)^{1/5},\qquad \mathcal A=48^{-1/5},
\]
as fixed in Ref.~\cite{Letter}. With the same variance profile,
bandwidth convention, and spectral normalization, the orientable
contributions retain the unitary normalization. No additional
constants are fitted in the orthogonal comparison. The leading
term $2\xi^{3/2}/(3\pi)$ is required by the common semicircle law
and is not an independent test of the correspondence.
Equation~\eqref{eq:Phi1allowed} then predicts, without
any free parameter,
\be
\sigma_1(\xi)=\frac{2}{3\pi}\,\xi^{3/2}+\frac{\sqrt6}{2\pi}\,\xi^{1/4}
+O\big(\xi^{-9/4}\big),\qquad \xi\to+\infty .
\label{eq:sigma1pred}
\ee

\paragraph*{The single nontrivial check: the projective-plane term.}

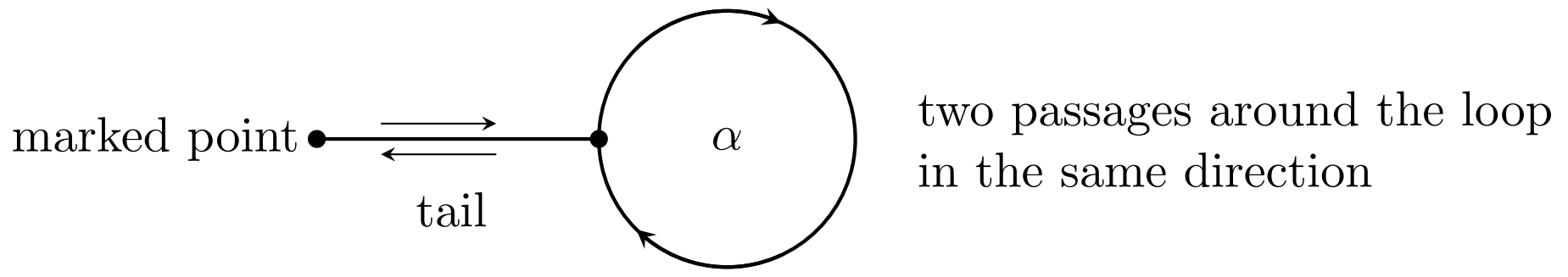
\begin{figure}[htbp]
\centering
\begin{tikzpicture}[>=stealth,scale=1]
  \draw[thick] (-2.2,0) -- (0,0);
  \fill (-2.2,0) circle (2pt);
  \node[left] at (-2.2,0) {marked point};
  \fill (0,0) circle (2pt);

  \draw[->] (-1.7,0.12) -- (-0.8,0.12);
  \draw[->] (-0.8,-0.12) -- (-1.7,-0.12);
  \node[below] at (-1.15,-0.28) {tail};

  \draw[thick] (1,0) circle (1);
  \node at (1,0) {$\alpha$};

  \draw[->,thick]
    (1,0) ++(135:1) arc[start angle=135,end angle=65,radius=1];
  \draw[->,thick]
    (1,0) ++(-65:1) arc[start angle=-65,end angle=-135,radius=1];

  \node[align=left,right] at (2.35,0)
    {two passages around the loop\\in the same direction};
\end{tikzpicture}
\caption{\footnotesize The leading non-orientable walk diagram.
Starting at the marked point, the walk follows the tail,
goes twice around the same loop in the same direction,
and returns along the tail. Reversing direction on the
second circuit would immediately retrace the last edge
of the first circuit, violating the non-backtracking
condition. The loop has length $\alpha$ and is traversed
twice, giving $2\alpha\le\tau$. The drawing shows the
underlying graph and walk; its ribbon thickening gives
the projective-plane contribution.}
\label{fig:RBM-projective}
\end{figure}

The leading non-orientable diagram consists of a marked
tail attached to a single loop, traversed twice in the
same direction; see Fig.~\ref{fig:RBM-projective}.
The non-backtracking condition excludes the opposite
orientation of the second passage, so this diagram
has no unitary counterpart. The diagram has one independent spatial loop.
For the uniform band profile, a step measured in units
of the bandwidth has variance
$\varsigma^2=\int_{-1}^{1}u^2\,\dd u/2=1/3$.
The continuum return probability for a loop of length
$\alpha$ is therefore
\[
p_\alpha(0)=\frac{1}{\sqrt{2\pi\varsigma^2\alpha}}.
\]
Although the walk traverses the loop twice, its spatial
closure imposes only one return condition, giving one
factor $p_\alpha(0)$. The two passages use total length
$2\alpha$; the remaining length is assigned to the
outward and return passages along the marked tail.
After eliminating the tail length, the admissible
range is $0<\alpha<\tau/2$. With the normalized-trace
factor $1/\tau$, the diagram contributes
\be
\phi_{1,\mathrm{proj}}(\tau)
=\frac1\tau\int_0^{\tau/2}p_\alpha(0)\,\dd\alpha
=\frac{1}{\tau\sqrt{2\pi\varsigma^2}}\,
  2\sqrt{\frac{\tau}{2}}
=\sqrt{\frac3\pi}\,\tau^{-1/2}.
\label{eq:RBMs1}
\ee
The diagram evaluated above is Sodin's $s=1$
diagram, with its continuum weight given by
Eq.~(4.6) of Ref.~\cite{Sodin2010}.
A term $B\,\xi_+^{1/4}$ in $\sigma_1$ contributes
$B\sqrt{2\pi}\,\tau^{-1/2}$ to
Eq.~\eqref{eq:phi1transform}. Thus the result
$\sqrt{3/\pi}\,\tau^{-1/2}$ agrees precisely with
the Riccati prediction $B=\sqrt6/(2\pi)$ in
Eq.~\eqref{eq:sigma1pred}, at the level of the
transformed edge asymptotics.
The corresponding $\xi^{1/4}$ correction to the
edge counting function was conjectured by
Sodin~\cite[Remark~7.2]{Sodin2010}.

\paragraph*{Symplectic prediction and status.}
For quaternion self-dual band matrices with independent Haar-random
unit-quaternion entries in the band, count each Kramers pair once.
Keeping the unitary rescaling in the doubled-size convention of
Sec.~\ref{sec:b1scaling}, the corresponding prediction in the natural
band coordinate $\xi=(2-\lambda)W^{4/5}$ is
\be
\sigma_4(\xi)=\frac{2}{3\pi}\xi^{3/2}
-\frac{\sqrt6}{4\pi}\xi^{1/4}
+O\big(\xi^{-9/4}\big),\qquad \xi\to+\infty.
\label{eq:sigma4pred}
\ee
The reversal of the first correction reflects the different
selection of the auxiliary solution for $\beta=4$.
This is a prediction; we do not perform an additional
quaternion band-matrix comparison here.

On the Riccati side the arguments are at the physics level, and on the
band-matrix side the single-loop contribution follows from the continuum counting
formulas of Refs.~\cite{Sodin2010,LiuZou2024}, while the passage from the
transform to the asymptotics of $\sigma_1$ itself is the step that Sodin
could establish only in a weak sense. The correspondence
$\Phi_1(c\xi)=\mathcal A\sigma_1(\xi)$ remains a conjecture, now supported
by the first symmetry-dependent allowed-side coefficient, with no
adjustment of the unitary rescaling constants. 

\section{Boundary admittance, absorption and scattering time delays}
\label{app:admittance-scattering}

This appendix collects the scattering interpretation of the Riccati
variable without interrupting the spectral derivation in the main text.
It distinguishes the internal norm for a prescribed boundary amplitude,
the dwell time referred to incident flux, and the Wigner--Smith delay.
It also records the high-energy connection with absorbing-waveguide
statistics and outlines the programme of Paper~II~\cite{PaperII}.
Throughout the boundary-response derivations, the sample occupies
$0<x<L$, with a Dirichlet boundary at $x=0$ and a lead, when present,
attached at $x=L$; the positive coordinate direction points from the
Dirichlet end towards the lead. Thus $\cT=-\partial_EZ$, consistently
with the main text. The semi-infinite sample is obtained by taking
$L\to\infty$.

\subsection{Boundary response, absorbed flux and internal norm}
\label{app:admittance-norm}

Replacing $E$ by $\cE=E+\ii\eta$, with $\eta>0$,
is equivalent to adding the absorbing potential
$-\ii\eta$ to $W$. Consider the sample $0<x<L$,
with a Dirichlet condition at $x=0$, and prescribe
the wave amplitude at the other end:
\[
\psi(0)=0,\qquad \psi(L)=b.
\]
If $\Psi$ is the matrix solution satisfying
$\Psi(0)=0$ and $\Psi'(0)=I$, then
\[
\psi(x)=F(x;\cE)b,\qquad
F(x;\cE):=\Psi(x;\cE)\Psi(L;\cE)^{-1}.
\]
In particular,
\[
\psi'(L)=Z(L;\cE)b.
\]
Thus $Z$ maps the boundary amplitude to the boundary
slope. This is its Dirichlet-to-Neumann interpretation;
with the appropriate physical normalization it is an
input admittance.

The quantity measured by the imaginary part of $Z$
follows directly from flux balance. Define
\[
j(x)=\Imn\bigl(\psi^\dagger\psi'\bigr),
\]
positive towards increasing $x$. Since
$\psi''=(W-\cE)\psi$ and $W$ is Hermitian,
\[
j'(x)=-\eta\,\psi^\dagger\psi.
\]
The Dirichlet boundary condition $\psi(0)=0$ implies
that no probability current flows through the boundary
at $x=0$, so $j(0)=0$. Writing $Z=A-\ii P$, with $A$ and $P$ Hermitian,
we therefore obtain
\[
-j(L)=b^\dagger Pb
=\eta\int_0^L\psi^\dagger(x)\psi(x)\,\dd x.
\]
The minus sign reflects the geometry: flux injected
at $x=L$ travels towards decreasing $x$.
Since this identity holds for every boundary vector
$b$, it gives the matrix identity
\[\;
\frac{P}{\eta}
=\int_0^L F(x;\cE)^\dagger F(x;\cE)\,\dd x.
\;
\]
Thus $P/\eta$ is exactly the internal-norm matrix
for a prescribed boundary amplitude, at any
$\eta>0$. Its positivity expresses the positivity
of absorption.

At real energy, away from the Dirichlet eigenvalues,
the same internal norm is obtained by differentiating
the boundary response with respect to energy.
The Wronskian identity gives
\[
\;
\cT(E):=-\partial_E Z(L;E)
=\int_0^L F(x;E)^\dagger F(x;E)\,\dd x.
\;
\]
Consequently,
\[
\lim_{\eta\to 0_+}\frac{P(E,\eta)}{\eta}
=\cT(E).
\]
The distinction at finite absorption is simple:
$P/\eta$ measures the norm of the solution at
$E+\ii\eta$, whereas $\cT$ measures that of the
real-energy solution. They coincide only in the
zero-absorption limit
(see Sec.~\ref{sec:realE}).

\paragraph*{Scalar local and integrated currents.}
For $N=1$, $Z=\psi'/\psi=T^{1/3}(u-\ii v)$. In the current
normalization used in the scalar warm-up,
\[
j(x)=\frac{\hbar}{m}|\psi(x)|^2\Imn Z(x)
=-\frac{\hbar}{m}T^{1/3}|\psi(x)|^2v(x).
\]
The local intensity appears because $\psi'=Z\psi$. The integrated
Wronskian identity gives the equivalent flux-balance expression
\[
|\psi(x)|^2\Imn Z(x)
=-\eta\int_0^x|\psi(s)|^2\,\dd s.
\]
The explicit factor $\hbar/m$ only restores the conventional
current normalization; $E$ and $\eta$ retain the rescaled
Schr\"odinger convention of the main text.

\subsection{Ideal-lead matching and the absorption deficit}
\label{app:admittance-reflection}

To connect this boundary response to scattering,
attach an ideal lead at $x=L$. For a common channel
wavenumber $q>0$, write the lead wave as
\[
\psi_{\rm lead}(x)
=e^{-\ii q(x-L)}a_{\rm in}
+e^{\ii q(x-L)}a_{\rm out},
\qquad
a_{\rm out}=r\,a_{\rm in}.
\]
Matching the amplitude and slope at $x=L$ gives
\[
r=-(Z-\ii qI)^{-1}(Z+\ii qI),
\qquad
b=(I+r)a_{\rm in}.
\]
Thus fixing the incident amplitude is different
from fixing $b$: reflection changes the amplitude
actually applied to the sample. Flux balance now reads
\[
\;
I-r^\dagger r
=\frac1q(I+r)^\dagger P(I+r).
\;
\]
This is the scattering form of the same absorption
identity. The factors $I+r$ convert the prescribed
incident amplitude into the boundary amplitude.

A closely related convention for absorbing scattering
systems is to write
\[
I-S_\Gamma^\dagger S_\Gamma=\Gamma Q_\Gamma,
\qquad \Gamma=2\eta.
\]
Here $Q_\Gamma$ is the finite-absorption delay matrix
\cite{SavinSommers2003,FyodorovSavinSommers2005,
ChenAnlageFyodorov2021}.
For the lead construction above, the corresponding
absorption-deficit matrix is
\[
\frac{I-r^\dagger r}{2\eta}
=\frac1{2q}(I+r)^\dagger
\frac{P}{\eta}(I+r).
\]
This formula makes explicit both the internal norm
and the conversion to incident-flux normalization.

\subsection{Dwell time and the Wigner--Smith matrix}
\label{app:admittance-delays}

We retain the same coordinate convention: the sample occupies
$0<x<L$, the Dirichlet boundary is at $x=0$, and the lead is
attached at $x=L$. In particular,
$\cT=-\partial_E Z(L;E)$, as in the main text.
All formulas in this subsection refer to real energy and a
lossless sample. The potential and the Dirichlet boundary condition
are independent of energy. We use units $\hbar=2m=1$.

With $F(x;E)=\Psi(x;E)\Psi(L;E)^{-1}$, the scattering state is
\[
\psi(x)=F(x;E)(I+r)a_{\rm in}.
\]
The integrated probability in the sample is therefore
\[
\int_0^L\psi^\dagger\psi\,\dd x
=a_{\rm in}^\dagger(I+r)^\dagger\cT(I+r)a_{\rm in}.
\]
Since the magnitude of the incident flux is
$2q\,a_{\rm in}^\dagger a_{\rm in}$, the dwell-time matrix is
\be
Q_{\rm dwell}=\frac1{2q}(I+r)^\dagger\cT(I+r).
\label{eq:app-dwell}
\ee
It is positive and converts the internal norm for a prescribed
boundary amplitude into a time referred to incident-flux normalization.

To verify the energy-derivative identity used here, differentiate
$-F''+WF=EF$ at real energy. The Wronskian relation becomes
\[
\partial_x\left[F^\dagger\partial_EF'
-F'^\dagger\partial_EF\right]=-F^\dagger F.
\]
At $x=0$ the bracket vanishes because $F(0;E)=0$ for all $E$.
At $x=L$, $F(L;E)=I$ and $\partial_EF(L;E)=0$, so the bracket
is $\partial_E Z(L;E)$. Hence
\[
-\partial_E Z(L;E)=\int_0^L F(x;E)^\dagger F(x;E)\,\dd x=\cT.
\]
No change of boundary orientation is involved.

The Wigner--Smith matrix differs from the dwell-time matrix
when the lead wavenumber depends on energy. In the present convention,
boundary matching gives
\[
Z(I+r)=\ii q(r-I),\qquad
r=-(Z-\ii qI)^{-1}(Z+\ii qI).
\]
At real energy $Z=Z^\dagger$, so $r^\dagger r=I$.
Using $\partial_E Z=-\cT$, differentiation gives
\[
(Z-\ii qI)\partial_Er
=\ii(\partial_Eq)(r-I)+\cT(I+r).
\]
The Cayley representation implies
\[
r^\dagger(Z-\ii qI)^{-1}
=-\frac{(I+r)^\dagger}{2\ii q}.
\]
Consequently, without assuming that $Z$ commutes with $\cT$,
\begin{align*}
Q_{\rm WS}:=-\ii r^\dagger\partial_Er
&=\frac1{2q}(I+r)^\dagger\cT(I+r)
+\frac{\ii\,\partial_Eq}{2q}(I+r)^\dagger(r-I)\\
&=Q_{\rm dwell}
+\frac{\ii\,\partial_Eq}{2q}(r-r^\dagger),
\end{align*}
where the last equality uses unitarity. Thus
\be
Q_{\rm WS}=Q_{\rm dwell}
+\frac{\ii\,\partial_Eq}{2q}(r-r^\dagger).
\label{eq:app-wigner-smith}
\ee
The additional term is the self-interference contribution associated
with the incoming and reflected waves in the lead and need not be
positive~\cite{HaugeStovneng1989,Winful2003}.
For a free lead, $E=q^2$ and $\partial_Eq=1/(2q)$, giving
\[
Q_{\rm WS}-Q_{\rm dwell}
=\frac{\ii}{4q^2}(r-r^\dagger).
\]
In the scalar case $r=e^{\ii\phi(E)}$, this reads
\[
\partial_E\phi=Q_{\rm dwell}-\frac{\sin\phi}{2q^2}.
\]
If the lead wavenumber is held fixed when differentiating with respect
to the sample energy, the correction vanishes. The zero-absorption
limit of the absorption-deficit matrix then agrees directly with the
Wigner--Smith matrix. With a dispersive physical lead, the contact
term above must be retained.

\subsection{Waveguide geometry and the high-energy reflection limit}
\label{app:admittance-highenergy}

The connection between random waveguide scattering
and hyperbolic geometry goes back to the classical
paper of Gertsenshtein and
Vasil'ev~\cite{GertsenshteinVasiliev1959},
an early treatment of one-dimensional wave localization.
A multichannel extension of such framework to Siegel's circle was
studied later in~\cite{ZakharItkin}, with closely related recent work~\cite{MenonYu2023}
constructing a Dyson-type stochastic evolution
on the Siegel half-space. Here it is also appropriate to mention the extremely successful 
characterization of {\it transport}  properties (transmission, conductance and other scattering characteristics) of $N$-channel system in the framework of the so-called DMPK \cite{Dorokhov1982,MelloPereyraKumar1988} approach, with many exact results obtained over the last three decades, see e.g. \cite{BeenakkerRejaei1994,Caselle1995, BruceChalker1996,BeenakkerPaasschensBrouwer1996,Ossipov2018a}.

\paragraph*{Microscopic versus effective diffusion.}
The invariant volume element also appears in the classical
description of random waveguide scattering by
Gertsenshtein and Vasil'ev~\cite{GertsenshteinVasiliev1959},
whose evolution is generated by the hyperbolic
Laplacian, with absorption represented by an
additional drift.
The microscopic Riccati dynamics~\eqref{eq:uv}
is not itself isotropic hyperbolic diffusion:
the noise acts only on $u$, while the equation
for $v$ contains no Brownian increment.
This distinction does not preclude an isotropic
effective description after rapid-phase averaging
in a suitable high-energy limit.
Indeed, the high-energy scaling of the stationary
distribution suggests a connection with the classical
reflection statistics of absorbing waveguides
\cite{BruceChalker1996,BeenakkerPaasschensBrouwer1996}.
The scaling limit below makes this connection explicit; its
detailed scattering interpretation will be developed in
Paper~II~\cite{PaperII}.

\paragraph*{High-energy stationary limit.}
Consider $E\to\infty$ with $\eta\sqrt E/T>0$ fixed.
The stationary law~\eqref{eq:lawZ} suggests an isotropic
limiting description in reflection coordinates.
Indeed, setting $E=k^2$, $H=A/k$, $Q=P/k$, and
$c=\eta k/T$, its exponent has, as $k\to\infty$ with $c>0$ fixed, the limiting form
\[
 -c\mbox{Tr}\bigl[(H^2+I)Q^{-1}+Q\bigr].
\]
The corresponding candidate limiting distribution is therefore
\[
 \mathrm d\mathbb P_\infty(H,Q)\propto
 \exp\!\left\{-c\mbox{Tr}
 [(H^2+I)Q^{-1}+Q]\right\}\mathrm d\mu(H,Q),
\]
where $\mathrm d\mu$ is the invariant volume
of the matrix half-space.
Under the Cayley transformation to reflection
coordinates, this weight becomes radial and leads
formally to
\[
 p_\infty(\lambda)\propto
 |\Delta(\lambda)|^\beta
 \exp\!\left[-4c\sum_{j=1}^N\lambda_j\right],
 \qquad
 \lambda_j=\frac{R_j}{1-R_j}\geq0,
\]
where $R_j$ are the reflection eigenvalues. This is the Laguerre form of the stationary
reflection-eigenvalue distribution previously obtained for semi-infinite absorbing waveguides in the isotropic scattering description
\cite{BruceChalker1996,BeenakkerPaasschensBrouwer1996}, with $N=1$ case considered earlier in \cite{FreilikherPustilnikYurkevich1994}.
The high-energy limit of our microscopic admittance distribution thus provides a direct connection with this
classical result. The matching of scattering and absorption conventions, and a detailed treatment
of the limiting reflection statistics, will be
given in Paper~II~\cite{PaperII}.

\subsection{Scattering statistics and the programme of Paper II}
\label{app:admittance-paperII}

Unlike the complex contour weight, which reproduces holomorphic
averages, the full stationary admittance law can be used to evaluate
non-holomorphic functions of $Z$ and $Z^\dagger$. With the lead
convention specified above, these include reflection eigenvalues,
absorbed-flux observables, and resonance-counting functions such as
$\log|\det r|^2$. Time delays additionally require the real-energy
limit and the distinction between the internal norm and the physical
Wigner--Smith matrix established above. Related scalar scattering
problems were studied long ago in \cite{FreilikherPustilnikYurkevich1994}
and more recently in Refs.~\cite{FyodorovMeibohm2026,FyodorovRes2026}, and a
$\sigma$-model treatment of reflection resonances is given in
Ref.~\cite{FyodorovSkvortsovTikhonov2024}.

The following programme concerns Paper~II~\cite{PaperII}; the
scattering-statistics calculations listed here are not derived in
Paper~I. Paper~II will use the full stationary law for a semi-infinite
absorbing wire attached to an ideal lead, in all three Dyson classes.
\begin{itemize}
\item {\it Admittance and absorbing DMPK.}
At high energy, the stationary admittance distribution
reduces to the Laguerre-$\beta$ ensemble of absorbing
DMPK theory (see Appendix~\ref{app:admittance-highenergy}). The exact distribution also determines
the corrections to this limit: a parity symmetry
cancels the first-order contribution to the
DMPK marginal, so its leading correction is of
order $(k\ell)^{-2}$, where $k$ is the wave number
and $\ell$ the mean free path.
\item {\it Reflection resonances.} Through a Poincar\'e--Lelong counting identity, the distribution gives
the density of resonances for all three classes. This includes Ward identities that make narrow resonances
$\beta$-independent, closed-form $\sigma$-model-type hard-edge densities for $\beta=1,4$, and the
broad-resonance tail.
\item {\it Time delays.} The zero-absorption law gives the joint
statistics of the boundary admittance and its energy derivative,
which will be used to determine the reflection and Wigner--Smith
time-delay statistics (incorporating the dispersive-lead correction
in Eq.~\eqref{eq:app-wigner-smith}).
\item {\it The band edge.} In the window $E\sim T^{2/3}$, where there is no DMPK regime, the
narrow-resonance coefficient is a density of states. For $N=1$ it is exactly Halperin's $\nu_1$; for
general $N$ it is conjecturally the $\nu_N$ of Eq.~\eqref{eq:main}.
\item {\it Experiment.} A concrete experimental realization using
disordered, lossy $LC$ ladders—where ohmic dissipation plays the role of
the uniform absorption $\eta$—will be developed in Paper~II. Measuring
the complex input impedance as frequency and loss are varied provides
direct access to the disorder statistics of the boundary admittance $Z$,
allowing quantitative comparison with the stationary law~\eqref{eq:law}
and its predicted crossover with absorption.
\end{itemize}

The finite-length problem in the unitary class, including the full
finite-$N$ solution of the absorbing DMPK equation, will be treated
separately~\cite{ReflB2}.

\section{Intersection numbers and the outgoing contour}
\label{app:intersection}
\label{sec:intersection}

For $\beta=2$, the formal expansion of the normalized Kontsevich
integral generates intersection numbers on moduli spaces of
curves~\cite{Witten1991,Kontsevich1992}. These numbers are geometric
coefficients obtained by integrating products of the classes $\psi_i$;
$\psi_i$ is the first Chern class of the cotangent line at the $i$-th
marked point. They are also the correlation functions of
two-dimensional topological gravity~\cite{Witten1991}.

Our wire problem supplies an outgoing contour prescription for this
integral. Admissible contours with the same dominant-saddle contribution
can share a formal perturbative expansion while differing by
exponentially small terms. The outgoing prescription fixes these
additional saddle contributions and hence selects a nonperturbative
completion of the formal series. The connection developed below is
that the tunnelling contribution responsible for the Lifshitz tail
also controls the large-order growth of the perturbative coefficients;
these coefficients, in turn, count gluings of triangles.

\paragraph*{Perturbative content.}
For $\beta=2$, shifting to a saddle and removing the saddle exponential
and Gaussian normalization identifies the perturbative expansion with
the scalar-external-field specialization of the Witten--Kontsevich
model. Its cubic vertices generate trivalent ribbon graphs with a
factor $N$ per face~\cite{BIZ1980,LandoZvonkin2004}. The coefficients
therefore give weighted sums of intersection numbers, rather than
resolving the individual numbers encoded by the full generating
function with a general external matrix.

For $\beta=1,4$, the analogous expansion also contains non-orientable
graphs~\cite{GouldenJackson1997,MulaseWaldron2003}. We do not establish
here an intersection-number interpretation of the non-orientable
coefficients of these outgoing integrals. Their explicit map
expansions provide benchmarks for possible comparisons with refined
($\mathfrak b$-deformed) topological
recursion~\cite{CDO2024,GFGO2025}.

\subsection{Large-order growth and random surface gluings}
\label{sec:largeordergluing}

Here large order means $k\to\infty$ at fixed $N$. Each coefficient
sums over maps of different topologies; it is not a fixed-genus
coefficient. For a connected orientable trivalent map with $2k$
vertices and $F$ faces, Euler's relation gives $k=2g-2+F$.
The partition function includes disconnected maps, whereas its
logarithm selects connected ones.

\paragraph*{The formal series.} Consider the Lifshitz tail regime $s\to+\infty$ of
Sec.~\ref{sec:Lifshitz}, and put all $N$ integration variables near the
dominant saddle point, $w_j=-\sqrt s+u_j$. Since
$\varphi(-\sqrt s+u)=\varphi(-\sqrt s)-\sqrt s\,u^2+u^3/3$ exactly, the
perturbative part of $\tau_N^{(\beta)}$, divided by its leading term, is
a formal average over a Gaussian $\beta$-ensemble. The exponential
is expanded before taking Gaussian moments; the unexpanded real
Gaussian integral with a cubic exponential is not convergent:
\be
P_N(\zeta)=\Big\langle e^{p_3/3}\Big\rangle_{\mathrm G,\beta}
=\sum_{k\ge0}a_k(N)\,\zeta^{-k},\qquad
\zeta=\tfrac23s^{3/2},\qquad
a_k(N)=\frac{\aver{p_3^{2k}}_{\mathrm G,\beta}\big|_{\sigma=1}}{9^k\,(2k)!\,12^k}.
\label{eq:Pseries}
\ee
Here $p_3=\sum_ju_j^3$, and the Gaussian variance parameter
is $\sigma=1/(2\sqrt s)$, so that $\sigma^{3k}=(12\zeta)^{-k}$.
The even-moment formula makes $a_k$ positive for all three symmetry
classes, including $\beta=4$, whose graph representation has signed
weights. The obstruction to Borel summation on the positive $s$-axis
is the positive-axis singularity associated with the tunnelling
saddle described next.

\paragraph*{Large-order behaviour.} At fixed $N$, $\tau_N^{(\beta)}$ is a
finite-dimensional exponential integral, and the large-order behaviour
of its expansion is controlled by the other saddle points. Their
relative actions, in units of $\zeta$, are $2\ell$ for the configuration
with $\ell$ eigenvalues moved to $w_+$. The nearest, $\ell=1$, is the
one-eigenvalue tunnelling configuration of Sec.~\ref{sec:Lifshitz}.
Because the series is not summable directly on the positive $s$-axis,
it can be resummed by approaching this axis from either side. The two
results differ by $\ii\,P_N\,\tau_N^{(1)}/\tau_N^{(0)}$, the full
one-eigenvalue contribution; on the axis itself it enters only with the
factor $\frac12$ found in Sec.~\ref{sec:Lifshitz}. The standard
dispersion argument~\cite{Lipatov1977,BerryHowls1991,MarinoSchiappaWeiss2008}
then gives, with $b=\frac{\beta(N-1)}2$, 
\be
a_k(N)\simeq\frac1{2\pi}\,\frac{N\,\Gamma(1+\beta/2)}{\Gamma(1+N\beta/2)}\,6^{b}\,2^{-k}
\Big[\Gamma(k+b)+\gamma^{(\beta)}_N\,\Gamma(k+b-1)+O\big(\Gamma(k+b-2)\big)\Big].
\label{eq:largegenus}
\ee
The coefficient of the first correction is
\be
\gamma^{(\beta)}_N=\frac43\,\kappa_N+\frac{K_\beta(N)}{108}
=\frac{(N-2)\big(4\beta^2N^2-43\beta^2N+18\beta N+39\beta^2-18\beta+20\big)}{144}.
\label{eq:gammaN}
\ee
It combines two contributions:
\begin{itemize}
\item the relative correction $\kappa_N$ to the instanton
ratio $\tau_N^{(1)}/\tau_N^{(0)}$ (Sec.~\ref{sec:Lifshitz});
\item the first perturbative coefficient
$a_1(N)=K_\beta(N)/216$, obtained from Eq.~\eqref{eq:genus}
at the saddle $w_-$.
\end{itemize}
It is $\kappa_N$, not $c^{(\beta)}_N$, that enters here:
$c^{(\beta)}_N$ contains in addition the term $-3\beta(N-1)/8$,
which comes from differentiating $\nu_N$.

This dispersion argument assumes that the nearest Borel singularity
is the one-eigenvalue tunnelling configuration, with the Stokes
constant $\ii$ whose half, $\ii/2$, appears in
Sec.~\ref{sec:Lifshitz}. The resulting large-order asymptotics is
therefore stated at the same physics level as that saddle analysis.
For $N=1$, Eq.~\eqref{eq:largegenus} gives
\[
a_k\simeq\frac{2^{-k}}{2\pi}
\left[\Gamma(k)-\frac5{36}\Gamma(k-1)\right],
\]
in agreement with the known Airy coefficients, since
$\frac5{36}=2u_1$ with $u_1=\frac5{72}$.
For $N=2$, Eq.~\eqref{eq:gammaN} gives
$\gamma^{(\beta)}_2=0$: the first subleading Gamma term vanishes.

The Lifshitz tail thus determines the large-order growth of the
generating functions of $\beta$-weighted trivalent maps:
\begin{itemize}
\item The large-order growth is controlled by a single-eigenvalue tunnelling saddle. The only remaining freedom is the direction in
channel space along which it does so, a unit vector up to a unit of $\F$. These directions form the
projective space $\mathbb{RP}^{N-1}$, $\mathbb{CP}^{N-1}$ or $\mathbb{HP}^{N-1}$, of real dimension
$\beta(N-1)$.
\item The number of colours $N$ enters the leading large-order constant {\it only} through the dimension and
the volume of this space of directions. In the Fubini--Study normalization of Sec.~\ref{sec:Lifshitz}, the constant in~\eqref{eq:largegenus} relative to $N=1$ is $\mathrm{Vol}(\F P^{N-1})(6/\pi)^{\beta(N-1)/2}$.
\end{itemize}
\paragraph*{What is counted.} Dually, a trivalent map with $2k$ vertices is a surface glued from
$n=2k$ triangles. Pair the $3n$ sides of the triangles at random and glue each pair.
\begin{itemize}
\item For $\beta=2$ every pair is glued so that the orientations
match, and the surface is orientable.
\item For $\beta=1$ each pair is glued straight or with a flip,
with probability $\frac12$ each, so non-orientable surfaces
(projective planes, Klein bottles, \dots) occur. The equal
weights follow from the real-symmetric covariance of
Sec.~\ref{sec:model}.
\end{itemize}
Let $F$ be the number of vertices of the glued
surface, i.e.\ of faces of the trivalent map; its Euler characteristic is $F-k$, so, for a connected orientable surface at fixed $k$, fewer vertices
means higher genus. Each vertex carries a matrix index, so the weight $N^F$ is the number of ways to
colour the vertices with $N$ colours. By Wick's theorem, exactly, for $\beta=1,2$,
\be
a_k(N)/a_k(1)=\mathbb E\big[N^F\big]
\label{eq:gluing}
\ee
over random gluings of $2k$ triangles (possibly disconnected). 

For $\beta=4$ the same twisted gluings enter with signed weights,
through the $\beta=1\leftrightarrow4$ duality~\cite{MulaseWaldron2003}:
\be
a^{(\beta=4)}_k(N)/a_k(1)=(-2)^k\,\mathbb E\big[(-2N)^F\big]_{\beta=1\ \rm gluings}.
\label{eq:gluing4}
\ee
Two facts fix this form:
\begin{itemize}
\item the duality acts on a graph with $2k$ vertices and $3k$
edges by rescalings, which produce a factor $c^k$;
\item $c=-2$ follows from the case $k=1$, since
$K_4(N)=-2K_1(-2N)$ in Eq.~\eqref{eq:Kbeta}.
\end{itemize}
\subsection{The Ewens comparison and twisted-gluing conjecture}
\label{sec:ewensgluing}

 With $\theta=\beta/2$ and $n=2k$, Eqs.~\eqref{eq:largegenus}
and~\eqref{eq:gluing} give, for $\beta=1,2$ and every integer $N\ge1$,
\be
\mathbb E\big[N^F\big]=\frac{\Gamma(\theta)}{\Gamma(N\theta)}\,6^{\theta(N-1)}\,
\frac{\Gamma\big(k+\theta(N-1)\big)}{\Gamma(k)}
\Big[1+\frac{\gamma^{(\beta)}_N+\frac5{36}}{k}+O(k^{-2})\Big]
\simeq\frac{\Gamma(\theta)}{\Gamma(N\theta)}\,(3n)^{\theta(N-1)} .
\label{eq:Ewens}
\ee

The leading form coincides with the large-$n$ asymptotics of the
generating function, at
$z=N$, of the number of cycles of a random permutation of $3n$
elements under the Ewens distribution with parameter
$\theta$~\cite{Ewens1972}. That generating function is
$\prod_{j=0}^{3n-1}(z\theta+j)/(\theta+j)$. The constant $\Gamma(\theta)/\Gamma(N\theta)$ is the Ewens
normalization; in the instanton language it is $\mathrm{Vol}(\F P^{N-1})/\pi^{\beta(N-1)/2}$.

The change from $\theta=1$ for orientable gluings to
$\theta=\tfrac12$ for gluings with random flips has a combinatorial
interpretation. For orientable gluings, the orientation induces a
permutation of the $3n$ sides whose cycles correspond to vertices.
For gluings with random flips, a useful description instead uses two
perfect matchings of the $6n$ side-endpoints: the fixed matching
$\alpha$ that joins the endpoints meeting at each triangle corner,
and the matching $\mu_{\rm glue}$ induced by the gluing. Each vertex
of the glued surface is one loop in their union.

The matching $\mu_{\rm glue}$ is constrained by
the pairing of whole sides. If it were uniformly
random among all $(6n-1)!!$ matchings, the resulting
loops would form a random cyclation in the sense
of Pippenger~\cite{Pippenger2013}. Their number
would then have the exact generating function
\[
\prod_{j=0}^{3n-1}\frac{z+2j}{1+2j},
\]
equal to that of the cycle count of an Ewens
permutation with parameter $\tfrac12$.

This comparison motivates the conjecture
below. Equation~\eqref{eq:Ewens} itself concerns fixed integer
$z=N\ge1$; extending its asymptotics to non-integer $z$ requires
additional justification.

For $\beta=4$, Eq.~\eqref{eq:Ewens} with $\theta=2$
holds for the signed sums~\eqref{eq:gluing4},
whose interpretation as a probability measure
on gluings is not evident.

Equation~\eqref{eq:Ewens} also predicts the $1/n$ corrections.
For $N=2$ it gives
\begin{itemize}
\item $\mathbb E[2^F]=6k+\frac56+O(k^{-1})$ for orientable gluings,
against $3n+1=6k+1$ for a uniform permutation;
\item $\mathbb E[2^F]=\sqrt{6\pi k}\,\big[1+\frac1{72k}+O(k^{-2})\big]$
for twisted gluings, against $\sqrt{6\pi k}\,\big[1+\frac1{48k}+O(k^{-2})\big]$
for two uniform matchings.
\end{itemize}
Gluings therefore agree with the uniform models only to
leading order.
\paragraph*{Relation to known results.} For $\beta=2$, the distribution of the vertex count has been studied rigorously. Pippenger and Schleich
conjectured $\mathbb E F=\log(3n)+\gamma$ and $\mathrm{Var}\,F=\log(3n)+\gamma-\pi^2/6$ from simulations, and
proved $\mathbb EF=\log n+O(1)$~\cite{PippengerSchleich2006}. Their conjectures were confirmed in
certain cases in Ref.~\cite{FlemingPippenger2010}. Chmutov and Pittel encoded the glued surface by a
permutation of the sides, whose cycles are the vertices, and showed that it is asymptotically uniform
on the alternating group or on its complement, depending on parity~\cite{ChmutovPittel2016}. They
deduced a local central limit theorem for the number of vertices and a Gaussian genus with variance
$\frac14\log(3n)$ for triangles.

For a uniform permutation of $3n$ elements, conditioning on parity
does not change the moment $\mathbb E[N^F]$ at an integer
$1\le N<3n$: the signed generating polynomial
$\sum_\pi\mathrm{sgn}(\pi)z^{K(\pi)}
=z(z-1)\cdots(z-3n+1)$ vanishes there. The uniform model therefore gives
\[
\frac{\Gamma(3n+N)}{\Gamma(N)\Gamma(3n+1)}
\simeq\frac{(3n)^{N-1}}{\Gamma(N)},
\]
precisely the leading expression predicted by
Eq.~\eqref{eq:Ewens} at $\theta=1$. This agreement is consistent
with asymptotic uniformity, but is not by itself an independent
proof of the instanton prefactor: transferring exponential moments
from the uniform model to the gluing model requires additional tail
control, beyond convergence in total variation.

For $\beta=1$ we have not found the corresponding statement in the
literature, and we formulate it as a conjecture.

{\bf  Conjecture (twisted gluings).} {\it Let $n$ be even. Glue $n$ triangles by a uniformly random pairing of their $3n$
sides, each pair glued straight or with a flip with probability $\frac12$, and let $F_n$ be the number
of vertices of the resulting surface. Then
\[
\mathbb E\big[z^{F_n}\big]=\frac{\Gamma(1/2)}{\Gamma(z/2)}\,(3n)^{(z-1)/2}\,\big(1+o(1)\big),
\]
uniformly for $z$ in a complex neighbourhood of $z=1$, and
pointwise for $z>0$. Thus the conjecture predicts the same leading
probability-generating-function asymptotics as the cycle count of an
Ewens($\frac12$) random permutation of $3n$ elements.}

For integer $z=N\ge1$ the leading asymptotics, together with its $1/n$
correction~\eqref{eq:Ewens}, follows from the spectral calculation at
the physics level; the conjectural part is the extension to non-integer
$z$. If the asymptotic form holds uniformly for $z$ near $1$, a standard
result on generating functions of this type~\cite{Hwang1998} yields
$\mathbb EF_n=\frac12\log(3n)+\frac12(\gamma+2\log2)+o(1)$ and
$\mathrm{Var}\,F_n=\frac12\log(3n)+\frac12(\gamma+2\log2)-\frac{\pi^2}8+o(1)$,
together with a central limit theorem for $F_n$. These are exactly the
values for a uniformly random second matching, for which the mean and
variance are $\sum_{j=1}^{3n}\frac1{2j-1}$ and
$\sum_{j=1}^{3n}\big[\frac1{2j-1}-\frac1{(2j-1)^2}\big]$.

The conjecture asserts that the side-pairing constraint on
$\mu_{\rm glue}$ becomes irrelevant for the leading loop-count
asymptotics. It is an analogue, at the level of these generating
functions, of the orientable near-uniformity results. The constraint
remains visible in the $1/n$ corrections displayed above. There is
no corresponding orientability-imposed parity constraint on the
vertex count in the twisted model.

This is a concrete prediction for randomly twisted gluings.
A direct combinatorial derivation, possibly using the methods of
Ref.~\cite{ChmutovPittel2016}, would test the spectral large-order
argument independently. Comparisons with refined recursion should
first identify the same map weights and specialization. Likewise,
comparison with large-genus intersection-number
asymptotics~\cite{EGGGL2023} requires matching the sum over marked
points and the scalar external-field specialization.

\subsection{Non-degenerate thresholds and the external matrix}
\label{sec:thresholdsKontsevich}
When each transverse mode $i$ has its own threshold $\epsilon_i$, the
scalar energy $E$ is replaced by the diagonal matrix
$\mathsf E=EI-\mathrm{diag}(\epsilon_1,\dots,\epsilon_N)$, which no
longer commutes with $Z$. The local identities in Secs.~\ref{sec:mechanism}
and~\ref{sec:kontsevich} extend directly: the equation for $P$ is
independent of $\mathsf E$, the drift identities use only
$\nabla_X\Tr(\mathsf EX)=\mathsf E$, and the factorization enters only
through $\Tr(\mathsf EY)$. Continuing the outgoing integration-surface
argument from equal thresholds at fixed $\eta>0$ gives the stationary
distribution~\eqref{eq:law} with $\Tr U_{\mathsf E}(X)=\Tr(\mathsf EX)+\frac13\Tr X^3$
and $\cZ_N\propto|\cJ_N(\mathsf E+\ii\eta I)|^2$.

For $\beta=2$, $\cJ_N$ reduces to Kontsevich's determinant
$\cJ_N\propto\det[\cA_-^{(j-1)}(s_i)]/\Delta(\bm s)$ with
$s_i=-(E-\epsilon_i+\ii\eta)/T^{2/3}$, and shifting
$W=-\Lambda+U$ with $\Lambda=\mathrm{diag}(\sqrt{s_1},\dots,\sqrt{s_N})$
recovers the external-field Kontsevich model with Miwa times
proportional to $\Tr\Lambda^{-(2j+1)}$. What is lost compared with equal
thresholds is everything relying on rotational invariance, which
$\Tr(\mathsf EY)$ breaks: the eigenvalue form~\eqref{eq:Yens} and the
$\beta=1,4$ reductions of Secs.~\ref{sec:explicit}
and~\ref{sec:beta4edge}. Physically, differentiating the integrated
density of states with respect to $\epsilon_i$ gives the
channel-resolved density of states $\rho_i=-\partial_{\epsilon_i}\nu_N$
by the Hellmann--Feynman theorem. For $\beta=1,4$, the corresponding
external-field integrals generate non-orientable
graphs~\cite{MulaseWaldron2003,BrezinHikami2008}, now equipped with the
outgoing contour prescription and a positive stationary normalization; we
leave any detailed comparison with unorientable gravity
models~\cite{StanfordWitten2019,WeberEtAl2024,DiUbaldoEtAl2025} open.


\begin{thebibliography}{99}

\bibitem{AllezDumaz2015} R.~Allez and L.~Dumaz, Random matrices in non-confining potentials, {\it J. Stat. Phys.} \textbf{160}, 681--714 (2015).

\bibitem{BBDY2024} A.~Barhoumi, P.~Bleher, A.~Dea\~no, and M.~Yattselev, ``On Airy solutions of P$_{\rm II}$ and the complex cubic ensemble of random matrices, II,'' arXiv:2403.03023.

\bibitem{BaurKratz1989}
G.~Baur and W.~Kratz,
``A general oscillation theorem for self-adjoint
differential systems with applications to
Sturm--Liouville eigenvalue problems and
quadratic functionals,''
{\it Rend.\ Circ.\ Mat.\ Palermo} (2) \textbf{38},
329--370 (1989).

\bibitem{BeenakkerRejaei1994} C.~W.~J. Beenakker and B.~Rejaei, Exact solution for the distribution of transmission eigenvalues in a disordered wire and comparison with random-matrix theory, {\it Phys. Rev. B} \textbf{49}, 7499--7510 (1994).

\bibitem{BeenakkerPaasschensBrouwer1996}
C.~W.~J.~Beenakker, J.~C.~J.~Paasschens and P.~W.~Brouwer,
``Probability of Reflection by a Random Laser,''
{\it Phys. Rev. Lett.} \textbf{76}, 1368--1371 (1996).

%

\bibitem{BerryHowls1991}
M.~V.~Berry and C.~J.~Howls, ``Hyperasymptotics for integrals with saddles,''
{\it Proc.\ R.\ Soc.\ Lond.\ A} \textbf{434}, issue 1892 (1991), doi:10.1098/rspa.1991.0119.

\bibitem{BIZ1980} D.~Bessis, C.~Itzykson, and J.-B.~Zuber, ``Quantum field theory techniques in graphical enumeration,'' Adv. Appl. Math. \textbf{1}, 109 (1980).

\bibitem{BleherDeano2013}
P.~M.~Bleher and A.~Dea\~no,
``Topological expansion in the cubic random matrix model,''
Int.\ Math.\ Res.\ Not.\ IMRN \textbf{2013}, 2699--2755 (2013).

\bibitem{BleherDeano2016} P.~M.~Bleher and A.~Dea\~no, ``Painlev\'e I double
scaling limit in the cubic random matrix model,'' Random Matrices Theory
Appl. \textbf{5}, 1650004 (2016).

\bibitem{BruceChalker1996}
N.~A.~Bruce and J.~T.~Chalker,
``Multiple scattering in the presence of absorption: a theoretical treatment for quasi one-dimensional systems,''
{\it J. Phys. A: Math. Gen.} \textbf{29}, 3761--3768 (1996).

\bibitem{Caselle1995} M.~Caselle, On the distribution of transmission eigenvalues in disordered wires, {\it Phys. Rev. Lett.} \textbf{74}, 2776--2779 (1995).

\bibitem{ChenAnlageFyodorov2021}
L.~Chen, S.~M.~Anlage, and Y.~V.~Fyodorov,
``Generalization of Wigner time delay to sub-unitary scattering systems,''
{\it Phys.\ Rev.\ E} \textbf{103}, L050203 (2021).

\bibitem{CDO2024}
N.~K.~Chidambaram, M.~Do{\l}{\fontencoding{T1}\selectfont\k{e}}ga,
and K.~Osuga, ``$\mathfrak b$-Hurwitz numbers from refined topological recursion,''
{\it Math.\ Ann.} \textbf{394}, 103 (2026)

\bibitem{ChmutovPittel2016} S.~Chmutov and B.~Pittel, ``On a surface formed by randomly gluing together polygonal discs,'' {\it Adv. Appl. Math.} \textbf{73}, 23 (2016).

\bibitem{ComtetTourigny2017}
A.~Comtet and Y.~Tourigny,``Impurity models and products of random matrices,''
 in {\it Stochastic Processes and Random Matrices: Lecture Notes of the Les Houches Summer School, Vol.~104, July 2015},
edited by G.~Schehr, A.~Altland, Y.~V.~Fyodorov, N.~O'Connell, and L.~F.~Cugliandolo
(Oxford University Press, Oxford, 2017), Chap.~11, pp.~474--547; see Sec.~11.6.2.
doi:10.1093/oso/9780198797319.003.0011.

\bibitem{CostinHuangTanveer2014}
O.~Costin, M.~Huang, and S.~Tanveer, ``Proof of the Dubrovin conjecture and analysis of the
tritronqu\'ee solutions of $P_I$,'' {\it Duke Math.\ J.} \textbf{163}, 665--704 (2014),
doi:10.1215/00127094-2429589.

\bibitem{DavyShiParkTianGenack2015}
M.~Davy, Z.~Shi, J.~Park, C.~Tian, and A.~Z.~Genack,
Universal structure of transmission eigenchannels
inside opaque media,
{\it Nat.~Commun.} {\bf 6}, 6893 (2015).

\bibitem{DavyShiWangChengGenack2015}
M.~Davy, Z.~Shi, J.~Wang, X.~Cheng, and A.~Z.~Genack,
Transmission eigenchannels and the densities of states
of random media,
{\it Phys.~Rev.~Lett.} {\bf 114}, 033901 (2015).

\bibitem{deBruijn1955} N.~G. de Bruijn, ``On some multiple integrals involving determinants,'' {\it J. Indian Math. Soc.} (N.S.) \textbf{19}, 133 (1955).

\bibitem{DiUbaldoEtAl2025}
G.~Di Ubaldo, A.~Etkin, F.~M.~Haehl, and M.~Rozali,
``Mind the crosscap: $\tau$-scaling in non-orientable gravity and
time-reversal-invariant systems,''
arXiv:2509.20448 (2025).

\bibitem{Dorokhov1982} 
O.~N. Dorokhov, Transmission coefficient and the localization length of an electron in $N$ bound disordered chains,
{\it  JETP Lett.} \textbf{36}, 318--321 (1982).

\bibitem{GoldsheidMolchanovPastur1977}
I.~Ya.~Gol'dsheid, S.~A.~Molchanov, and L.~A.~Pastur,
A pure point spectrum of the stochastic one-dimensional Schr\"odinger operator,
{\it Funct. Anal. Appl.} \textbf{11}, 1--8 (1977).

\bibitem{DumazLabbe2020Localization} L.~Dumaz and C.~Labb\'e, Localization of the continuous Anderson Hamiltonian in one dimension, {\it Probab. Theory Relat. Fields} \textbf{176}, 353--419 (2020).

\bibitem{DumazLabbe2023Delocalized} L.~Dumaz and C.~Labb\'e, The delocalized phase of the Anderson Hamiltonian in one dimension, {\it Ann. Probab.} \textbf{51}, 805--839 (2023).

\bibitem{DumazLabbe2024Crossover} L.~Dumaz and C.~Labb\'e, Localization crossover for the continuous Anderson Hamiltonian in 1-d, {\it Invent. Math.} \textbf{235}, 345--440 (2024).

\bibitem{DumazLabbe2024WhiteNoise} L.~Dumaz and C.~Labb\'e, Anderson localization for the 1-d Schr\"odinger operator with white noise potential, J. Funct. Anal. \textbf{286}, 110191 (2024).

\bibitem{Dyson1962} F.~J.~Dyson,
``A Brownian-Motion Model for the Eigenvalues of a Random Matrix,''
{\it J.\ Math.\ Phys.} \textbf{3}, 1191--1198 (1962).

\bibitem{Ewens1972} W.~J. Ewens, ``The sampling theory of selectively neutral alleles,'' {\it Theor. Popul. Biol.} \textbf{3}, 87 (1972).

\bibitem{EGGGL2023}
B.~Eynard, E.~Garcia-Failde, A.~Giacchetto, P.~Gregori, and D.~Lewa\'nski,
``Resurgent large genus asymptotics of intersection numbers,'' {\it Forum Math.\ Pi} \textbf{14} (2026),
doi:10.1017/fmp.2026.10037.

\bibitem{BrezinHikami2008}
E.~Br\'ezin and S.~Hikami,
``Intersection numbers from the antisymmetric Gaussian matrix model,''
arXiv:0804.4531 (2008).

\bibitem{FlemingPippenger2010} K.~Fleming and N.~Pippenger, ``Large deviations and moments for the Euler characteristic of a random surface,'' 
{\it Random Struct. Algorithms} \textbf{37}, 465 (2010).

\bibitem{Forman1987}
R.~Forman,
``Functional determinants and geometry,''
{\it Invent.\ Math.} \textbf{88}, 447--493 (1987).

\bibitem{Forrester2010} P.~J. Forrester, {\it Log-Gases and Random Matrices}, London Mathematical Society Monographs \textbf{34} (Princeton University Press, 2010).

\bibitem{FreilikherPustilnikYurkevich1994}
V.~Freilikher, M.~Pustilnik, and I.~Yurkevich,
Effect of absorption on the wave transport in the strong localization regime,
\emph{Phys. Rev. Lett.} \textbf{73}, 810--813 (1994).

\bibitem{FrischLLoyd1960} H.~L. Frisch and S. P. Lloyd, Electron levels in a one-dimensional random lattice, {\it Phys. Rev.} {\bf 120} (1960),  1175.

\bibitem{FyodorovMirlin1991} Y.~V. Fyodorov and A.~D. Mirlin, Scaling properties of localization in random band matrices: a $\sigma$-model approach,
{\it  Phys. Rev. Lett.} \textbf{67}, 2405--2409 (1991).

\bibitem{FyodorovMirlin1994}
Y.~V.~Fyodorov and A.~D.~Mirlin,
``Statistical properties of eigenfunctions of random quasi-1D
one-particle Hamiltonians,''
{\it Int.\ J.\ Mod.\ Phys.\ B} \textbf{8}, 3795--3842 (1994).

\bibitem{FyodorovSavinSommers2005}
Y.~V.~Fyodorov, D.~V.~Savin, and H.-J.~Sommers,
``Scattering, reflection and impedance of waves in chaotic and
disordered systems with absorption,''
{\it J.\ Phys.\ A: Math.\ Gen.} \textbf{38}, 10731--10760 (2005),
doi:10.1088/0305-4470/38/49/017.

\bibitem{FyodorovSkvortsovTikhonov2024}
Y.~V.~Fyodorov, M.~A.~Skvortsov and K.~S.~Tikhonov,
``Resonances in a single-lead reflection from a disordered medium: $\sigma$-model approach,''
\emph{Ann. Phys.} \textbf{460}, 169568 (2024).

\bibitem{ChiralRBM}
Y.~V.~Fyodorov, The spectral edge of chiral random band matrices: an exactly solvable multichannel Dyson chain and the Painlev\'e~I edge,
{\it in preparation} (2026).

\bibitem{FLDO2025} Y.~V. Fyodorov, P.~Le Doussal, and A.~Ossipov, Large deviations of spectral determinants of matrix-valued random Schr\"odinger operators and Dyson Brownian motion in cubic potentials, {\it J. Math. Phys.} {\bf 67}, 103301 (2026) [arXiv:2511.00954]

\bibitem{FyodorovAD} Y.~V. Fyodorov, ``Stationary distribution of the cubic Dyson gas with explosions at arbitrary $\beta$ and finite $N$: Riesz cones, $\beta$-addition and Kontsevich integrals,'' {\it in preparation} (2026).

\bibitem{FyodorovMeibohm2026}
Y.~V.~Fyodorov and J.~Meibohm,
``Density of reflection resonances in one-dimensional disordered Schr\"odinger operators,''
{\it New J. Phys.} \textbf{28}, 034603 (2026).

\bibitem{PaperII} Y.~V. Fyodorov, ``The $N$-channel white-noise Schr\"odinger operator in all three Dyson classes. II.  Statistics of reflection, resonances and time delays,'' {\it in preparation} (2026).

\bibitem{FyodorovRes2026} Y.~V.~Fyodorov.
    Reflection Resonances in the One-Dimensional Anderson Localization: Finite-Length Statistics, Wigner Time Delay, and Boundary Eigenfunctions
{\it  arXiv:2608.15562}
    
%

\bibitem{Letter} Y.~V. Fyodorov and P.~Le Doussal, ``Exact spectral integrability of a finite-channel random Schr\"odinger operator: from fermionization of the Dyson--Riccati gas to Kontsevich integral,'' {\it arXiv:2610.27154}

\bibitem{ReflB2} Y.~V. Fyodorov, ``Reflection resonances and time delays in disordered $N$-channel wires: exact finite-$N$ solution of the absorbing class-A DMPK problem,'' {\it in preparation}.

\bibitem{GFGO2025} E.~Garcia-Failde, P.~Gregori, and K.~Osuga, ``Volumes of moduli spaces of bordered Klein surfaces,'' arXiv:2511.21986.

\bibitem{GertsenshteinVasiliev1959}
M.~E.~Gertsenshtein and V.~B.~Vasil'ev,
``Waveguides with random inhomogeneities and Brownian
motion in the Lobachevsky plane,''
{\it Theory Probab. Appl.} \textbf{4}, 391--398 (1959).


\bibitem{GuillaumeRemi2026}
P.~Ghosal and G.~Remy, Quantum unique ergodicity and the critical crossover for 1-d Gaussian band matrices,
{\it arXiv:2611.15085} (2026).

\bibitem{GouldenJackson1997} I.~P. Goulden and D.~M. Jackson, ``Maps in locally orientable surfaces and integrals over real symmetric surfaces,'' {\it Can. J. Math.} \textbf{49}, 865 (1997).


\bibitem{Halperin1965} B.~I. Halperin, Green's functions for a particle in a one-dimensional random potential, {\it Phys. Rev} \textbf{139}, A104--A117 (1965).

\bibitem{HaugeStovneng1989}
E.~H.~Hauge and J.~A.~St{\o}vneng,
``Tunneling times: a critical review,''
{\it Rev.\ Mod.\ Phys.} \textbf{61}, 917 (1989).

\bibitem{HuangMaGenackGopar2021}
Y.~Huang, X.~Ma, A.~Z.~Genack, and V.~A.~Gopar,
Characterizing random one-dimensional media with an
embedded reflector via scattered waves,
{\it Phys.~Rev.~B} {\bf 104}, 104204 (2021).

\bibitem{Hwang1998}
H.-K.~Hwang, ``On convergence rates in the central limit theorems for combinatorial structures,''
{\it European J.\ Combin.}, 329--343 (1998)

\bibitem{IvanovOstrovskySkvortsov2009}
D.~A.~Ivanov, P.~M.~Ostrovsky, and M.~A.~Skvortsov,
Correlations of the local density of states in
quasi-one-dimensional wires,
{\it Phys.~Rev.~B} {\bf 79}, 205108 (2009).

\bibitem{IvanovSkvortsovOstrovskyFominov2012}
D.~A.~Ivanov, M.~A.~Skvortsov, P.~M.~Ostrovsky,
and Ya.~V.~Fominov,
Hybridization of wave functions in one-dimensional
localization,
{\it Phys.~Rev.~B} {\bf 85}, 035109 (2012).

\bibitem{JoshiKitaev2001}
N.~Joshi and A.~V.~Kitaev, ``On Boutroux's tritronqu\'ee solutions of the first Painlev\'e equation,''
{\it Stud.\ Appl.\ Math.} \textbf{107}, 253--291 (2001)

\bibitem{Kapaev2004}
A.~A.~Kapaev,
``Quasi-linear Stokes phenomenon for the Painlev\'e first equation,''
{\it J.\ Phys.\ A: Math.\ Gen.} \textbf{37}, 11149--11167 (2004).

\bibitem{KKS}
M.~Kieburg, A.~B.~J.~Kuijlaars, and D.~Stivigny,
Singular value statistics of matrix products with truncated unitary matrices, {\it International Mathematics Research Notices} \textbf{2016}, 3392--3424 (2016). See Theorem 2.3 and Eq.~(2.8).

\bibitem{Kolokolov1993}
I.~V.~Kolokolov,
The method of functional integration for one-dimensional localization,
higher correlators, and the average current flowing in a mesoscopic ring
in an arbitrary magnetic field, {\it JETP} \textbf{76}, 1099--1109 (1993).

\bibitem{Kontsevich1992} M.~Kontsevich, Intersection theory on the moduli space of curves and the matrix Airy function, {\it Commun. Math. Phys.} \textbf{147}, 1--23 (1992).

\bibitem{KotaniSimon1988}
S.~Kotani and B.~Simon,
Stochastic Schr\"odinger operators and Jacobi matrices on the strip,
{\it Commun. Math. Phys.} \textbf{119}, 403--429 (1988).

\bibitem{LandoZvonkin2004}
S.~K.~Lando and A.~K.~Zvonkin,
{\it Graphs on Surfaces and Their Applications},
with an appendix by D.~B.~Zagier
(Springer, Berlin, 2004).

\bibitem{LifshitsGredeskulPastur} I.~M. Lifshits, S.~A. Gredeskul, and L.~A. Pastur, \emph{Introduction to the Theory of Disordered Systems} (Wiley, New York, 1988).

\bibitem{Lipatov1977} L.~N.~Lipatov, ``Divergence of the perturbation-theory series
and the quasi-classical theory,'' {\it Sov.\ Phys.\ JETP} \textbf{45}, 216--223 (1977).

\bibitem{LiuZou2024} D.-Z.~Liu and G.~Zou, Edge statistics for random band matrices, arXiv:2401.00492 (2024), revised version.

\bibitem{LiuZou2026} D.-Z.~Liu and G.~Zou, Edge universality for inhomogeneous random matrices II: Markov chain comparison and critical statistics, arXiv:2604.20215 (2026).

\bibitem{MarinoSchiappaWeiss2008} M.~Mari{\~n}o, R.~Schiappa, and M.~Weiss,
``Nonperturbative effects and the large-order
behavior of matrix models and topological strings,'' {\it Commun.\ Number Theory Phys.}\ \textbf{2},
349--419 (2008).

\bibitem{MelloPereyraKumar1988} P.~A. Mello, P.~Pereyra, and N.~Kumar, Macroscopic approach to multichannel disordered conductors, 
{\it Ann. Phys.} \textbf{181}, 290--317 (1988).

\bibitem{MenonYu2023}
G.~Menon and T.~Yu,
``Siegel Brownian motion,''
{\it Anal. Math. Phys.} \textbf{15}, Article~144 (2025).
arXiv:2309.04299.

\bibitem{MulaseWaldron2003}
M.~Mulase and A.~Waldron,
``Duality of Orthogonal and Symplectic Matrix
Integrals and Quaternionic Feynman Graphs,''
{\it Commun.\ Math.\ Phys.} \textbf{240},
553--586 (2003),
arXiv:math-ph/0206011.

\bibitem{Pena_etal_2014}
A.~Pe\~na, A.~Girschik, F.~Libisch, S.~Rotter,
and A.~A.~Chabanov,
The single-channel regime of transport through random media,
{\it Nat.~Commun.} {\bf 5}, 3488 (2014).

\bibitem{DLMFAiry}
NIST Digital Library of Mathematical Functions,
Chapter 9, Airy and Related Functions, Secs.~9.5, 9.10 and 9.11,
\url{https://dlmf.nist.gov/9}.

\bibitem{OsipovSommersZyczkowski2010}
V.~Al.~Osipov, H.-J.~Sommers and K.~\.Zyczkowski,
``Random Bures mixed states and the distribution
of their purity,'' {\it J. Phys. A: Math. Theor.}
\textbf{43}, 055302 (2010).


\bibitem{Ossipov2018} A.~Ossipov, Gelfand--Yaglom formula for functional determinants in higher dimensions,
{\it J. Phys. A: Math. Theor.} \textbf{51}, 495201 (2018).

\bibitem{Ossipov2018a}
A.~Ossipov,
``Scattering approach to Anderson localization,'' {\it Phys. Rev. Lett.} \textbf{121}, 076601 (2018).

\bibitem{PippengerSchleich2006} N.~Pippenger and K.~Schleich, ``Topological characteristics of random triangulated surfaces,'' 
{\it Random Struct. Algorithms} \textbf{28}, 247 (2006).

\bibitem{Pippenger2013}
N.~Pippenger,
``Random Cyclations,''
{\it Electron. J. Combin.} \textbf{20}(4),
P9 (2013).

\bibitem{SavinSommers2003}
D.~V.~Savin and H.-J.~Sommers,
``Delay times and reflection in chaotic cavities with absorption,''
{\it Phys.\ Rev.\ E} \textbf{68}, 036211 (2003).

\bibitem{ShiGenack2012}
Z.~Shi and A.~Z.~Genack,
Transmission eigenvalues and the bare conductance
in the crossover to Anderson localization,
{\it Phys.~Rev.~Lett.} {\bf 108}, 043901 (2012).

\bibitem{Silvestrov1997} P.~G. Silvestrov, Summing graphs for random band matrices, {\it Phys. Rev. E} \textbf{55}, 6419--6432 (1997)

\bibitem{Sodin2010} S.~Sodin, The spectral edge of some random band matrices, {\it Ann. Math.} (2) \textbf{172}, 2223--2251 (2010).

\bibitem{StanfordWitten2019}
D.~Stanford and E.~Witten,
``JT gravity and the ensembles of random matrix theory,''
arXiv:1907.03363 (2019).

\bibitem{Thouless1972} D.~J. Thouless, A relation between the density of states and range of localization for one dimensional random systems, {\it J. Phys. C}  \textbf{5}, 77--81 (1972).

\bibitem{TracyWidom1996}
C.~A.~Tracy and H.~Widom, ``On Orthogonal and Symplectic Matrix Ensembles,''
{\it Commun.\ Math.\ Phys.} \textbf{177}, 727--754 (1996).

\bibitem{WeberEtAl2024}
T.~Weber, J.~Tall, F.~Haneder, J.~D.~Urbina, and K.~Richter,
``Unorientable topological gravity and orthogonal random matrix
universality,''
{\it J.\ High Energy Phys.} \textbf{2024}(7), 267 (2024).


\bibitem{Widom1999}
H.~Widom, ``On the Relation Between Orthogonal, Symplectic and Unitary Matrix Ensembles,''
{\it J.\ Stat.\ Phys.} \textbf{94}, 347--364 (1999).

\bibitem{Winful2003}
H.~G.~Winful,
``Delay time and the Hartman effect in quantum tunneling,''
{\it Phys.\ Rev.\ Lett.} \textbf{91}, 260401 (2003).

\bibitem{Witten1991} E.~Witten, ``Two dimensional gravity and intersection theory on moduli space,'' {\it Surveys in Diff. Geom.} \textbf{1}, 243 (1991).

\bibitem{ZakharItkin}
M.~Kh.~Zakhar-Itkin,
``A multi-wave transmission line with random
inhomogeneities and Brownian motion in Siegel's circle,''
{\it Theory Probab. Appl.} \textbf{15}, 282--294 (1970).

\end{thebibliography}
\end{document}